\documentclass[review]{elsarticle}

\usepackage[colorlinks,citecolor=blue,linktoc=all,linkcolor=cyan]{hyperref}
\usepackage{graphicx}

\usepackage[T1]{fontenc}
\usepackage{dsfont}               
\usepackage{mathrsfs}             
\usepackage{slashed}              
\usepackage{amsmath}
\usepackage{amssymb}
\usepackage{amsbsy}
\usepackage{amsfonts}

\usepackage{graphicx,subfigure}

\DeclareMathOperator{\sech}{sech}

\numberwithin{equation}{section}
\numberwithin{table}{section}
\numberwithin{figure}{section}

\journal{Progress in Particle and Nuclear Physics}

\usepackage{titlesec}
\usepackage{sectsty}
\titleformat{\section}{\normalfont\Large\bfseries}{\thesection}{1em}{}
\titleformat{\subsection}{\normalfont\large\bfseries}{\thesubsection}{1em}{}
\titleformat{\subsubsection}{\normalfont\normalsize\bfseries}{\thesubsubsection}{1em}{}

\begin{document}
	
	\begin{frontmatter}
		
		\title{Hot QCD Phase Diagram From Holographic Einstein-Maxwell-Dilaton Models}

		\author[1staddress]{Romulo Rougemont\corref{mycorrespondingauthor}}
		\cortext[mycorrespondingauthor]{Corresponding author}
		\ead{rougemont@ufg.br}

		\author[2ndaddress]{Joaquin Grefa}
		\author[3rdaddress]{Mauricio Hippert}
		\author[3rdaddress]{Jorge Noronha}
		\author[3rdaddress]{Jacquelyn Noronha-Hostler}
		\author[2ndaddress]{Israel Portillo}
		\author[2ndaddress]{Claudia Ratti}
		
		\address[1staddress]{Instituto de F\'{i}sica, Universidade Federal de Goi\'{a}s, Av. Esperan\c{c}a - Campus Samambaia, CEP 74690-900, Goi\^{a}nia, Goi\'{a}s, Brazil}
		\address[2ndaddress]{Physics Department, University of Houston, Houston TX 77204, USA}
		\address[3rdaddress]{Illinois Center for Advanced Studies of the Universe, Department of Physics, University of Illinois at Urbana-Champaign, Urbana, IL 61801, USA}
		
		\begin{abstract}
		\hspace{0.42cm}  In this review, we provide an up-to-date account of quantitative bottom-up holographic descriptions of the strongly coupled quark-gluon plasma (QGP) produced in relativistic heavy-ion collisions, based on the class of gauge-gravity Einstein-Maxwell-Dilaton (EMD) effective models. 
  The holographic approach is employed to tentatively map the QCD phase diagram at finite temperature onto a dual theory of charged, asymptotically Anti-de Sitter (AdS) black holes living in five dimensions. With a quantitative focus on the hot QCD phase diagram, the nonconformal holographic EMD models reviewed here are  adjusted to describe first-principles lattice results for the finite-temperature QCD equation of state, with $2+1$ flavors and physical quark masses, at zero chemical potential and vanishing electromagnetic fields.  
  We review the evolution of such effective models and the corresponding improvements produced in quantitative holographic descriptions of the deconfined hot QGP phase of QCD.
The predictive power of holographic EMD models is tested by quantitatively comparing their predictions for the hot QCD equation of state at nonzero baryon density and the corresponding state-of-the-art lattice QCD results. Hydrodynamic transport coefficients such as the  shear and bulk viscosities predicted by these EMD constructions are also compared to the corresponding profiles favored by the latest phenomenological multistage models simultaneously describing different types of heavy-ion data.
We briefly report preliminary results from a Bayesian analysis using EMD models, which provide systematic evidence that lattice QCD results at finite temperature and \textit{zero} baryon density strongly constrains the free parameters of such bottom-up holographic constructions. Remarkably, the set of parameters constrained by lattice results at vanishing chemical potential turns out to produce EMD models in quantitative agreement with lattice QCD results also at finite baryon density.  
We also review 
results for equilibrium and transport properties from magnetic EMD models, which  effectively describe the hot and magnetized QGP at finite temperatures and magnetic fields with zero chemical potentials.   
Finally, we provide a critical assessment of the main limitations and drawbacks of the holographic models reviewed in the present work, and point out some perspectives we believe are of fundamental importance for future developments.
		\end{abstract}
		
		\begin{keyword}
			QCD phase diagram \sep critical point \sep quark-gluon plasma \sep gauge-gravity duality \sep equations of state
			
		\end{keyword}
		
	\end{frontmatter}
	
	\newpage
	
	\thispagestyle{empty}
	\tableofcontents
	


	\newpage
	\section{Introduction}
         \label{sec:intro}

\hspace{0.42cm} Quantum chromodynamics (QCD) is the quantum field theory (QFT) 
responsible for the 
sector of the standard model of particle physics associated with the strong interaction. At the most fundamental level, it comprises quarks and gluons (collectively called  \textit{partons}) as particles of the corresponding fermionic and non-Abelian gauge vector fields, respectively \cite{Peskin:1995ev,Schwartz:2014sze}. 
A rich and complex diversity of phases and regimes is possible for QCD matter, depending on the conditions to which partons are subjected \cite{Kogut:2004su,Alford:2007xm,Shuryak:2008eq,Shuryak:2014zxa}. These different regimes have been intensively investigated in the last five decades, conjuring simultaneous efforts from theory, experiments, astrophysical observations, and large computational simulations \cite{Gross:2022hyw,Dexheimer:2020zzs,Lovato:2022vgq,MUSES:2023hyz,Achenbach:2023pba,Arslandok:2023utm}.

At the microscopic level, QCD is fundamentally responsible for two of the most important aspects of ordinary baryonic matter in our universe, namely: i) the stability of nuclei due to the effective exchange of pions binding the nucleons (protons and neutrons), with the most fundamental interaction between the composite hadronic particles being mediated via gluon exchange between quarks; ii) 
most of its mass, thus generating the vast majority of the mass of ordinary matter in our universe, as a result of the dynamical breaking of chiral symmetry at low energies --- for instance, at low temperatures compared to the typical scale $T_c\sim 150$ MeV of the QCD deconfinement crossover transition at zero baryon density  \cite{Aoki:2006we,Borsanyi:2013bia,Borsanyi:2016ksw}. In fact, about $\gtrsim 98\%$ of the mass of the nucleons (and, consequently, also the mass of atoms and the ordinary macroscopic structures of the universe built upon them) comes from strong interactions, 
with the tiny rest being actually due to the current quark masses generated by the Higgs mechanism \cite{Peskin:1995ev,Schwartz:2014sze,Weinberg:1995mt,Weinberg:1996kr,Griffiths:2008zz}. Intrinsically related to the two aforementioned facts, QCD also presents what is called \textit{color confinement}, which generically refers to the fact that quarks and gluons, as degrees of freedom carrying color charge under the non-Abelian gauge group $SU(N_c=3)$ of QCD, are never observed in isolation as asymptotic states in experiments, being confined inside color-neutral hadrons \cite{Greensite:2011zz}.

Relying on various properties of QCD, we can determine its degrees of freedom at specific energy scales. 
Due to the number of colors, $N_c=3$, and quark flavors, $N_f=6$, QCD is an asymptotically free non-Abelian gauge theory \cite{Gross:1973id,Politzer:1973fx}. That is, the $\beta$-function for the QCD coupling constant is negative, implying that it is a decreasing function of the renormalization group energy scale, vanishing at asymptotically high energies. Conversely, QCD becomes a strongly coupled non-perturbative QFT at energy scales below or around the QCD dimensional transmutation scale, $\Lambda_\textrm{QCD}\sim 200$ MeV, indicating the failure of perturbative QFT methods when applied to low energy QCD phenomena (e.g. quark confinement). Indeed, due to quark confinement, one expects a hadron gas resonance (HRG) phase at low energies and temperatures, 
while, due to asymptotic freedom, a deconfined phase of quarks and gluons  
called the quark-gluon plasma (QGP) is expected at high energies. Because of its asymptotic freedom, the latter could naively be expected to be a weakly interacting medium. In fact, at high enough temperatures, as attained in the quark epoch (where the cosmic background radiation temperature varied from hundreds of GeV to hundreds of MeV within a time window of microseconds), and before the QCD phase transition in the early universe, the QGP was a weakly coupled fluid. As a clear comparison, hard thermal loop (HTL) perturbation theory in QCD seems to provide a reasonable description of some thermodynamic observables computed non-perturbatively in lattice QCD (LQCD) simulations for temperatures $T\gtrsim 300$ MeV \cite{Ghiglieri:2020dpq}. However, at temperatures below that approximate threshold, the agreement between perturbative QCD (pQCD) and non-perturbative LQCD results is generally lost, which approximately sets the temperature window $T_c \sim 150\,\, \textrm{MeV} < T < 2T_c \sim 300$ MeV (at zero baryon density) for which the QGP is a strongly coupled fluid \cite{Shuryak:2014zxa}. This is just within the range of temperatures probed by relativistic heavy-ion collision experiments conducted e.g. at the \textit{Relativistic Heavy Ion Collider} (RHIC) \cite{BRAHMS:2004adc,PHENIX:2004vcz,PHOBOS:2004zne,STAR:2005gfr} and at the \textit{Large Hadron Collider} (LHC) \cite{ALICE:2010suc,ATLAS:2010isq}.

         \subsection{Some phenomenological results from heavy-ion collisions}
         \label{sec:preHIC}

\hspace{0.42cm} The strongly coupled nature of the QGP produced in heavy-ion collisions is not only deduced from  thermodynamic observables but also from hydrodynamic transport coefficients. These coefficients are  typically inferred from the analysis of phenomenological models simultaneously describing several types of heavy-ion data \cite{Shuryak:2014zxa,Heinz:2013th,Luzum:2013yya,Ryu:2015vwa,Bernhard:2019bmu,JETSCAPE:2020shq}.

\begin{figure}[h]
\begin{centering}
\includegraphics[scale=0.7]{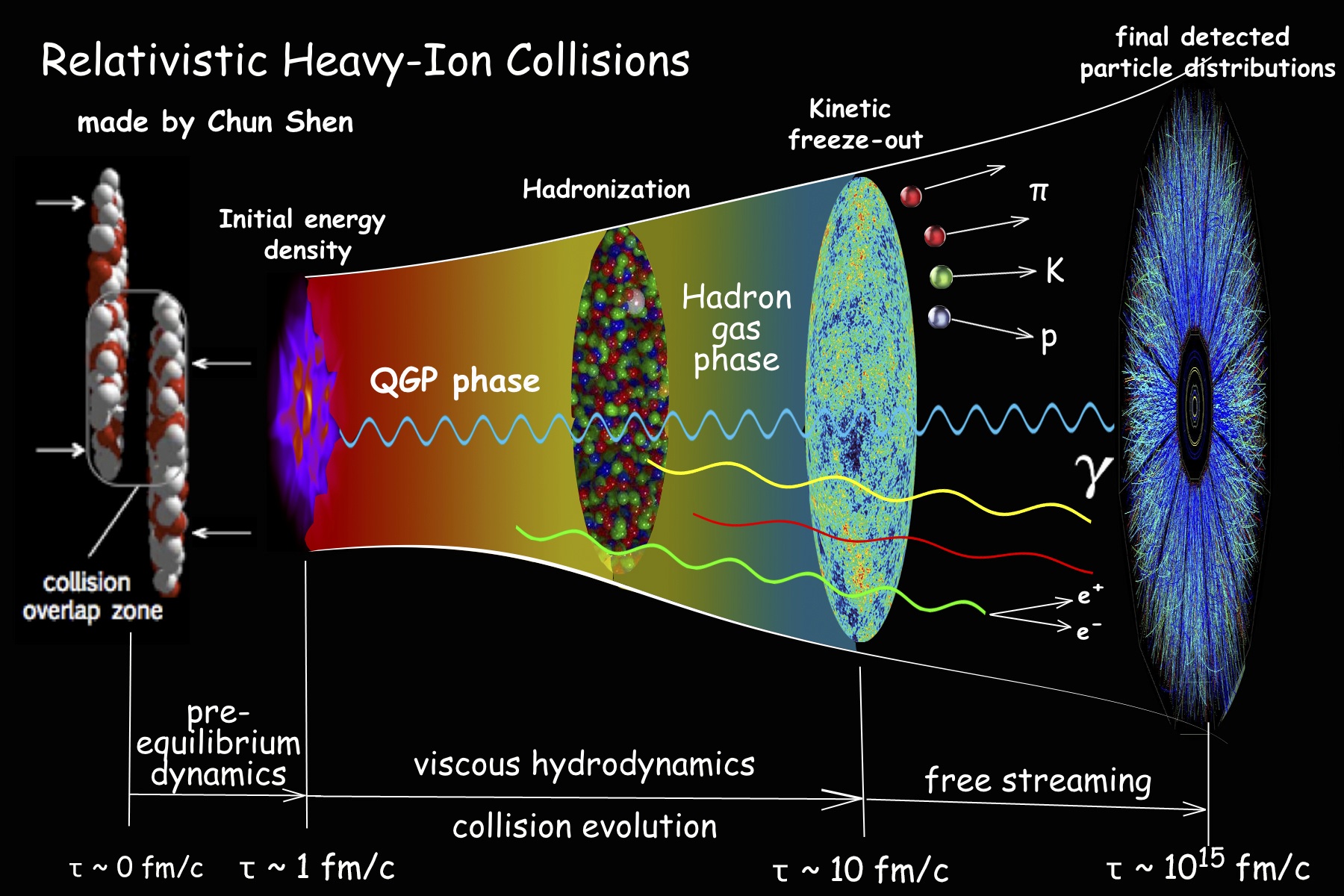}
\par\end{centering}
\caption{An artistic illustration (made by C.~Shen) regarding the expected evolution of the medium produced in relativistic heavy-ion collisions. From: \url{https://u.osu.edu/vishnu/2014/08/06/sketch-of-relativistic-heavy-ion-collisions/} 
\label{fig:HICs}}
\end{figure}

The hot and dense medium produced in relativistic heavy-ion collisions is commonly believed to pass through several different stages during its space and time evolution, as sketched in Fig. \ref{fig:HICs}. Initially, two heavy ions are accelerated to speeds close to the speed of light, and at very high energies, the gluon density inside those nuclei grows until reaching a saturation value, forming the so-called \textit{color glass condensate} (CGC) \cite{McLerran:2001sr,Iancu:2003xm,Weigert:2005us,Gelis:2010nm}, which is a typical source of initial conditions for the medium produced after the collision. For a characteristic time interval $\lesssim 1$ fm/c after the collision\footnote{Notice that 1 fm/c $\approx 3.33564\times 10^{-24}$ s, so that the characteristic time scales involved in heavy-ion collisions are extremely short.}, in the pre-equilibrium stage, the system is expected to be described by a turbulent medium composed by highly coherent gluons. Therefore, this stage is dominated by the dynamics of classical chromodynamic fields forming the so-called \textit{glasma}, a reference to the fact that this is an intermediate stage between the color \emph{glass} condensate and the quark-gluon \emph{plasma} \cite{Gelis:2012ri}. 
As the glasma expands and cools, it begins to decohere towards a state of QCD matter which possesses an effective description in terms of relativistic viscous hydrodynamics \cite{Teaney:2000cw,Kolb:2000sd,Florkowski:2017olj,Romatschke:2017ejr} and whose physically relevant degrees of freedom correspond to deconfined, but still strongly interacting quarks and gluons formed around $\gtrsim 1$ fm/c after the collision. 
As the QGP keeps expanding and cooling, it eventually hadronizes by entering into the QGP-HRG crossover region of the QCD phase diagram \cite{Aoki:2006we,Borsanyi:2016ksw}. The next stage of the space and time evolution of the system comprise the so-called chemical freeze-out \cite{Andronic:2005yp}, when inelastic collisions between the hadrons cease and the relative ratio between the different kinds of particles in the hadron gas is kept fixed. Afterwards, there is the thermal or kinetic freeze-out, when the average distance between the hadrons is large enough to make the short-range residual strong nuclear interaction between them effectively negligible. This fixes the momentum distribution of the hadrons. After that, the produced hadrons are almost free and the particles resulting from their decays reach the experimental detectors, providing information on the previous stages in the evolution of the system.

Of particular relevance for the topics to be approached in the present review are the shear, $\eta$, and bulk viscosities, $\zeta$. These hydrodynamic transport coefficients cannot be directly measured in heavy-ion collision experiments and are typically employed as free functions (of temperature and eventually also of other possible variables, such as chemical potentials and/or electromagnetic fields) in phenomenological hydrodynamic models, which are then fixed by comparison to heavy-ion data (for example, using Bayesian inference methods \cite{Bernhard:2019bmu,JETSCAPE:2020shq,Nijs:2020roc,Parkkila:2021tqq}).

\begin{figure}[h]
\begin{centering}
\includegraphics[scale=0.55]{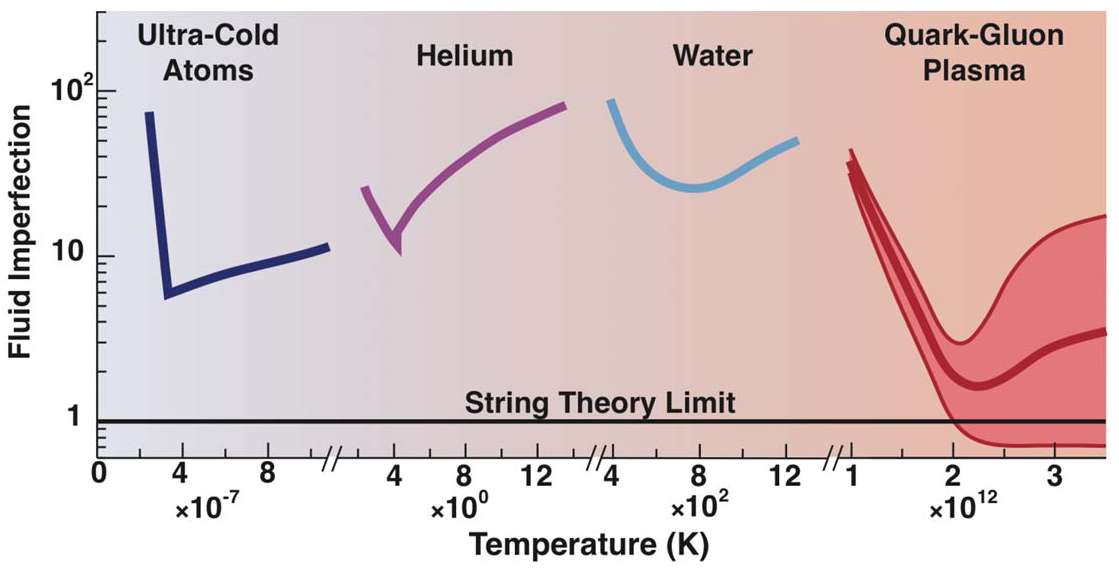}
\par\end{centering}
\caption{A schematic illustration comparing the rescaled specific shear viscosity --- the ``fluid imperfection'' index, $4\pi\eta/s$, --- for different fluids in nature. Notice also the different temperature scales on the horizontal axis: the QGP is the hottest\protect\footnotemark $\,$ and the less viscous (or almost perfect) fluid ever created in nature. The string theory limit refers to the  holographic benchmark value, $\eta/s=1/4\pi$. From: \href{https://www.asc.ohio-state.edu/physics/ntg/6805/readings/2013_NSAC_Implementing_the_2007_Long_Range_Plan.pdf}{R. Tribble (Chair), A. Burrows \textit{et al.}, \textit{Report to the Nuclear Science Advisory Committee, Implementing the 2007 Long Range Plan}, January 31, 2013.} 
\label{fig:shearcomp}}
\end{figure}

From such an approach, it is generally found that, around the QGP-HRG crossover region at zero baryon density in the QCD phase diagram, $\eta/s$ (where $s$ is the entropy density of the medium) should be of the same order of magnitude (in natural units with $c=\hbar=k_B=1$) of $1/4\pi$ (which, as we shall discuss in section \ref{sec:purpose}, is a benchmark value for strongly coupled quantum fluids coming from a very broad class of holographic models \cite{Kovtun:2004de,Policastro:2001yc,Buchel:2003tz}), being at least one order of magnitude smaller than perturbative calculations \cite{Arnold:2003zc,Ghiglieri:2018dib,Danhoni:2022xmt}. The small value of the shear viscosity to entropy density ratio, $\eta/s$, inferred for the QGP produced in heavy-ion collisions is physically interpreted as a clear manifestation of its nearly-perfect fluidity, as sketched in Fig. \ref{fig:shearcomp}.

\footnotetext{As a reference, in the QGP-HRG crossover window, where the QGP temperature is low enough to make the medium hadronize, $T_c\sim 150\,\textrm{MeV}\,\sim 1.72\times 10^{12}\,\textrm{K}\,\sim 10^5\, T_\textrm{center of sun}$ (see e.g. \href{https://solarscience.msfc.nasa.gov/interior.shtml}{NASA/Marshall Solar Physics}). In heavy-ion collisions realized in particle accelerators, the QGP attains temperatures at most 2 - 3 times $T_c$ while much higher temperatures were achieved in the early universe.}

\begin{figure}[h]
\begin{centering}
\includegraphics[scale=0.9]{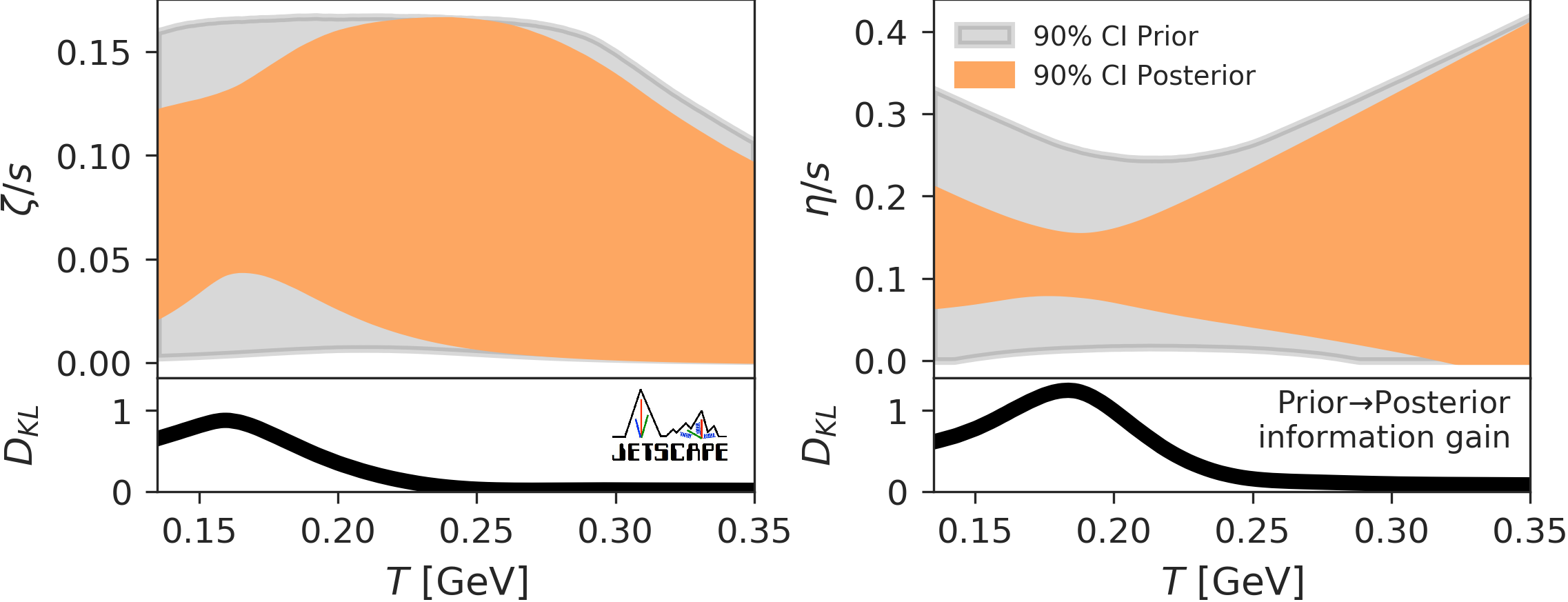}
\par\end{centering}
\caption{\textbf{From Ref. \cite{JETSCAPE:2020shq}.} The left and right plots display, respectively, the specific bulk and shear viscosities inferred for the QGP in heavy-ion collisions from model-to-data comparisons. The gray-shaded regions represent the $90\%$ credible intervals (CI) for the priors, and the orange-shaded regions describe the Bayesian model-averaged posteriors. Also shown is the corresponding information gain of the posteriors relative to the priors, associated with the Kullback-Leibler divergence, $D_\textrm{KL}$: the larger its value, the more the model results are constrained by the experimental data. \label{fig:bulkjetscape}}
\end{figure}

Besides $\eta/s$, also the bulk viscosity to entropy density ratio $\zeta/s$ plays a prominent role in the phenomenological description of heavy-ion data \cite{Noronha-Hostler:2013gga,Noronha-Hostler:2014dqa,Ryu:2015vwa}. For instance, in Ref. \cite{JETSCAPE:2020shq} the JETSCAPE Collaboration developed a state-of-the-art phenomenological multistage model for heavy-ion collisions, which was employed to simultaneously describe several hadronic measurements from different experiments at RHIC and LHC. Their results favor the temperature-dependent profiles (at zero baryon density) for $\zeta/s$ and $\eta/s$ shown in Fig. \ref{fig:bulkjetscape}. These phenomenological results for the hydrodynamic viscosities will be compared to quantitative microscopic holographic calculations and predictions in section \ref{sec:transport}.

By varying the conditions under which heavy-ion collisions take place in particle accelerators, it is possible to experimentally probe some aspects and regions of the QCD phase diagram at finite temperature and nonzero baryon density. For instance, for heavy-ion collisions at the LHC operating at the center of mass energies of $\sqrt{s_{\textrm{NN}}} = 2.76 - 5.02$ TeV, the energy of the collisions is so large that \emph{average} effects due to a nonzero baryon chemical potential $\mu_B$ become negligible (note that \emph{fluctuations} of conserved charges do still play a role at these energies \cite{Ratti:2018ksb,Pratt:2018ebf,Carzon:2019qja}). On the other hand, the Beam Energy Scan (BES) program at RHIC scans out lower collision energies spanning the interval $\sqrt{s_{\textrm{NN}}} = 7.7 - 200$ GeV \cite{STAR:2010vob}, where the baryon chemical potential reached within the QGP is of the same order of magnitude of the temperature, allowing experimental access to some regions of the QCD phase diagram at nonzero $\mu_B$. Furthermore, fixed-target experiments at RHIC \cite{Cebra:2014sxa,Meehan:2016qon,Meehan:2017cum}, and also experiments with lower collision energies at HADES \cite{HADES:2019auv}, and FAIR \cite{Tahir:2005zz,Friese:2006dj,PANDA:2009yku,Ablyazimov:2017guv,Durante:2019hzd,Almaalol:2022xwv},  aim at experimentally probing the structure of the QCD phase diagram in the $(T,\mu_B)$-plane at higher baryon densities. One of the main purposes of such experiments is to determine the location of the conjectured critical endpoint (CEP) of the line of first-order phase transition which, from several different model calculations, is expected to exist in the QCD phase diagram at high-baryon densities \cite{Stephanov:1998dy,Bzdak:2019pkr}.

         \subsection{Lattice QCD results}
         \label{sec:preLQCD}

\hspace{0.42cm} An important limitation of phenomenological multistage models is that 
several physical inputs are not calculated from  self-consistent microscopic models or systematic effective field theories. 
As mentioned above, these inputs can be constrained by experimental data (and some underlying phenomenological model assumptions). However, such a phenomenological approach cannot explain \textit{why and how} certain transport and equilibrium properties arise from QCD.

The strongly coupled nature of QCD at low energies renders the systematic methods of pQCD not applicable to describe a wide range of physically relevant phenomena that can be probed by experiments in high-energy particle accelerators 
and also by astrophysical observations. 
However, at vanishing or small chemical potentials $\mu_B$, 
another first-principles method for investigating equilibrium phenomena (such as the behavior of several thermodynamic observables) in QCD is available, namely, LQCD simulations.

The general reasoning behind this method, originally developed by Kenneth Wilson \cite{Wilson:1974sk}, amounts to discretizing the Euclidean, imaginary-time version of the background spacetime. Matter fields, such as the fermion fields of the quarks, are defined at the sites of the resulting discretized grid, while gauge fields, such as the gluons, are treated as link variables connecting neighboring sites \cite{Creutz:1983njd,Montvay:1994cy,DeGrand:2006zz,Gattringer:2010zz,Smit:2002ug,Rothe:1992nt}. The Euclidean path integral, defined in the imaginary-time Matsubara formalism for finite-temperature statistical systems, can then be performed using Monte Carlo methods. Continuum QCD can formally be recovered by taking the limit in which the lattice spacing between neighboring sites goes to zero. In practice, due to the large increase in the computational cost of numerical simulations with decreasing lattice spacing, the formal continuum limit is approached by extrapolating a sequence of calculations with progressively decreasing lattice spacings, which are nonetheless still large enough to be computationally manageable \cite{Ratti:2018ksb}.
Some very remarkable achievements of LQCD relevant to this review include the first principles calculation of light hadron masses, like pions and nucleons, compatible with experimental measurements \cite{Durr:2008zz}, and mainly the determination of the nature of the transition between the HRG and QGP phases of QCD at zero baryon density, which turns out to be a broad continuous crossover \cite{Aoki:2006we,Borsanyi:2016ksw}.

However, despite its notable successes, LQCD calculations also feature some important limitations, in particular: i) the difficulties in performing numerical simulations at nonzero baryon density, due to the so-called sign problem of lattice field theory \cite{deForcrand:2009zkb,Philipsen:2012nu}, and ii) the issues in calculating non-equilibrium transport observables associated with the real-time dynamics of the system. The former is an algorithmic issue that arises from the fermion determinant of the quarks becoming a complex quantity at real nonzero $\mu_B$, which implies that it cannot be employed to define a probabilistic measure to be used in importance sampling --- thus spoiling the direct evaluation of the LQCD path integral by means of Monte Carlo methods. The latter is due to difficulties in analytically continuing the Euclidean correlators calculated in the lattice at imaginary times to real-time intervals in  a spacetime with Minkowski signature \cite{Meyer:2011gj}.

\begin{figure}[h]
\begin{centering}
\includegraphics[scale=0.8]{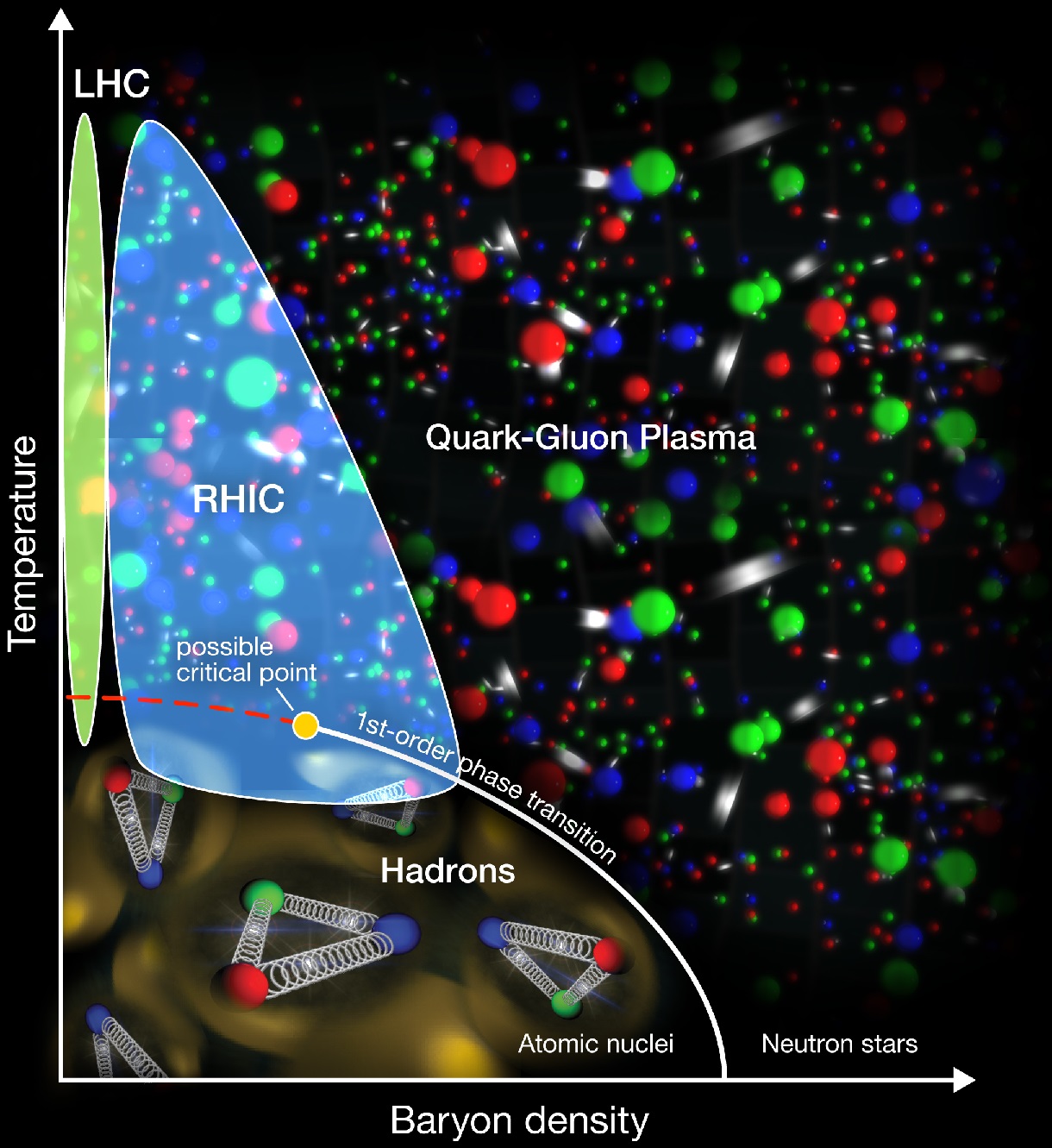}
\par\end{centering}
\caption{An artistic illustration of the QCD phase diagram at finite temperature and baryon density. From: \url{https://www.bnl.gov/newsroom/news.php?a=24473} (
Courtesy of Brookhaven National Laboratory). \label{fig:QCDphases}}
\end{figure}

Nonetheless, in recent years several different techniques have been developed and applied to calculate in LQCD the equation of state at finite temperature and moderate values of baryon chemical potential, and also to estimate the behavior of some transport coefficients at finite temperature and zero baryon density, as reviewed in Refs. \cite{Ratti:2018ksb,Ding:2015ona}. In fact, state-of-the-art lattice simulations for the continuum-extrapolated QCD equation of state with $2+1$ flavors and physical values of the quark masses are now available up to $\mu_B/T\le 3.5$ \cite{Borsanyi:2021sxv} from a novel expansion scheme, and up to $\mu_B/T\le 3$ from a traditional Taylor expansion \cite{Bollweg:2022fqq}. Some of these LQCD results for thermodynamic observables at finite $(T,\mu_B)$ will be compared to quantitative microscopic holographic calculations and predictions in section \ref{sec:EoS}.

         \subsection{Some basic aspects of the holographic gauge-gravity duality}
         \label{sec:preholo}

\hspace{0.42cm} The  limitations of present-day lattice simulations mentioned above prevent first-principles QCD calculations to be employed in the investigation of strongly interacting QCD matter at higher baryon densities, where an actual phase transition 
between confining hadronic and deconfined partonic degrees of freedom may exist, as depicted in the sketch displayed in Fig. \ref{fig:QCDphases}. Also, LQCD simulations of QCD transport properties are considerably difficult already at $\mu_B=0$ \cite{Meyer:2011gj}, let alone at finite baryon density. In such cases, it is customary to resort to effective models and other alternative theoretical approaches to obtain some qualitative insight and even some quantitative predictions for the behavior of QCD matter under such extreme conditions.

One such alternative approach, which is the theoretical tool considered in the present review, is what is broadly called the \textit{holographic gauge-gravity duality} (also known, under more restricted conditions, as the \textit{AdS-CFT correspondence}) \cite{Maldacena:1997re,Gubser:1998bc,Witten:1998qj,Witten:1998zw}. The holographic gauge-gravity duality is motivated by the framework of string theory, which originally had an old and curious relationship with the strong interaction. Indeed, (non-supersymmetric) string theory was originally developed as an S-matrix theory for the strong nuclear force between hadrons, which were empirically known to fall into linear Regge trajectories relating their total angular momentum $J$ to their mass squared $m^2$, in what is known as the Chew-Frautschi plots \cite{Chew:1961ev}. By modeling a meson as a relativistic open string spinning around its center, it is possible to reproduce the observed Chew-Frautschi relation, $J=\alpha_0+\alpha'm^2$, where the relativistic string tension is given in terms of the measured slope of the linear Regge trajectory, $\sigma=\left(2\pi\alpha'\right)^{-1}\approx \left(440\,\textrm{MeV}\right)^2$ \cite{Greensite:2011zz}. The slope is approximately the same for the different Regge trajectories defined by the different measured values of the Regge intercept, $\alpha_0$ (which is known to depend on the flavor quantum numbers of the hadrons considered --- hadrons with the same flavor quantum numbers fall into the same Regge trajectory, and can be viewed as resonances of this trajectory with different values of mass and angular momentum). However, since this simple string model also predicts results in striking contradiction with hadronic experiments (e.g. a wrong, soft exponential falloff for the associated Veneziano scattering amplitude in the high energy limit of hard scattering for hadrons at fixed angles), it has been abandoned as a model for hadrons, being superseded by the advent of QCD, with its theoretical and experimental successes as the fundamental description of the strong interaction.

Later, the theoretical interest in string theory greatly resurfaced, although within a very different context, with the so-called first and second superstring revolutions, which correspond, respectively: 1) to the discovery of five different consistent supersymmetric quantum string theories in 10 spacetime dimensions (superstring theories of Type I, Type IIA, Type IIB, Heterotic $SO(32)$ and Heterotic $E_8\otimes E_8$); and also, 2) the latter discovery that these five superstring theories in 10 dimensions are related through a web of duality transformations, besides being also related to a theory of membranes defined in 11 spacetime dimensions called M-theory, whose low energy limit corresponds to a unique 11-dimensional theory of supergravity. A remarkable common feature of all superstring theories is that all of them possess a tensorial spin 2 massless particle in their spectrum, which is the graviton, the hypothetic vibrational string mode responsible for mediating the gravitational interaction at the quantum level. Due to that reason, and also due to the fundamental fact that at low energies superstring reduces to supergravity, therefore containing general relativity as the low energy, classical description of gravity, superstring theory is an interesting candidate for a theory of quantum gravity \cite{Green:1987sp,Green:1987mn,Polchinski:1998rq,Polchinski:1998rr,Becker:2006dvp}. There is also some expectation that the standard model would emerge as a low-energy sector in string theory with 6 of its 10 dimensions compactified in some appropriate manifold, which should be chosen in a very specific way in order to  generate the observed phenomenology of particle physics in our universe. This way, string theory could be seen as a ``theory of everything'', in the sense of possibly describing all the particles and fundamental interactions in nature.

Regardless of whether string theory is the unifying theory of all the fundamental interactions of nature \cite{Vafa:2005ui,Obied:2018sgi,Aoki:2021ckh} or not, it is undeniable that new effective approaches and applications, directly inspired by string theory and aimed towards the strong interaction, flourished with the advent of the holographic gauge-gravity duality. Before discussing some of their phenomenological aspects in regard to the physics of the hot and baryon dense strongly-coupled QGP in section \ref{sec:holo}, we discuss below some basic general aspects of the holographic correspondence.

The original formulation of the so-called AdS-CFT correspondence \cite{Maldacena:1997re,Gubser:1998bc,Witten:1998qj,Witten:1998zw}, relates Type IIB superstring theory defined on the product manifold between a 5-dimensional Anti-de Sitter (AdS) spacetime and a 5-dimensional sphere, AdS$_5\otimes S^5$, to a conformal quantum field theory (CFT) corresponding to $\mathcal{N}=4$ Supersymmetric Yang-Mills (SYM) theory with gauge group $SU(N_c)$,\footnote{$\mathcal{N}=4$ refers to the number of different supersymmetries of the theory.} defined on the conformally flat 4-dimensional boundary of AdS$_5$. Two other early realizations of the AdS-CFT duality comprise also the relation between M-theory defined on AdS$_4\otimes S^7$ and the Aharony-Bergman-Jafferis-Maldacena (ABJM) superconformal field theory defined on the 3-dimensional boundary of AdS$_4$, besides the relation between M-theory defined on AdS$_7\otimes S^4$ and the so-called $6D\,\, (2,0)$ superconformal field theory defined on the 6-dimensional boundary of AdS$_7$. In a very naive and imprecise way, one could in principle think of the first example of the $\mathcal{N}=4$ SYM theory as a ``toy model'' for QCD, while the second example regarding the ABJM theory could be taken as a ``toy model'' for low-dimensional condensed matter systems. However, this is inadequate from a realistic phenomenological perspective, both at the quantitative and qualitative levels, as we shall discuss in section \ref{sec:purpose}.

Before doing that, let us first comment a little bit more on the original proposal (see e.g. the discussion in section 3 of the standard review \cite{Aharony:1999ti}, and also other works such as \cite{Petersen:1999zh,Nastase:2007kj,Natsuume:2014sfa,Ramallo:2013bua} for details). We take for definiteness the example relating Type IIB superstring theory compactified on AdS$_5\otimes S^5$ and $\mathcal{N}=4$ SYM theory living on the boundary of AdS$_5$. One first considers Type IIB string theory in flat $\mathbb{R}^{1,9}$ Minkowski spacetime and a collection of $N_c$ coincident parallel D3-branes in this background.\footnote{An endpoint of an open string must satisfy either Dirichlet or Neumann boundary conditions. If one considers Neumann boundary conditions on $p$ spatial dimensions plus time, then the remaining $D-p-1$ dimensions must satisfy Dirichlet boundary conditions. Since for Dirichlet boundary conditions a string endpoint is fixed in space, while for Neumann boundary conditions it must move at the speed of light, then with Neumann boundary conditions on $p+1$ dimensions, the open string endpoints are constrained to move within a $(p+1)$-dimensional hypersurface, which is a dynamical object called Dp-brane. Dp-branes are shown to be related to black p-branes \cite{Polchinski:1995mt,Polchinski:1996na}, which are solutions of higher dimensional (super)gravity which generalize the concept of black holes by having extended event horizons which are translationally invariant through $p$ spatial dimensions. They actually provide different descriptions of a single object, which in a perturbative string regime is accurately described by Dp-branes not backreacting on the background spacetime, while at low energies (corresponding to take $\alpha'\equiv l_s^2$ to be small, where $l_s$ is the fundamental string length, so that massive string states can be neglected) and large gravitational fields, the backreaction of the Dp-branes on the background produces a black p-brane geometry \cite{Polchinski:2014mva}.} The perturbative string theory excitations in this system correspond to vibrational modes of both, closed strings, and also open strings with their ends attached to the D3-branes. If we consider the system defined at low energies compared to the characteristic string scale, $\left(\alpha'\right)^{-1/2}\equiv\left(l_s\right)^{-1}$, only massless string modes can be excited which, for closed strings give a gravity supermultiplet and, for the open strings with their ends attached to the $(3+1)$-dimensional worldvolume of the $N_c$ coincident D3-branes, give a $\mathcal{N}=4$ vector supermultiplet with gauge group $SU(N_c)$. A low energy effective action for these massless string excitations in the background considered can be schematically written by integrating out the massive string modes,
\begin{align}
S_\textrm{eff} = S_{\mathbb{R}^{1,9}\,\textrm{bulk}} + S_{\mathbb{R}^{1,3}\,\textrm{brane}} + S_\textrm{int},
\end{align}
where $S_{\mathbb{R}^{1,9}\,\textrm{bulk}}$ is the low energy action for the gravity supermultiplet, corresponding to Type IIB supergravity (SUGRA) in $\mathbb{R}^{1,9}$ plus higher order derivative corrections coming from the integration of the string massive modes; $S_{\mathbb{R}^{1,3}\,\textrm{brane}}$ is the low energy action for the $\mathcal{N}=4$ vector supermultiplet living on the $\mathbb{R}^{1,3}$ worldvolume of the $N_c$ coincident D3-branes, corresponding to $\mathcal{N}=4$ SYM theory with gauge group $SU(N_c)$ plus higher order derivative corrections coming from the integration of the string massive modes; and $S_\textrm{int}$ is an interaction term between the bulk and brane modes.

The higher order derivative corrections for the bulk and brane actions coming from the integration of massive string modes are proportional to positive powers of $\alpha'$, while the interaction action is proportional to positive powers of the square root of the $10D$ Newton's gravitational constant, $\kappa_{10}\equiv\sqrt{8\pi G_{10}}\sim g_s\alpha'\,^2$, where $g_s$ is the string coupling, so that by considering the so-called decoupling limit where $\alpha'\equiv l_s^2\to 0$ with fixed $N_c,g_s$, one has $S_{\mathbb{R}^{1,9}\,\textrm{bulk}}\to S_{\mathbb{R}^{1,9}\,\textrm{IIB SUGRA}}$, $S_{\mathbb{R}^{1,3}\,\textrm{brane}}\to S_{\mathbb{R}^{1,3}\,\mathcal{N}=4\,\,\textrm{SYM}}$, and $S_\textrm{int}\to 0$, so that we end up with two decoupled actions,
\begin{align}
\lim_{\alpha'\to 0\,(\textrm{fixed}\, N_c,g_s)} S_\textrm{eff} = S_{\mathbb{R}^{1,9}\,\textrm{IIB SUGRA}} + S_{\mathbb{R}^{1,3}\,\mathcal{N}=4\,\,\textrm{SYM}}.
\end{align}
For a given number $N_c$ of coincident D3-branes, the `t Hooft coupling effectively controlling the strength of the interactions in the $\mathcal{N}=4$ SYM $SU(N_c)$ gauge theory is given by $\lambda_t\equiv N_c g_\textrm{SYM}^2= N_c g_s$.\footnote{The relation $g_\textrm{SYM}^2= g_s$ can be inferred from the fact that a closed string, governed by the $g_s$ coupling, can be formed from the collision between the endpoints of two open strings moving on the D3-branes, with $g_\textrm{SYM}$ being the coupling of the non-Abelian gauge field corresponding to the massless mode of the open strings on these branes \cite{Nastase:2007kj}.} This picture holds for any value of $\lambda_t$ (and since the SYM theory is a CFT, its `t Hooft coupling remains constant for any value of energy so that one actually has infinitely many different SYM theories, each one of them defined at some given value of $\lambda_t$).

Another perspective for the same system can be considered as follows. The effective gravitational field generated by the collection of $N_c$ coincident D3-branes is $\sim N_c g_s (l_s/r)^4$ \cite{Natsuume:2014sfa,Ramallo:2013bua}, and by considering a very large $N_c$ such that $\lambda_t = N_c g_s\gg 1$ even for small values of $g_s$ (so that one can ignore quantum string loop contributions in the bulk), very close to the D3-branes for $r\to 0$ the gravitational field is very intense and its backreaction on the background spacetime highly distorts its geometry, producing a curved manifold. In this limit it is necessary to replace the perturbative string description of D3-branes in flat Minkowski spacetime with the associated black 3-brane supergravity solution, whose near-horizon (i.e. near-black brane) geometry approaches precisely that of AdS$_5(L)\otimes S^5(L)$, with the same curvature radius $L$ for the AdS$_5$ and $S^5$ manifolds.\footnote{For the other two early examples of the AdS-CFT correspondence mentioned before, one obtains: AdS$_4(L/2)\otimes S^7(L)$ and AdS$_7(2L)\otimes S^4(L)$ (see e.g. \cite{Petersen:1999zh}).} On the other hand, far away from the black brane the background geometry is still that of Minkowski $\mathbb{R}^{1,9}$. In both regions (near and far from the black brane), since we considered that the string coupling $g_s$ is small (so that string loops may be discarded), by taking the decoupling limit as before, with $l_s\to 0$ and fixed $N_c,g_s$, the bulk spacetime is inhabited only by Type IIB SUGRA fields.

By comparing the two perspectives above for the same system, when defined in the same regime corresponding to low energies, low string coupling, large $N_c$, and strong `t Hooft coupling ($\alpha'\equiv l_s^2\to 0$ with fixed $N_c,g_s$, but such that $g_s$ is small, $N_c$ is large and $\lambda_t = N_c g_\textrm{SYM}^2 = N_c g_s\gg 1$), one notices that in both views there is a common element, which is Type IIB SUGRA defined on $\mathbb{R}^{1,9}$, and it is then conjectured that the remaining pieces in each perspective should be dual to each other: strongly coupled, large $N_c$, $\mathcal{N}=4$ SYM theory with gauge group $SU(N_c)$, defined on $\mathbb{R}^{1,3}$ (which is equivalent, up to a conformal factor, to the boundary of AdS$_5$), and classical, weakly coupled Type IIB SUGRA defined on AdS$_5(L)\otimes S^5(L)$. The duality involved in this comparison actually conveys a detailed mathematical dictionary translating the evaluation of physical observables in a classical SUGRA theory defined at weak coupling on top of a background given by the product of an AdS spacetime and a compact manifold, to the calculation of other observables in a different, conformal quantum gauge field theory defined at strong coupling and with a large number of colors on top of the conformally flat boundary of the AdS manifold. Then, the notion of the hologram comprised in the AdS-CFT duality refers to the fact that the gravitational information of a higher dimensional bulk spacetime can be encoded in its boundary.

This is the weakest form of the holographic AdS-CFT correspondence, and a particular case of the broader gauge-gravity duality, being largely supported by a plethora of independent consistency checks (see e.g. \cite{Aharony:1999ti,Petersen:1999zh,Nastase:2007kj,Natsuume:2014sfa,Ramallo:2013bua}). The strongest version of the AdS-CFT conjecture, corresponding to a particular case of the so-called gauge-string duality (which is more general than the gauge-gravity duality, which can be seen as a low-energy limit of the latter), proposes that the duality should be valid for all values of $g_s$ and $N_c$, therefore relating $\mathcal{N}=4$ SYM theory on $\mathbb{R}^{1,3}$ with arbitrary `t Hooft coupling and an arbitrary number of colors for the gauge group $SU(N_c)$, and full quantum Type IIB superstring theory generally formulated in a nonperturbative way on AdS$_5(L)\otimes S^5(L)$ (instead of just its classical low energy limit corresponding to Type IIB SUGRA). It is also posited that high derivative/curvature corrections in the bulk correspond to the inverse of `t Hooft coupling corrections in the dual CFT, since according to the detailed holographic dictionary, $\alpha'/L^2=\left\{l_s/\left[l_s \left(N_c g_s\right)^{1/4}\right]\right\}^2 =1/\sqrt{\lambda_t}$, and that quantum string loop corrections in the bulk correspond to the inverse of $N_c$ corrections in the dual CFT, since, $g_s \left(l_s/L\right)^4 = g_s \left(l_s/\left[l_s \left(N_c g_s\right)^{1/4}\right]\right)^4 = 1/N_c$.

The conjectured holographic AdS-CFT duality has a very clear attractive feature, which is the fact that complicated nonperturbative calculations in a strongly coupled quantum CFT can be translated, through the detailed mathematical holographic dictionary, into much simpler (although not necessarily easy) calculations involving weakly coupled classical gravity in higher dimensions.

More generally, the broader holographic gauge-gravity duality\footnote{The even broader gauge-string duality is very difficult to handle in practice, due to the present lack of a detailed and fully nonperturbative definition of string theory on asymptotically AdS spacetimes. Consequently, we focus in this review only on its low-energy manifestation corresponding to the gauge-gravity duality, which is the framework where the vast majority of the calculations are done in the literature regarding the holographic correspondence.} is not restricted to bulk AdS spacetimes and dual boundary CFTs. Indeed, for instance, by considering the backreaction of effective $5D$ massive fields living on AdS$_5$, which are associated with the Kaluza-Klein (KK) reduction on $S^5$ of the originally $10D$ massless modes of SUGRA, the background AdS$_5$ metric is generally deformed within the bulk, and the effective $5D$ bulk spacetime geometry becomes just asymptotically AdS, with the metric of AdS$_5$ being recovered asymptotically near the boundary of the bulk spacetime. Generally, there is also a corresponding deformation of the dual QFT theory at the boundary of the asymptotically AdS spacetime induced by the consideration of relevant or marginal operators, which may break conformal symmetry and supersymmetry and whose scaling dimension is associated through the holographic dictionary to the masses of the effective $5D$ bulk fields. In this sense, one has a broader holographic gauge-gravity duality relating a strongly coupled QFT (not necessarily conformal or supersymmetric) living at the boundary of a higher dimensional asymptotically AdS spacetime, whose geometry is dynamically determined by a classical gravity theory interacting with different matter fields in the bulk. In the holographic gauge-gravity duality, the extra dimension connecting the bulk asymptotically AdS spacetime to its boundary plays the role of a geometrization of the energy scale of the renormalization group flow in the QFT living at the boundary \cite{deBoer:1999tgo}, with low/high energy processes in the QFT being mapped into the deep interior/near-boundary regions of the bulk spacetime, respectively.

Since its original proposal by Maldacena in 1997 \cite{Maldacena:1997re}, the holographic gauge-gravity duality has established itself as one of the major breakthroughs in theoretical physics in the last few decades, being applied to obtain several insights into the nonperturbative physics of different strongly coupled quantum systems, comprising studies in the context of the strong interaction \cite{Erdmenger:2007cm,Brodsky:2014yha,CasalderreySolana:2011us,Adams:2012th,Kim:2012ey,DeWolfe:2013cua,Hoyos:2021uff}, condensed matter systems \cite{Hartnoll:2009sz,Herzog:2009xv,McGreevy:2009xe,Horowitz:2010gk,Sachdev:2011wg,Cai:2015cya} and, more recently, also quantum entanglement and information theory \cite{Nishioka:2009un,VanRaamsdonk:2016exw,Chen:2021lnq,Chapman:2021jbh}.

         \subsection{Main purpose of this review}
         \label{sec:purpose}

\hspace{0.42cm} Holographic gauge-gravity models are generally classified as being either i) \emph{top-down} constructions when the bulk supergravity action comes from known low-energy solutions of superstrings and the associated holographic dual at the boundary is precisely determined, ii) or \emph{bottom-up} constructions when the bulk effective action is generally constructed by using phenomenological inputs and considerations with the purpose of obtaining a closer description of different aspects of some real-world physical systems, but the exact holographic dual, in this case, is not precisely known. Actually, for bottom-up holographic models, one assumes or conjectures that the main aspects of the gauge-gravity dictionary inferred from top-down constructions remain valid under general circumstances, such that for a given asymptotically AdS solution of Einstein field equations coupled to other fields in the bulk, some definite holographic dual QFT state at the boundary should exist.\footnote{This putative bottom-up holographic dual does not need to (and generally will not) coincide with the exact QFT taken as a target to be described in the real world. Instead, one will generally obtain some holographic dual of a QFT which is close to some aspects of the target QFT, but which differs from the latter in many other regards. In a general sense, this is not different, for instance, from the reasoning employed to construct several non-holographic effective models for QCD, where a given effective model is used to produce approximate results for some but not all aspects of QCD. In fact, if an exact holographic dual of real-word QCD (with gauge group $SU(3)$, 6 flavors and physical values of the quark masses) does exist, its dual bulk formulation will likely comprise not merely a gravity dual, but instead some complicated nonperturbative full string dual whose formulation is currently unknown.} In order to be useful in practice for different phenomenological purposes, such an assumption for bottom-up holographic models should provide explicit examples where the target phenomenology is indeed well reproduced by the considered bulk gravity actions, which should furthermore be able to provide new and testable predictions. In fact, as we are going to discuss in this review, one can construct holographic bottom-up models which are able to provide quantitative results and predictions in compatibility with first principles LQCD simulations and with some phenomenological outputs inferred from heavy-ion collisions, besides providing new predictions for thermodynamic and transport quantities in regions of the QCD phase diagram currently not amenable to first principles analysis due to the limitations discussed in the preceding sections.

Let us first analyze thermal SYM theory\footnote{That the SYM theory is completely inadequate as a holographic model for the confined phase of QCD is immediately obvious from e.g. the fact that SYM is a CFT and QCD is a nonconformal QFT with a mass gap in the spectrum. Even if one considers a comparison of SYM with just pure YM theory (i.e. the pure gluon sector of QCD without dynamical quarks), the issues remain since YM features linear confinement between static, infinitely heavy probe quarks (corresponding to an area law for the Wilson loop \cite{Greensite:2011zz}) and a mass gap in the spectrum.} as a possible ``proxy'' for the strongly coupled deconfined QGP, as it has been commonly considered within a considerable part of the holographic literature for years. It is often said that SYM theory has some qualitative features in common with QCD at the typical temperatures attained by the QGP in heavy-ion collisions, namely: within the considered temperature window, both theories are strongly coupled, deconfined, with non-Abelian vector fields corresponding to gluons transforming in the adjoint representation of the gauge group, and their $\eta/s$ have comparable magnitude.

Although the points above are true, they are insufficient to establish a reliable connection between SYM and QCD. Indeed, there are infinitely many different holographic theories with the same properties listed above. In fact, all gauge-gravity duals are strongly coupled and all isotropic and translationally invariant Einstein's\footnote{That is, with the kinetic term for the metric field in the bulk action given by the usual Einstein-Hilbert term with two derivatives.} gauge-gravity duals have a specific shear viscosity given by the ``(quasi)universal holographic'' result $\eta/s=1/4\pi$ \cite{Kovtun:2004de,Policastro:2001yc,Buchel:2003tz}, which is actually a clear indication that even for nonconformal gauge-gravity duals with running coupling (which is not the case of SYM theory, since it is a CFT), the effective coupling of the holographic theory remains large at all temperature scales. Consequently, classical gauge-gravity duals lack asymptotic freedom, featuring instead a strongly coupled ultraviolet fixed point, being asymptotic safe but not asymptotic free. Moreover, there are infinitely many different holographic duals with deconfined phases at high temperatures. In the face of this infinite degeneracy of holographic gauge-gravity duals with the very same generic features often employed to ``justify'' the use of SYM theory as a ``proxy'' for the QGP, one may be led to conclude that such a choice is not well-defined. One may argue that this choice is  more related to the fact that SYM theory is the most well-known and one of the simplest examples of gauge-gravity duality, than to any realistic phenomenological connection between the SYM plasma and the real-world QGP.

In order to take steps towards lifting the  infinite degeneracy of holographic models to describe (some aspects of) the actual QGP, one needs to look at the behavior of more physical observables than just $\eta/s$. In this regard, the SYM plasma is easily discarded as a viable phenomenological holographic model for the QGP due to several reasons, among which we mention mainly the following. The SYM plasma is a CFT, while the QGP is highly nonconformal within the window of temperatures probed by heavy-ion collisions, and this fact makes the equation of state for the SYM plasma completely different from the one obtained for the QGP in LQCD simulations, not only quantitatively, but also qualitatively \cite{Rougemont:2016etk}. Indeed, dimensionless ratios for thermodynamic observables such as the normalized pressure ($P/T^4$), energy density ($\epsilon/T^4$), entropy density ($s/T^3$), the speed of sound squared ($c_s^2$), and the trace anomaly ($I/T^4=(\epsilon-3P)/T^4$, which is identically zero for a CFT), are all given by constants in the SYM plasma, while they display nontrivial behavior as functions of the temperature in the QGP. Furthermore, the bulk viscosity vanishes for the conformal SYM plasma, while it is expected to possess nontrivial behavior as a function of the temperature in the QGP, playing an important role in the description of heavy-ion data, as inferred from phenomenological multistage models (see the discussion in section \ref{sec:preHIC} and Fig. \ref{fig:bulkjetscape}). Therefore, when considering  thermodynamic equilibrium observables and  transport coefficients, the SYM plasma is not a realistic model for the QGP both at the quantitative and qualitative levels.

On the other hand, the holographic duality \emph{can} be indeed employed to construct effective gauge-gravity models which make it possible to actually \emph{calculate} several thermodynamic and transport observables, displaying remarkable quantitative agreement with state-of-the-art LQCD simulations at zero and finite baryon density, while simultaneously possessing transport properties very close to those inferred in state-of-the-art phenomenological multistage models for heavy-ion collisions. Additionally, such holographic models also provide quantitative predictions for the QGP in regions of the QCD phase diagram which are currently out of the reach of first-principles calculations. The main purpose of the present paper is to review these results, mainly obtained through specific bottom-up constructions engineered within the so-called Einstein-Maxwell-Dilaton class of holographic models, discussing the main reasoning involved in their formulation, and also pointing out their phenomenological limitations and drawbacks, in addition to their successful achievements. This will be done in the course of the next sections, with holographic applications to the hot and baryon dense strongly coupled QGP being discussed in section \ref{sec:holo}. We will also review some applications to the hot and magnetized QGP (at zero chemical potential) in section \ref{sec:magnetic}. In the concluding section \ref{sec:outlook}, we provide an overview of the main points discussed through this review and list important perspectives for the future of phenomenological holographic model applications to the physics of the QGP.

In this review, unless otherwise stated, we make use of natural units where $c=\hbar=k_B=1$, and adopt a mostly plus metric signature.

         \newpage
	\section{Holographic models for the hot and baryon dense quark-gluon plasma}
         \label{sec:holo}

\hspace{0.42cm} In this section, we review the construction and the main results obtained from phenomenologically-oriented bottom-up holographic models aimed at a quantitative description of the strongly coupled QGP at finite temperature and baryon density. We focus on a class of holographic constructions called Einstein-Maxwell-Dilaton (EMD) gauge-gravity models, which has provided up to now the best quantitative holographic models for describing equilibrium thermodynamic and hydrodynamic transport properties of the hot and baryon dense QGP produced in heavy-ion collisions. We also discuss different predictions for the structure of the QCD phase diagram, comprising at high baryon chemical potential a line of first-order phase transition ending at a CEP, which separates the phase transition line from the smooth crossover observed at low baryon densities.

	\subsection{Holographic Einstein-Maxwell-Dilaton models}
         \label{sec:EMD}

\hspace{0.42cm} In order to possibly obtain a quantitative holographic model for the QGP (and also quantitative holographic constructions for other strongly coupled physical systems in the real world), one necessarily needs to break conformal symmetry in the holographic setting. However, breaking conformal symmetry alone is not sufficient to reproduce several QCD results, since one needs to obtain a holographic modeling of specific phenomenological properties, and not just an arbitrary or generic nonconformal model. Therefore, the conformal symmetry-breaking pattern needs to be driven in a phenomenologically-oriented fashion.

One possible approach to obtain a nonconformal system is a bottom-up holographic construction where the free parameters of the model are constrained by existing results from LQCD in some specific regime. Once the parameters are fixed, one can then use this model to make predictions. 
Of course, as in any effective theory construction, the functional form of the bulk action and also the ansatze for the bulk fields must be previously chosen based on some symmetry and other physically relevant considerations, taking into account a given set of observables from the target phenomenology and the basic rules for evaluating these observables using holography.

The seminal works of \cite{Gubser:2008ny,Gubser:2008yx,Gubser:2008sz,DeWolfe:2010he, DeWolfe:2011ts} laid down a remarkably simple and efficient way of constructing quantitative holographic models for the strongly coupled QGP in equilibrium. The general reasoning originally developed in these works may be schematically structured as follows:
\begin{enumerate}[i.]
\item The focus is on constructing an approximate holographic dual or emulator for the equation of state of the strongly coupled QGP in the deconfined regime of QCD, without trying to implement confinement (e.g. Regge trajectories for hadrons), chiral symmetry breaking at low temperatures, asymptotic freedom at asymptotically high temperatures, nor an explicit embedding into string theory.
In this construction, the QCD equation of state (and the second-order baryon susceptibility for the case of finite baryon densities, see section \ref{sec:EoS}) is used to fix the free parameters at finite temperature and \emph{vanishing} chemical potentials. Note that only these specific LQCD data are used to fix the free parameters of the model. All other resulting thermodynamic quantities or transport coefficients are then predictions of the holographic construction;

\item The dynamical field content and the general functional form of the bulk gravity action is taken to be the simplest possible in order to accomplish the above. 
One considers a bulk metric field  (holographically dual to the boundary QFT energy-momentum tensor) plus a Maxwell field with the boundary value of its time component providing the chemical potential at the dual QFT. Additionally, a real scalar field (called the dilaton) is used to break conformal symmetry in the holographic setting, emulating the QGP equation of state at zero chemical potential. The dilaton field also relates string and Einstein frames, as used e.g. in the holographic calculation of parton energy loss (some results in this regard will be briefly reviewed in section \ref{sec:transport});

\item 
The general functional form for the bulk action constructed with the dynamical field content features at most two derivatives of the fields. The bulk action includes the Einstein-Hilbert term with a negative cosmological constant (associated with asymptotically AdS$_5$ spacetimes) for the metric field $g_{\mu\nu}$, the kinetic terms for the Abelian gauge field $A_\mu$ and the dilaton field $\phi$, an almost arbitrary potential (free function) $V(\phi)$ for the dilaton, and an interaction term between the Maxwell and the dilaton fields, which features another free function of the dilaton field, $f(\phi)$. 
The free functions, $V(\phi)$ and $f(\phi)$, the effective $5D$ Newton's constant, $G_5$, and the characteristic energy scale of the nonconformal model, $\Lambda\propto L^{-1}$, need to be dynamically fixed by holographically matching the specific set of LQCD results mentioned in the first item above. Note that these parameters comprise the entire set of free parameters of the bottom-up EMD construction.

\item The effects of the dynamical quarks in the medium are assumed to be effectively encoded in the form of the bottom-up model parameters fixed to holographically match the QCD equation of state and second-order baryon susceptibility obtained from LQCD simulations at zero chemical potential (no explicit flavor-branes are employed for this purpose in the holographic EMD models reviewed in the present paper).
\end{enumerate}

More details on the  procedure mentioned above will be discussed in section \ref{sec:EoS}. Let us now comment on the main limitations of such an approach, some of which are fairly general and refer to all classical gauge-gravity models.

First, gauge-gravity models such as the one mentioned above lack asymptotic freedom. This is expected from the original AdS-CFT correspondence since classical  gravity in the bulk lacks the contributions coming both from massive string states and quantum string loops. By discarding such contributions in the bulk, one obtains a strongly coupled dual QFT at the boundary with a large number of degrees of freedom (large $N_c$). 
The consideration of deformations of the bulk geometry given by asymptotic (but not strictly) AdS solutions of classical gravity does not seem enough to claim that such deformations could in principle describe asymptotic freedom in the dual gauge theory at the boundary. 
The fact that $\eta/s=1/4\pi$ for any value of temperature (and chemical potentials) in isotropic and translationally invariant gauge-gravity models with two derivatives of the metric field, conformal or not, is a clear indication that such models are strongly coupled at all energy scales.  Therefore, these models miss asymptotic freedom in the ultraviolet regime.
It is then clear that the ultraviolet regime of such models is in striking contradiction with perturbative QCD (expected to be relevant at high temperatures), where $\eta/s$ is an order of magnitude larger than $ 1/4\pi$. 
One possible way of improving this situation has been discussed in Ref. \cite{Cremonini:2012ny}. 
There they consider the effects of higher curvature corrections to the metric field in the bulk (i.e., higher derivative corrections to Einstein's gravity) in the presence of a dilaton field, which allows for a temperature-dependent  $\eta/s$. 
Higher derivative corrections for the bulk action are associated with  contributions coming from massive string states, which are expected to lead to a reduction of the effective coupling of the boundary QFT theory. However, consistently including higher derivative curvature corrections for an EMD model, taking into account the full dynamical backreaction of the higher curvature terms into the background geometry, is a very challenging task that has yet to be done.

Another general limitation of gauge-gravity models for QCD is that a realistic holographic description of thermodynamic and hydrodynamic observables in the HRG confining phase seems unfeasible. Standard gauge-gravity models describe large $N_c$ systems. However, the pressure of the QCD medium in the confining hadronic phase goes as $\sim N_c^0 = \mathcal{O}\left(1\right)$, while in the deconfined QGP phase it goes as $\sim N_c^2$. Therefore, the pressure in hadron thermodynamics is $N_c^{-2}$ suppressed relative to the pressure in the QGP phase in a large $N_c$ expansion.  Formally, the hadron phase requires string loop corrections in the bulk in order to have a feasible holographic dual description at the boundary. Such a quantum string loop corrected holographic dual would be much more complicated than simple classical gauge-gravity models.

The two above limitations are common to all gauge-gravity models aimed at realistically describing QCD. Further limitations are related to the EMD constructions reviewed here. We have already alluded to the fact that such models are not intended to describe chiral symmetry breaking, confinement, and thus, hadron spectroscopy.  These points, together with the intrinsic limitations of gauge-gravity models regarding the description of hadron thermodynamics and asymptotic freedom, clearly restrict the target phenomenology of such EMD models to be the hot deconfined phase of QCD matter corresponding to the strongly coupled QGP produced in heavy-ion collisions.

Another phenomenological limitation of EMD models is that they only describe a single conserved charge (i.e. only one finite chemical potential is considered; it is possible to consider in holography more than one conserved charge and different global symmetry patterns by working with more than one Maxwell field or by considering a Yang-Mills field in the bulk, however, in this review we focus on simple EMD models --- perspectives to extend the holographic phenomenological approaches reviewed here to more general bottom-up constructions will be briefly mentioned in the conclusions). Typically, finite baryon chemical potential $\mu_B$ is considered (see section\ \ref{sec:EoS}).  
However, the hot and baryon dense QGP produced in relativistic heavy-ion collisions at low energies actually comprises three chemical potentials ($\mu_B$,  the electric charge chemical potential $\mu_Q$, and the strangeness chemical potential $\mu_S$). In equilibrium, these chemical potentials can be related to each other through the global strangeness neutrality condition realized in such collisions, due to the fact that the colliding nuclei do not carry net strangeness. The strangeness neutrality condition is 
\begin{equation}
    \langle S\rangle = \langle N_{\bar{S}}-N_S\rangle = VT^3\hat{\chi}_1^S = 0,
\end{equation}
where $N_S$ is the number of strange quarks, $N_{\bar{S}}$ is the number of strange antiquarks, and $\hat{\chi}_1^S\equiv\partial\left(P/T^4\right)/\partial(\mu_S/T)$ is the reduced strangeness density. 

Additionally, $\mu_Q$ can also be constrained by the charge to baryon number ratio of the colliding nuclei.  
There is a small isospin imbalance for lead-lead (Pb+Pb) collisions at the LHC and gold-gold (Au+Au) collisions at RHIC,
\begin{equation}
    \langle Q\rangle/\langle B\rangle = \langle N_Q - N_{\bar{Q}}\rangle/\langle N_B - N_{\bar{B}}\rangle = \hat{\chi}_1^Q / \hat{\chi}_1^B = Z/A \approx 0.4,
\end{equation}
where $Z$ is the atomic number and $A$ is the mass number of the colliding nuclei. Thus, from strangeness neutrality and charge conservation, one can then determine $\mu_Q=\mu_Q(T,\mu_B)$ and $\mu_S=\mu_S(T,\mu_B)$ \cite{Karsch:2010ck,Bazavov:2012vg,Borsanyi:2013hza,Borsanyi:2014ewa,Bazavov:2017dus,Borsanyi:2022qlh,Rennecke:2019dxt}. These phenomenological constraints from heavy-ion collisions are not implemented in the holographic EMD constructions reviewed here, where one simply sets $\mu_Q=\mu_S=0$.

We finish these introductory comments on phenomenological bottom-up holographic EMD models for the QGP by remarking that these models are partially inspired by, but not actually derived from string theory. Therefore, the actual applicability of the holographic dictionary for such constructions, and more generally, for any bottom-up gauge-gravity model, may be questioned. Indeed, the phenomenological viability of bottom-up holographic models can be checked by direct comparison with the results of the target phenomenology. 
The degree of agreement between holographic EMD results and several first principles LQCD calculations as well as hydrodynamic viscosities inferred from phenomenological multistage models describing several heavy-ion data, provides compelling evidence that the holographic dictionary works in practice for these models.

The general reasoning outlined above may be systematically adapted to successfully describe different aspects of phenomenology, indicating that at least some of the entries in the holographic dictionary may have a  broad range of validity. 
For instance, one could consider using gauge-gravity models to describe pure YM theory without dynamical quarks.
Bottom-up dilatonic gauge-gravity models with specific functional forms for the dilaton potential may be engineered to quantitatively describe the thermodynamics of a deconfined pure gluon plasma with a first-order phase transition (although the thermodynamics of the confining phase corresponding to a gas of glueballs cannot be described by classical gauge-gravity models), besides describing also glueball spectroscopy \cite{Gubser:2008ny,Gubser:2008yx,Gursoy:2008bu,Gursoy:2010fj}.

	\subsubsection{Holographic equations of state}
         \label{sec:EoS}

\hspace{0.42cm} A gauge-gravity model is usually defined by its action on the classical gravity side of the holographic duality, while different dynamic situations for its dual QFT, living at the boundary of the asymptotically AdS bulk spacetime, are related to different ansatze and boundary conditions for the bulk fields. For instance, given some bulk action, the vacuum state in the dual QFT is associated with solutions of the bulk equations of motion with no event horizon, which is accomplished by an ansatz for the metric field with no blackening function. Thermal states in equilibrium for the same dual QFT are often associated with equilibrium black hole (or more generally, black brane) solutions of the bulk equations of motion, which now require a blackening function in the ansatz for the metric field. Hydrodynamic transport coefficients and characteristic equilibration time scales may be evaluated from the spectra of quasinormal modes \cite{Horowitz:1999jd,Kovtun:2005ev,Berti:2009kk} of these black hole solutions slightly disturbed out of thermal equilibrium, while different far-from-equilibrium dynamics may be simulated by taking into account boundary conditions and ansatze for the bulk fields with nontrivial dependence on spacetime directions parallel to the boundary \cite{Chesler:2013lia}.

The main bottom-up holographic models reviewed in the present manuscript are specified by actions of the EMD class, whose general form in the bulk is given below \cite{DeWolfe:2010he,DeWolfe:2011ts},\footnote{Since the dilaton is a real scalar field, being thus uncharged under the Abelian gauge symmetry associated to the Maxwell field, there is no minimal coupling between those two fields. Instead, in the action \eqref{eq:EMDaction} the form of the Maxwell-dilaton coupling involving the function $f(\phi)$ is inspired from top-down low-energy string theory solutions compactified to $5D$, as in the one R-charge black hole model also discussed e.g. in Ref. \cite{DeWolfe:2011ts}. However, contrary to bottom-up constructions where the $5D$ gravitational constant and the potentials $V(\phi)$ and $f(\phi)$ are taken as free parameters and functions of the holographic setup (to be fixed by some appropriately chosen phenomenological inputs), in top-down models those potentials and parameter are fixed by the kind of low-energy string construction considered. In such top-down constructions it is common to appear exponential terms in the potentials, which also serves as a motivation to employ e.g. hyperbolic functions in the parametrization of potentials in bottom-up models, as it will be used e.g. in Eqs. \eqref{eq:EMDV} and \eqref{eq:EMDf}.}
\begin{align}
S=\int_{\mathcal{M}_5} d^5x\,\mathcal{L} &= \frac{1}{2\kappa_5^2}\int_{\mathcal{M}_5} d^5x\,\sqrt{-g}\left[R-\frac{(\partial_\mu\phi)^2}{2} -V(\phi) -\frac{f(\phi)F_{\mu\nu}^2}{4}\right],
\label{eq:EMDaction}
\end{align}
where $\kappa_5^2\equiv 8\pi G_5$ is the $5D$ gravitational constant. The bulk action \eqref{eq:EMDaction} is supplemented by two boundary terms: i) the Gibbons-Hawking-York (GHY) boundary action \cite{York:1972sj,Gibbons:1976ue}, which in a manifold $\mathcal{M}_5$  with a boundary (as in the case of asymptotically AdS spacetimes) is required in the formulation of a well-defined variational problem with a Dirichlet boundary condition for the metric field,\footnote{By the variational principle, the variation of the gravity action must vanish for arbitrary variations $\delta g_{\mu\nu}$ of the metric field in the bulk. In the case of spacetime manifolds with a boundary, in calculating the variation of the metric tensor in the bulk, integration by parts in directions transverse to the boundary leads to a boundary term that is nonvanishing even by imposing the Dirichlet boundary condition that the metric is held fixed at the boundary, $\delta g_{\mu\nu}|_{\partial\mathcal{M}_5}=0$. This boundary term is exactly canceled out by the variation of the GHY action (see e.g. chapter 4 of \cite{Poisson:2009pwt}), allowing for the variation of the total gravity action to vanish in compatibility with Einstein's equations in a bulk spacetime with a boundary.} and ii) a boundary counterterm action employed to remove the ultraviolet divergences of the on-shell action by means of the holographic renormalization procedure \cite{deHaro:2000vlm,Bianchi:2001kw,Skenderis:2002wp,Papadimitriou:2011qb,Lindgren:2015lia,Elvang:2016tzz}. Although needed in order to write the full holographic renormalized on-shell action, those two boundary terms do not contribute to the bulk equations of motion and are not strictly required in the calculations reviewed in the present work. Therefore, we shall not write their explicit form here.

It is important to make some general remarks at this point. The holographic renormalized on-shell action is generally employed in the evaluation of the pressure and energy density (the diagonal entries in the expectation value of the energy-momentum tensor) of the medium defined in the dual QFT at the boundary, also for the calculation of hydrodynamic transport coefficients extracted from perturbations of the bulk fields, and for the analysis of far-from-equilibrium dynamics. However, here we will not consider far-from-equilibrium calculations. Regarding the equilibrium pressure of the medium, its calculation can also be done by integrating over temperature the entropy evaluated through the Bekenstein-Hawking relation for black hole thermodynamics \cite{Bekenstein:1973ur,Hawking:1975vcx}, which does not require holographic renormalization. Moreover, for the holographic calculation of the specific hydrodynamic transport coefficients reviewed in this work, which are related through Kubo formulas to the imaginary part of thermal retarded correlators of the relevant dual QFT operators, holographic renormalization can also be bypassed through the use of radially conserved fluxes extracted from the equations of motion for the relevant bulk perturbations --- see e.g. \cite{DeWolfe:2011ts} and also \cite{Gubser:2008sz,Iqbal:2008by,Critelli:2016ley}.

The holographic renormalization procedure is generally a very laborious task and the aforementioned shortcuts are surely convenient in order to have an alternative and easier access to some physical observables through the holographic machinery. On the other hand, without implementing the holographic renormalization procedure, one has to face some limitations. Besides the ones already mentioned above, another relevant limitation is the following: although one can calculate the energy density by using the thermodynamic relation \eqref{eq:EnDens}, it would be also important to provide an explicit check that Eq. \eqref{eq:EnDens} holds when calculating the pressure and the energy density through the holographic renormalized on-shell action, with the entropy density calculated independently by using Bekenstein-Hawking's relation \eqref{eq:s} and the charge density evaluated independently by using the boundary value of the  radial momentum conjugate to the bulk Maxwell field, as in Eq. \eqref{eq:rhoB}. Although holographic renormalization has not been implemented yet for the EMD models of Refs. \cite{DeWolfe:2010he,DeWolfe:2011ts,Rougemont:2015wca,Rougemont:2015ona,Finazzo:2015xwa,Rougemont:2017tlu,Critelli:2017oub,Grefa:2021qvt,Grefa:2022sav,Rougemont:2018ivt,Knaute:2017opk}, very recently it has been implemented for the EMD model of Refs. \cite{Cai:2022omk,Li:2023mpv}, with the aforementioned consistency check being successfully performed. Moreover, a nontrivial check of thermodynamic consistency has been also performed for the EMD model of Refs. \cite{Critelli:2017oub,Grefa:2021qvt,Grefa:2022sav,Rougemont:2018ivt} by using the Gibbs-Duhem equation to evaluate the pressure via the temperature integral of the Bekenstein-Hawking's entropy density at fixed chemical potential, and checking that it coincides with the chemical potential integral of the baryon density at fixed temperature (with an additive integration constant computed from the temperature integral of the entropy density at zero chemical potential). This consistency check is important since the entropy and baryon charge densities are two different entries of the holographic dictionary evaluated at the two opposite sides of the bulk geometry.

The set of free parameters and functions $\{G_5,\Lambda,V(\phi),f(\phi)\}$ comprised in the bottom-up EMD setup can be fixed by taking as phenomenological inputs some adequate lattice results on QCD thermodynamics at finite temperature and zero chemical potentials (and vanishing electromagnetic fields), where $\Lambda$ is a characteristic energy scale of the nonconformal holographic model employed to express in powers of MeV dimensionful observables in the dual QFT, which are calculated in the gravity side of the holographic correspondence in powers of the inverse of the asymptotic AdS radius $L$. In practice, we simply set $L=1$ and trade it off as a free parameter by the energy scale $\Lambda$, without changing the number of free parameters of the model \cite{Rougemont:2015wca,Critelli:2017oub}. The set $\{G_5,\Lambda,V(\phi)\}$ can be fixed by the LQCD equation of state evaluated at vanishing chemical potential, while $f(\phi)$ may be fixed, up to its overall normalization, by the LQCD second order baryon susceptibility, also evaluated at zero chemical potential \cite{DeWolfe:2010he,Rougemont:2015wca,Critelli:2017oub}.\footnote{However, as we are going to discuss afterward in this section, and more deeply in section \ref{sec:bayes}, available LQCD results cannot constrain the set of free parameters of the EMD model to be fixed in a unique way.}

In order to do this, one first needs to specify the adequate ansatze for the bulk EMD fields such as to describe isotropic and translationally invariant thermal states at the dual boundary quantum gauge theory (as in LQCD simulations). Since we are going to consider, in general, also the description of thermal states at finite baryon chemical potential, we take the form below for the bulk fields corresponding to isotropic and translationally invariant charged EMD black hole backgrounds in equilibrium \cite{DeWolfe:2010he,Rougemont:2015wca,Critelli:2017oub},
\begin{align}
ds^2 = g_{\mu\nu}dx^\mu dx^\nu = e^{2A(r)}[-h(r)dt^2+d\vec{x}^2]+\frac{dr^{2}}{h(r)}, \qquad
\phi = \phi(r), \qquad A_{\mu}dx^{\mu}=\Phi(r)dt,
\label{eq:ansatz}
\end{align}
where $r$ is the holographic radial coordinate, with the boundary at $r\to\infty$ and the black hole horizon at $r=r_H$, and $r_H$ being the largest root of the blackening function, $h(r_H)=0$. The set of general EMD equations of motion obtained by extremizing the bulk action \eqref{eq:EMDaction} with respect to the EMD fields can be written in the following form \cite{Critelli:2017euk},
\begin{align}
R_{\mu\nu}-\frac{g_{\mu\nu}}{3}\left[V(\phi)-\frac{f(\phi)}{4}F_{\alpha\beta}^2\right]-\frac{1}{2}\partial_\mu\phi\partial_\nu\phi-\frac{f(\phi)}{2}g^{\alpha\beta}F_{\mu\alpha}F_{\nu\beta}&=0,\label{eq:EinsteinEqs}\\
\partial_\mu\left(\sqrt{-g}f(\phi)g^{\mu\alpha}g^{\nu\beta}F_{\alpha\beta}\right)&=0,\label{eq:MaxwellEqs}\\
\frac{1}{\sqrt{-g}}\partial_\mu\left(\sqrt{-g}g^{\mu\nu}\partial_\nu\phi\right)-\frac{\partial V(\phi)}{\partial\phi}-\frac{F_{\mu\nu}^2}{4}\frac{\partial f(\phi)}{\partial\phi}&=0,\label{eq:DilatonEq}
\end{align}
which, for the isotropic ansatze for the EMD fields in equilibrium given in Eqs. \eqref{eq:ansatz}, reduce to the following set of coupled ordinary differential equations of motion,
\begin{align}
\phi''(r)+\left[\frac{h'(r)}{h(r)}+4A'(r)\right]\phi'(r)-\frac{1}{h(r)}\left[\frac{\partial V(\phi)}{\partial\phi}-\frac{e^{-2A(r)}\Phi'(r)^{2}}{2}\frac{\partial f(\phi)}{\partial\phi}\right]&=0,\label{eq:EoM1}\\
\Phi''(r)+\left[2A'(r)+\frac{d[\ln{f(\phi)}]}{d\phi}\phi'(r)\right]\Phi'(r)&=0,\label{eq:EoM2}\\
A''(r)+\frac{\phi'(r)^{2}}{6}&=0,\label{eq:EoM3}\\
h''(r)+4A'(r)h'(r)-e^{-2A(r)}f(\phi)\Phi'(r)^{2}&=0,\label{eq:EoM4}\\
h(r)[24A'(r)^{2}-\phi'(r)^{2}]+6A'(r)h'(r)+2V(\phi)+e^{-2A(r)}f(\phi)\Phi'(r)^{2}&=0,\label{eq:EoM5}
\end{align}
where Eq. \eqref{eq:EoM5} is a constraint. These equations of motion are discussed in detail in Refs. \cite{DeWolfe:2010he,Rougemont:2015wca,Critelli:2017oub}. They must be solved numerically, and different algorithms have been developed through the years to accomplish this task with increasing levels of refinement  \cite{DeWolfe:2010he,Rougemont:2015wca,Critelli:2017oub,Grefa:2021qvt}. Two different sets of coordinates are used in this endeavor: the so-called \textit{standard coordinates} (denoted with a tilde), in which the blackening function goes to unity at the boundary, $\tilde{h}(\tilde{r}\to\infty)=1$, and also $\tilde{A}(\tilde{r}\to\infty)\to\tilde{r}$, such that holographic formulas for the physical observables are expressed in standard form; and the so-called \textit{numerical coordinates} (denoted without a tilde), corresponding to rescalings of the standard coordinates used to specify definite numerical values for the radial location of the black hole horizon and also for some of the initially undetermined infrared expansion coefficients of the background bulk fields close to the black hole horizon, which is required to start the numerical integration of the bulk equations of motion from the black hole horizon up to the boundary.\footnote{Notice that the part of the bulk geometry within the interior of the black hole horizon is causally disconnected from observers at the boundary.} In fact, with such rescalings, all the infrared coefficients are determined in terms of just two initially undetermined coefficients, $\phi_0$ and $\Phi_1$, which are taken as the ``initial conditions'' (in the holographic radial coordinate, $r$) for the system of differential equations of motion. Those correspond, respectively, to the value of the dilaton field and the value of the radial derivative of the Maxwell field evaluated at the black hole horizon.

For the holographic calculation of physical observables at the boundary QFT, one also needs to obtain the ultraviolet expansion coefficients of the bulk fields near the boundary, far from the horizon. For the evaluation of the observables reviewed in this paper, it suffices to determine four ultraviolet expansion coefficients of the bulk fields, namely, $h_0^{\textrm{far}}$ coming from the blackening function $h(r)$ of the metric field, $\Phi_0^{\textrm{far}}$ and $\Phi_2^{\textrm{far}}$ coming from the nontrivial component of the Maxwell field $\Phi(r)$, and $\phi_A$ coming from the dilaton field $\phi(r)$, with the functional forms of the ultraviolet expansions being derived by solving the asymptotic forms of the equations of motion near the boundary \cite{DeWolfe:2010he}. In order to determine the numerical values of the  ultraviolet coefficients for a given numerical solution generated by a given choice of the pair of initial conditions $(\phi_0,\Phi_1)$, one matches the full numerical solution for the bulk fields to the functional forms of their corresponding ultraviolet expansions near the boundary. While the values of $h_0^{\textrm{far}}$, $\Phi_0^{\textrm{far}}$ and $\Phi_2^{\textrm{far}}$ can be easily obtained, the evaluation of $\phi_A$ is much more subtle and delicate due to the exponential decay of the dilaton close to the boundary \cite{DeWolfe:2010he,Rougemont:2015wca}. In Refs. \cite{Rougemont:2015wca,Critelli:2017oub}, different algorithms were proposed to extract $\phi_A$ in a reliable and numerically stable way from the near-boundary analysis of the numerical solutions for the dilaton field, with progressively increasing levels of accuracy and precision. Moreover, in Ref. \cite{Grefa:2021qvt}, a new algorithm for choosing the grid of initial conditions $(\phi_0,\Phi_1)$ was devised in order to cover the phase diagram of the dual QFT in the $(T,\mu_B)$-plane in a much more efficient and broader way than in earlier works, like e.g. \cite{Rougemont:2015wca,Rougemont:2015ona,Finazzo:2015xwa,Rougemont:2017tlu}. Together with more precise fittings to LQCD results at zero chemical potential, which led to the construction of an improved version of the EMD model at finite temperature and baryon density in Ref. \cite{Critelli:2017oub}, all the algorithmic upgrades mentioned above allowed to obtain predictions from this improved EMD model not only for the location of the CEP \cite{Critelli:2017oub}, but also for the location of the line of first-order phase transition and the calculation of several thermodynamic \cite{Grefa:2021qvt} and transport \cite{Grefa:2022sav} observables in a broad region of the $(T,\mu_B)$-plane, including the phase transition regions, where the numerical calculations are particularly difficult to perform due to the coexistence of competing branches of black hole solutions and the manifestation of significant noise in the numerical solutions.

Before comparing some thermodynamic results from some different versions of the EMD model in the literature, displaying the aforementioned improvements and discussing some of their consequences for the holographic predictions regarding the structure of the QCD phase diagram in the $(T,\mu_B)$-plane, we provide below the relevant formulas for their calculation on the gravity side of the holographic duality. The numerical solutions for the EMD fields in thermal equilibrium generated by solving the bulk equations of motion for different pairs of initial conditions $(\phi_0,\Phi_1)$ are associated through the holographic dictionary with definite thermal states at the boundary QFT, where the temperature $T$, the baryon chemical potential $\mu_B$, the entropy density $s$, and the baryon charge density $\rho_B$ of the medium are given by \cite{DeWolfe:2010he,Rougemont:2015wca},\footnote{We provide the formulas in the standard coordinates (with a tilde) and in the numerical coordinates (in terms of which the numerical solutions are obtained and the relevant ultraviolet coefficients are evaluated). It is worth mentioning that \cite{DeWolfe:2010he} introduced three extra free parameters in the holographic model, corresponding to different energy scaling parameters for $\mu_B$, $s$, and $\rho_B$, besides the one for $T$. These parameters are unnecessary as they artificially augment the number of free parameters of the bottom-up construction without a clear physical motivation. In the holographic formulas reviewed in this paper there is just a single energy scale $\Lambda$ associated with the nonconformal nature of the EMD model \cite{Rougemont:2015wca,Critelli:2017oub,Grefa:2021qvt,Rougemont:2015ona,Finazzo:2015xwa,Rougemont:2017tlu,Grefa:2022sav}, as mentioned above. In this context, if an observable has energy dimension $p$, its formula in the gravity side of the holographic duality gets multiplied by $\Lambda^p$ in order to express the corresponding result in the dual QFT at the boundary in physical units of MeV$^p$.}

\begin{align}
T &= \left.\frac{\sqrt{-g'_{\tilde{t}\tilde{t}}g^{\tilde{r}\tilde{r}}\,'}}{4\pi}\right|_{\tilde{r}=\tilde{r}_{H}} \!\!\!\!\!\!\!\!\!\!\!\!\Lambda=\frac{e^{\tilde{A}(\tilde{r}_{H})}}{4\pi}|\tilde{h}'(\tilde{r}_{H})|\Lambda = \frac{1}{4\pi\phi_{A}^{1/\nu}\sqrt{h_{0}^{\textrm{far}}}}\Lambda, \label{eq:T}\\
\mu_B &= \lim_{\tilde{r}\rightarrow\infty}\tilde{\Phi}(\tilde{r})\Lambda = \frac{\Phi_{0}^{\textrm{far}}}{\phi_{A}^{1/\nu}\sqrt{h_{0}^{\textrm{far}}}}\Lambda, \label{eq:muB}\\
s &= \frac{S}{V}\Lambda^3=\frac{A_{H}}{4G_{5}V}\Lambda^3=\frac{2\pi}{\kappa_{5}^{2}}e^{3\tilde{A}(\tilde{r}_{H})}\Lambda^{3} = \frac{2\pi}{\kappa_{5}^{2}\phi_{A}^{3/\nu}}\Lambda^{3}, \label{eq:s}\\
\rho_B &= \lim_{\tilde{r}\rightarrow\infty}\frac{\partial\mathcal{L}}{\partial(\partial_{\tilde{r}}\tilde{\Phi})}\Lambda^{3} = -\frac{\Phi_{2}^{\textrm{far}}}{\kappa_{5}^{2}\phi_{A}^{3/\nu}\sqrt{h_{0}^{\textrm{far}}}}\Lambda^{3}, \label{eq:rhoB}
\end{align}
where $A_H$ is the area of the black hole event horizon, the prime denotes radial derivative, and $\nu \equiv d-\Delta$, with $d=4$ being the number of spacetime dimensions of the boundary and with $\Delta=(d+\sqrt{d^{2}+4m^{2}L^2})/2$ being the scaling dimension of the (relevant) QFT operator dual to the bulk dilaton field $\phi(r)$, which has a mass $m$ obtained from the form of the dilaton potential $V(\phi)$, to be discussed in a moment.

The dimensionless ratio
\begin{equation}
    \hat{\chi}_2^B\equiv\frac{\chi_2^B}{T^2}\equiv \frac{\partial^2(P/T^4)}{\partial(\mu_B/T)^2}
\end{equation}
 corresponds to the reduced second order baryon susceptibility.  When evaluated at $\mu_B=0$, $\hat{\chi}_2^B$ has an integral expression given by \cite{DeWolfe:2010he,Rougemont:2015wca}
\begin{equation}
\hat{\chi}_{2}^B(T,\mu_{B}=0)=\frac{1}{16\pi^{2}}\frac{s}{T^{3}}\frac{1}{f(0)\int_{r_{H}}^{\infty}dr\ e^{-2A(r)}f(\phi(r))^{-1}},
\label{eq:chi2B0}
\end{equation}
which is to be evaluated over EMD backgrounds generated with the initial condition $\Phi_{1}=0$.\footnote{Although the holographic mapping $(\phi_0,\Phi_1)\mapsto(T,\mu_B,s,\rho_B)$ is highly nontrivial \cite{DeWolfe:2010he,Rougemont:2015wca,Grefa:2021qvt}, choosing $\Phi_1=0$ automatically provides only EMD backgrounds with $\mu_B=0$.} In numerical calculations \cite{Rougemont:2015wca,Critelli:2017oub,Grefa:2021qvt}, one actually takes the following substitutions in Eq. \eqref{eq:chi2B0}, $r_{H}\rightarrow r_{\textrm{start}}$ and $\infty\rightarrow r_{\textrm{max}}$, where $r_{\textrm{start}}$ is some small number (typically $r_{\textrm{start}}\sim 10^{-8}$) employed to avoid the singular point of the EMD equations of motion at the rescaled numerical horizon $r_H=0$, and $r_{\textrm{max}}$ is a numerical parametrization of the radial position of the boundary, which is ideally at $r\to\infty$. Of course, it is not possible to use infinity in numerical calculations, and in practice, $r_{\textrm{max}} \sim 2 - 10$ is typically enough for the numerical EMD backgrounds to reach, within a small numerical tolerance, the ultraviolet fixed point of the holographic renormalization group flow associated with the AdS$_5$ geometry. It must be also emphasized that Eq. \eqref{eq:chi2B0} is not valid at $\mu_B\neq 0$. In fact, to calculate  the second order baryon susceptibility at finite $\mu_B$, we take in practice 
\begin{equation}
\hat{\chi}_2^B=\partial(\rho_B/T^3)/\partial(\mu_B/T)
\end{equation}
where $\rho_B$ is the baryon density.

For holographic models where the holographic renormalization procedure is still not implemented, one cannot extract the pressure (and the energy density) directly from the renormalized on-shell boundary action, since such a quantity is still not available.\footnote{Notice, however, that holographic renormalization has been already successfully implemented for the EMD model of Refs. \cite{Cai:2022omk,Li:2023mpv}.} Nevertheless, in such a case, one may approximate the pressure of the dual QFT fluid as follows (for fixed values of $\mu_B$),
\begin{equation}
P(T, \mu_{B})\approx \int_{T_{\textrm{low}}}^{T} dT \,s(T,\mu_{B}),
\label{eq:Papprox}
\end{equation}
where $T_{\textrm{low}}$ is the lowest value of temperature available for all solutions with different values of $\mu_B$ within the set of EMD black hole backgrounds generated with the grid of initial conditions considered. Eq. \eqref{eq:Papprox} ceases to be a good approximation for the pressure for values of $T\sim T_{\textrm{low}}$.\footnote{The reason for taking a finite $T_{\textrm{low}}$ instead of zero as the lower limit in the temperature integral of the entropy density in Eq. \eqref{eq:Papprox} is that it is numerically difficult to obtain solutions of the EMD equations of motion at very low temperatures. For instance, $T_{\textrm{low}}=2$ MeV for the calculations done in Ref. \cite{Grefa:2021qvt}. By varying the value of $T_{\textrm{low}}$ it is possible to numerically check the window of values for which the approximate results for the pressure remain stable within a given numerical tolerance.} The energy density of the medium can be calculated from the thermodynamic relation,
\begin{align}
\epsilon(s,\rho_{B}) = Ts(T,\mu_{B})-P(T,\mu_{B})+\mu_{B}\rho_{B}(T,\mu_{B}).
\label{eq:EnDens}
\end{align}
The trace anomaly of the energy-momentum tensor (also known as the interaction measure) of the dual QFT at the boundary is given by,
\begin{align}
I(T,\mu_{B}) = \epsilon(T,\mu_{B})-3P(T,\mu_{B}).
\label{eq:trace}
\end{align}
The square of the speed of sound in the medium calculated along different trajectories of constant entropy over baryon number in the $(T,\mu_B)$-plane is defined as $c_s^2=\left(d P/d \epsilon\right)_{s/\rho_B}$. For phenomenological applications in the context of heavy-ion collisions, one can rewrite this $c_s^2$ in terms of derivatives of $(T,\mu_B)$ \cite{Floerchinger:2015efa,Parotto:2018pwx},
\begin{align}
\left[c_{s}^{2}(T,\mu_B)\right]_{s/\rho_B}=\frac{\rho_{B}^{2}\partial_{T}^{2}P-2s\rho_{B}\partial_{T}\partial_{\mu_{B}}P +s^{2}\partial_{\mu_{B}}^{2}P}{(\epsilon+P)[\partial_{T}^{2}P\partial_{\mu_{B}}^{2}P-(\partial_{T}\partial_{\mu_{B}}P)^{2}]}
\label{eq:cs2}
\end{align}
that provides a much more convenient formula since most equations of state use $(T,\mu_B)$ as the free variables.

The above expressions allow the calculation of the main thermodynamic observables characterizing the equilibrium state of the QGP. Particularly, in order to fix the free parameters of the EMD model, we take as phenomenological inputs state-of-the-art continuum extrapolated results from first principles LQCD simulations with $2+1$ flavors and physical values of the quarks masses, regarding the QCD equation of state \cite{Borsanyi:2013bia} and the second order baryon susceptibility \cite{Bellwied:2015lba}, both evaluated at finite temperature and zero chemical potential. In fact, the choice of an adequate susceptibility is what seeds the bottom-up EMD model with phenomenological information concerning the nature of the controlling state variable(s) of the medium besides the temperature.\footnote{For instance, while the baryon susceptibility is used in the present section, the magnetic susceptibility will be employed in section \ref{sec:EMDmag} within the context of the magnetic EMD model at finite temperature and magnetic field, but with zero chemical potential.} In this way, it was constructed in Ref. \cite{Critelli:2017oub}, and latter also used in Refs. \cite{Grefa:2021qvt,Grefa:2022sav,Rougemont:2018ivt}, a second-generation improved version of the EMD model (relative to previous constructions in the literature, namely, the original one in Refs. \cite{DeWolfe:2010he,DeWolfe:2011ts}, and the first generation improved EMD model of Refs. \cite{Rougemont:2015wca,Rougemont:2015ona,Finazzo:2015xwa,Rougemont:2017tlu}), which is defined by the bulk action \eqref{eq:EMDaction} with the following set of holographically fixed bottom-up parameters and functions,
\begin{align}
V(\phi) &= -12\cosh(0.63\,\phi)+0.65\,\phi^{2}-0.05\,\phi^{4}+0.003\,\phi^{6}, \qquad \kappa_{5}^{2} = 8\pi G_{5}=8\pi(0.46), \qquad \Lambda=1058.83\, \textrm{MeV}, \label{eq:EMDV}\\ 
f(\phi) &= \frac{\sech(-0.27\,\phi+0.4\,\phi^{2})+1.7\,\sech(100\,\phi)}{2.7}.
\label{eq:EMDf}
\end{align}
A number of observations are in order concerning the forms fixed above for the dilaton potential $V(\phi)$ and the Maxwell-dilaton coupling function $f(\phi)$.

First, regarding the dilaton potential, since from the ultraviolet asymptotic expansions for the EMD fields the dilaton is known to vanish at the boundary for relevant QFT deformations \cite{DeWolfe:2010he}, the boundary value $V(0)=-12 \,\,\dot{=}\,\, 2\Lambda_{\textrm{AdS}_{5}}$ is required in order to recover the value of the negative cosmological constant of AdS$_5$ in the ultraviolet regime, as $\Lambda_{\textrm{AdS}_{d+1}}=-d(d-1)/2L^2$ is equal to $-6$ for $d=4$ and $L=1$ (recall that we set here the asymptotic AdS radius to unity).\footnote{We remark that, in spite of the similar notation, the cosmological constant $\Lambda_{\textrm{AdS}_{5}}=-6$ has no relation with the nonconformal energy scale $\Lambda$ in \eqref{eq:EMDV}.} One notices from \eqref{eq:EMDV} that for this EMD model, the dilaton field has a mass squared given by $m^2=\partial_\phi^2V(0)\approx -3.4628$, which satisfies the Breitenlohner-Freedman (BF) stability bound \cite{Breitenlohner:1982jf,Breitenlohner:1982bm} for massive scalar fields in asymptotically AdS backgrounds, $m^2 > m^2_{\textrm{BF}} = -d^2/4L^2 = -4$. Also, since the scaling dimension of the QFT operator dual to the dilaton is $\Delta=(d+\sqrt{d^{2}+4m^{2}L^2})/2\approx 2.73294 < d = 4$ (which implies that $\nu\equiv d-\Delta \approx 1.26706$), as anticipated, this is a relevant operator triggering a renormalization group flow from the AdS$_5$ ultraviolet fixed point towards a nonconformal state as one moves from the ultraviolet to the infrared regime of the dual QFT, or correspondingly, as one moves from the near-boundary to the interior of the bulk in the gravity side of the holographic duality. In fact, if one wishes to introduce a relevant deformation in the dual QFT away from the conformal regime asymptotically attained in the ultraviolet, and simultaneously satisfy the BF stability bound, then one should engineer the dilaton potential such as to have $\Delta_{\textrm{BF}} = 2 < \Delta < d = 4$, or equivalently, $m^2_{\textrm{BF}} = -4 < m^2 < 0$. Moreover, the dilaton potential in \eqref{eq:EMDV} monotonically decreases from its maximum at the boundary to the deep infrared of the bulk geometry, such that there are no singular points (associated with local extrema of the potential) in the bulk equations of motion between the boundary and the black hole horizon, and also, Gubser's criterion for admissible classical gravitational singularities \cite{Gubser:2000nd}, $V(\phi(r_H))\le V(\phi(r\to\infty)=0)=-12$, is satisfied.

\begin{figure}[htp!]
    \centering
    \includegraphics[width=0.45\textwidth]{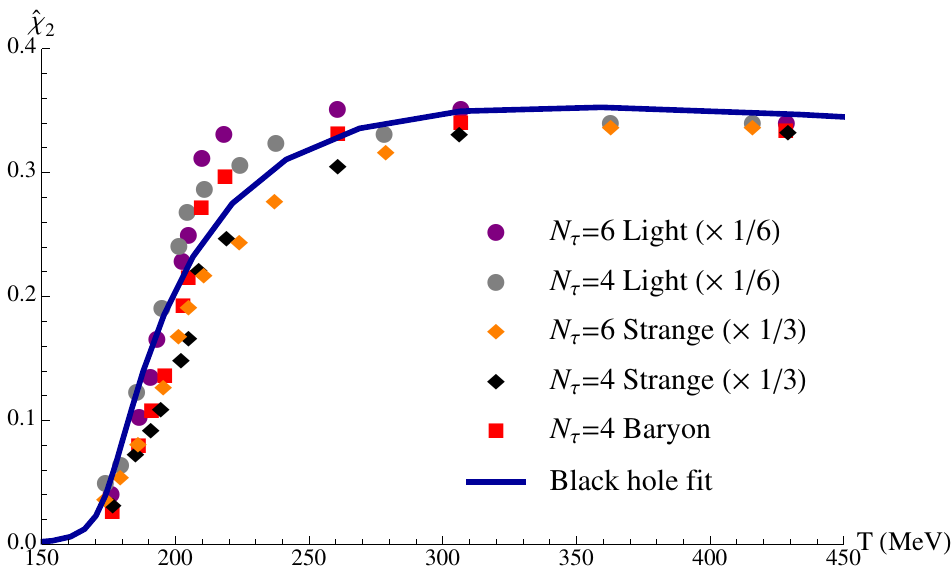}
    \includegraphics[width=0.35\textwidth]{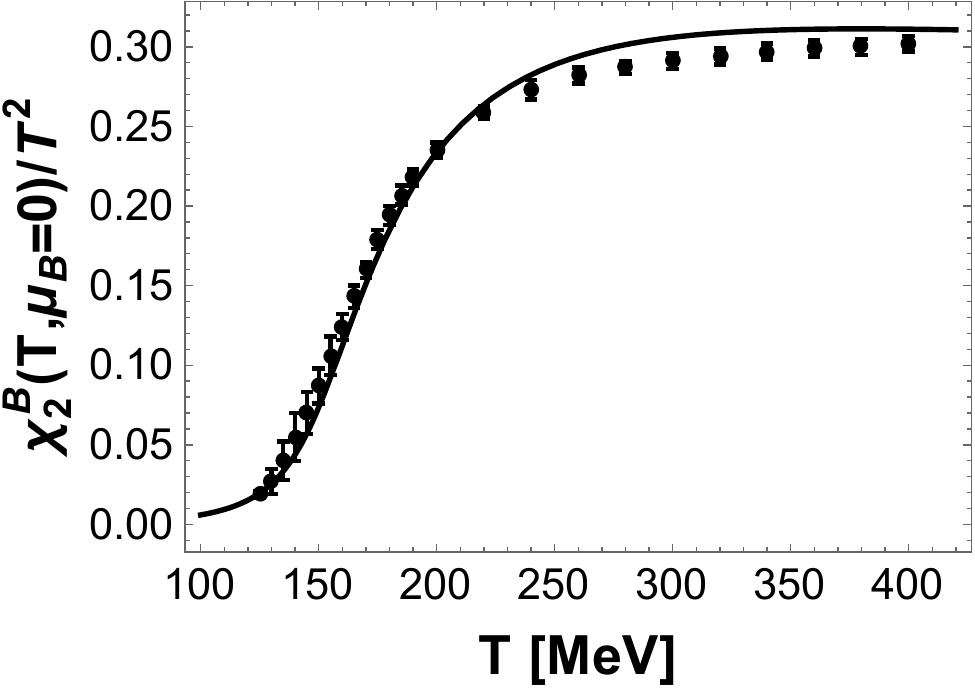}
    \includegraphics[width=0.45\textwidth]{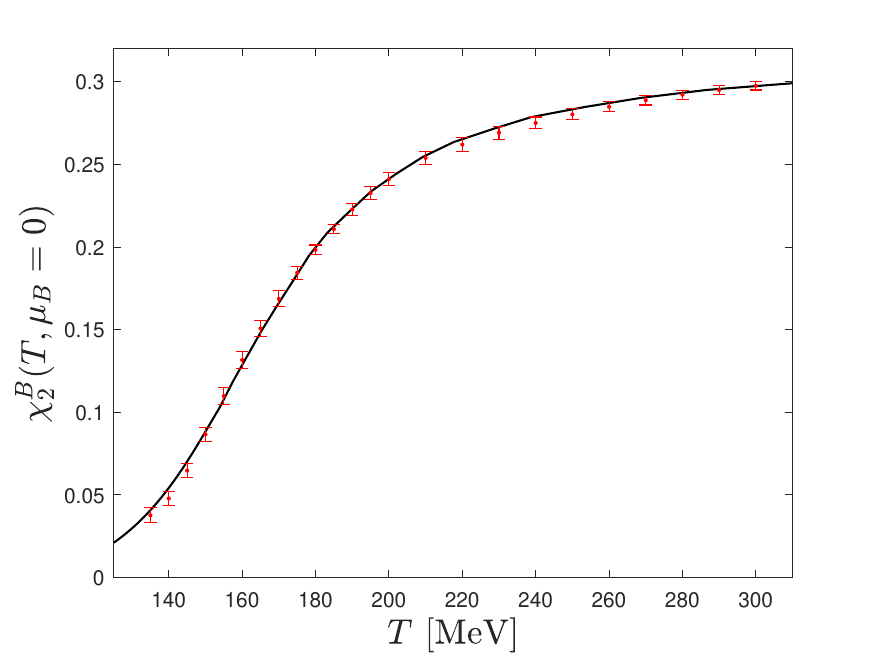}
    \caption{Holographic fits from different EMD models to different LQCD results for the reduced second order baryon susceptibility, $\hat{\chi}_2^B=\chi_2^B/T^2$, evaluated at zero chemical potential and with $2+1$ flavors. This is employed to fix the Maxwell-dilaton coupling, $f(\phi)$ (while the LQCD equation of state at zero chemical potential is used to fix the dilaton potential). \textbf{Top left panel (from Ref. \cite{DeWolfe:2010he}):} The original EMD model of Refs. \cite{DeWolfe:2010he,DeWolfe:2011ts} (which has $\Delta\approx 3.93$) used older LQCD results from \cite{Karsch:2007dp} as phenomenological inputs, which had almost physical values of the quark masses but were  not yet extrapolated to the continuum ($N_\tau\equiv (aT)^{-1}$ is the temporal extent of the lattice and $a$ is the lattice spacing). \textbf{Top right panel (from Ref. \cite{Rougemont:2015wca}):} The first generation improved EMD model of Refs. \cite{Rougemont:2015wca,Rougemont:2015ona,Finazzo:2015xwa,Rougemont:2017tlu} used continuum extrapolated LQCD data as inputs, and results are shown for $\hat{\chi}_2^B$ with physical values of the quark masses from \cite{Borsanyi:2011sw} (the LQCD equation of state at $\mu_B=0$ used to fix $V(\phi)$ was that of \cite{Borsanyi:2012cr}, and $\Delta\approx 3$). \textbf{Bottom panel (from Refs. \cite{Critelli:2017oub,Grefa:2021qvt}):} The second generation improved EMD model of Refs. \cite{Critelli:2017oub,Grefa:2021qvt,Grefa:2022sav,Rougemont:2018ivt} used state-of-the-art continuum extrapolated LQCD results for $\hat{\chi}_2^B$ with physical values of the quark masses from \cite{Bellwied:2015lba} as inputs (the LQCD equation of state at $\mu_B=0$ used to fix $V(\phi)$ was that of \cite{Borsanyi:2013bia}, and $\Delta\approx 2.73294$).}
    \label{fig:chi2B0}
\end{figure}

Second, concerning the Maxwell-dilaton coupling function, one should note from Eq. \eqref{eq:chi2B0} that the baryon susceptibility calculated at zero chemical potential cannot fix the overall normalization of $f(\phi)$. In \eqref{eq:EMDf} this overall normalization was chosen such that $f(0)=1$, as originally proposed in \cite{DeWolfe:2010he}.\footnote{In practice, this choice for the overall normalization of $f(\phi)$ can be motivated by the fact that it allows a quantitative description of LQCD results at nonzero $\mu_B$, as we are going to see later in this review.} Moreover, by also following \cite{DeWolfe:2010he}, we choose $f(\phi)$ such that it asymptotically goes to zero for large $\phi(r)$, in the infrared regime of the theory. However, differently from \cite{DeWolfe:2010he}, in order to obtain a quantitative description of this observable at zero chemical potential one seems to be forced to engineer a functional form for $f(\phi)$ such that it presents a very fast variation close to the boundary (i.e., for $\phi(r\to\infty)\to 0$).\footnote{This is the practical reason for the term $\sim \sech(100\,\phi)$ in \eqref{eq:EMDf} (the numerical factor of 100 can be substituted by some other `large number' without considerably affecting the results).} This peculiar feature has been also observed in other bottom-up EMD constructions with different functional forms for $f(\phi)$ and which had been proved to quantitatively describe $\hat{\chi}_2^B(T,\mu_B=0)$ from LQCD simulations with $2+1$ flavors and physical values of the quark masses \cite{Knaute:2017opk,Cai:2022omk}.

\begin{figure}[htp!]
\begin{center}
\begin{tabular}{c}
\subfigure[]{\includegraphics[width=0.4\textwidth]{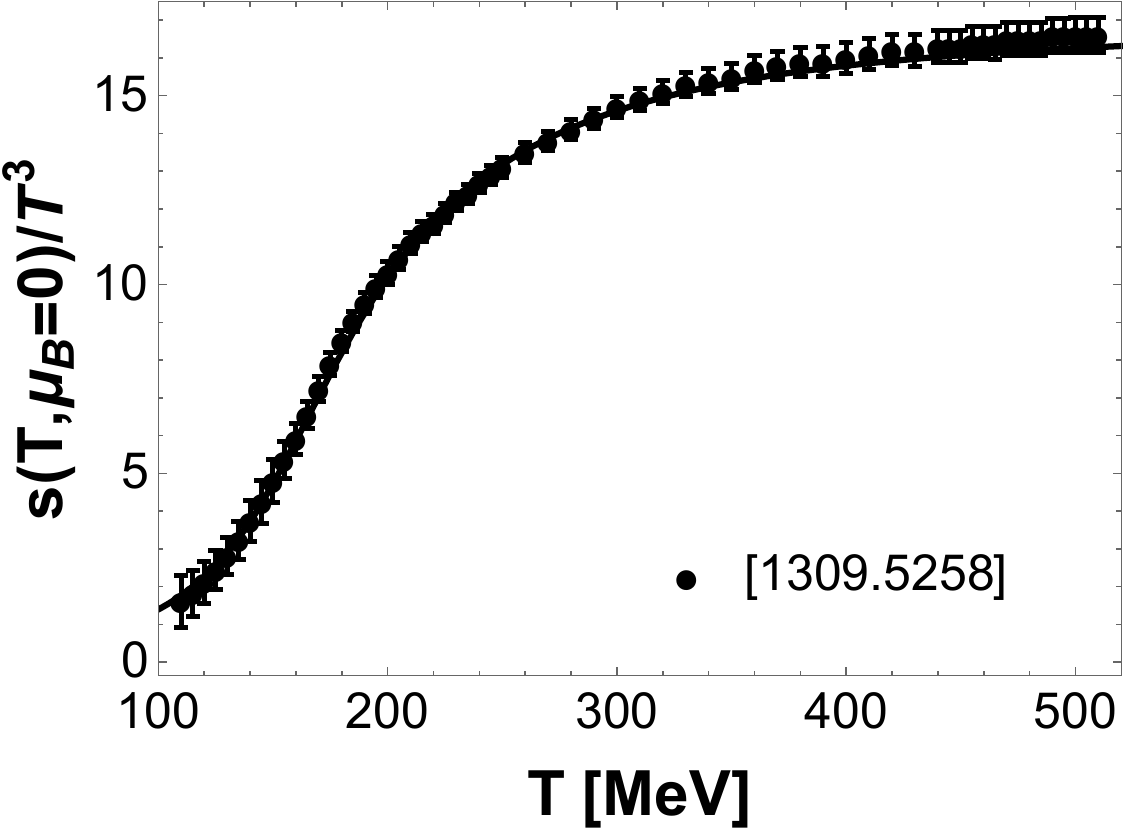}} 
\end{tabular}
\begin{tabular}{c}
\subfigure[]{\includegraphics[width=0.4\textwidth]{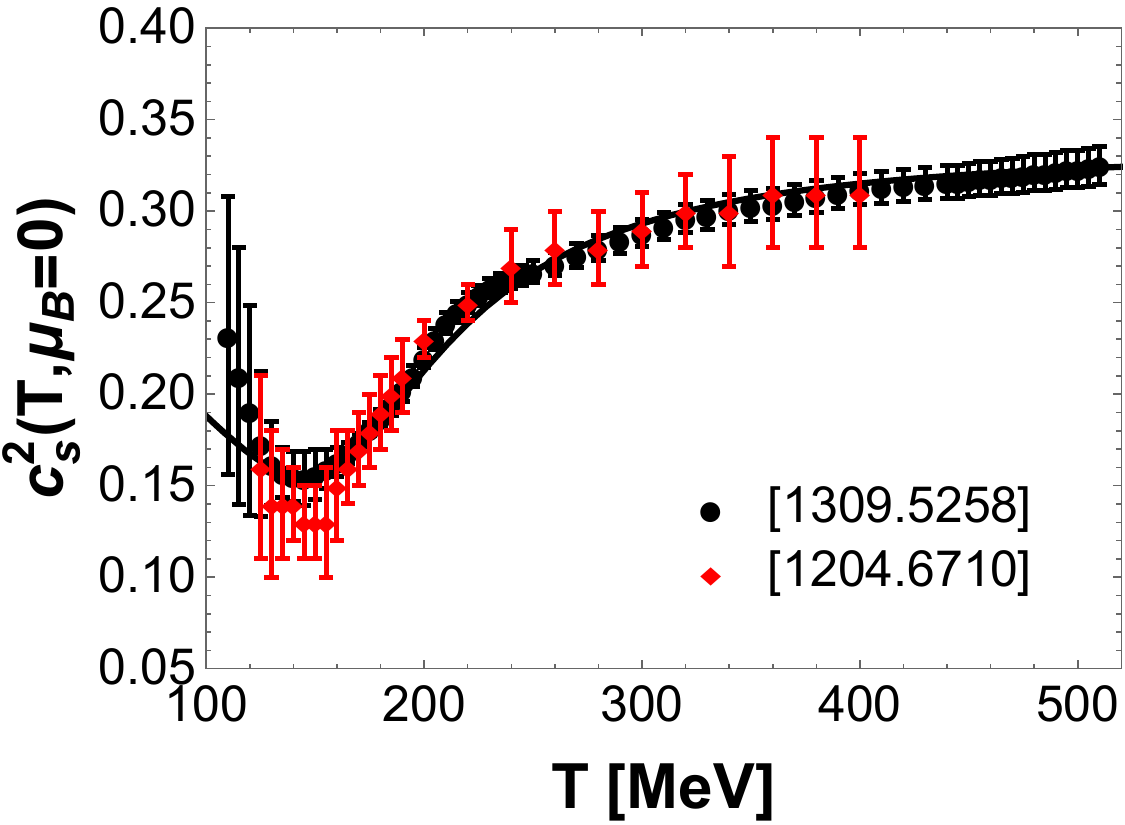}} 
\end{tabular}
\begin{tabular}{c}
\subfigure[]{\includegraphics[width=0.4\textwidth]{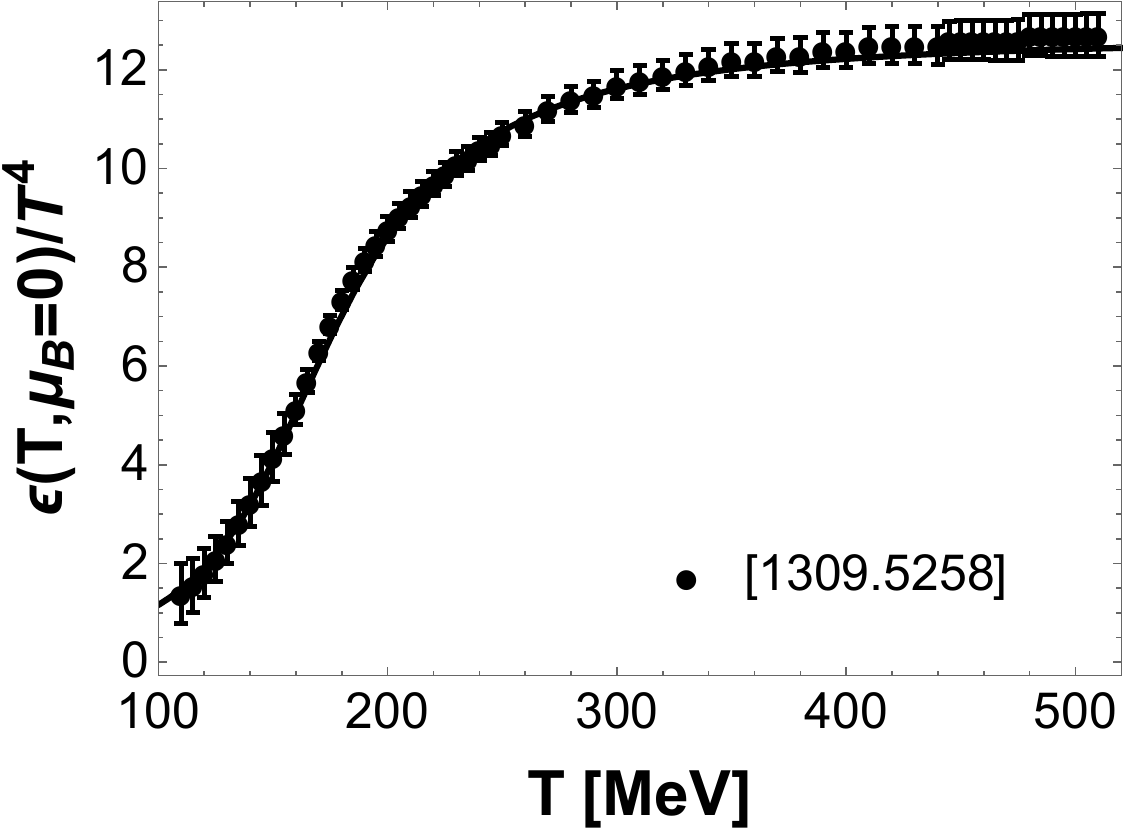}} 
\end{tabular}
\begin{tabular}{c}
\subfigure[]{\includegraphics[width=0.4\textwidth]{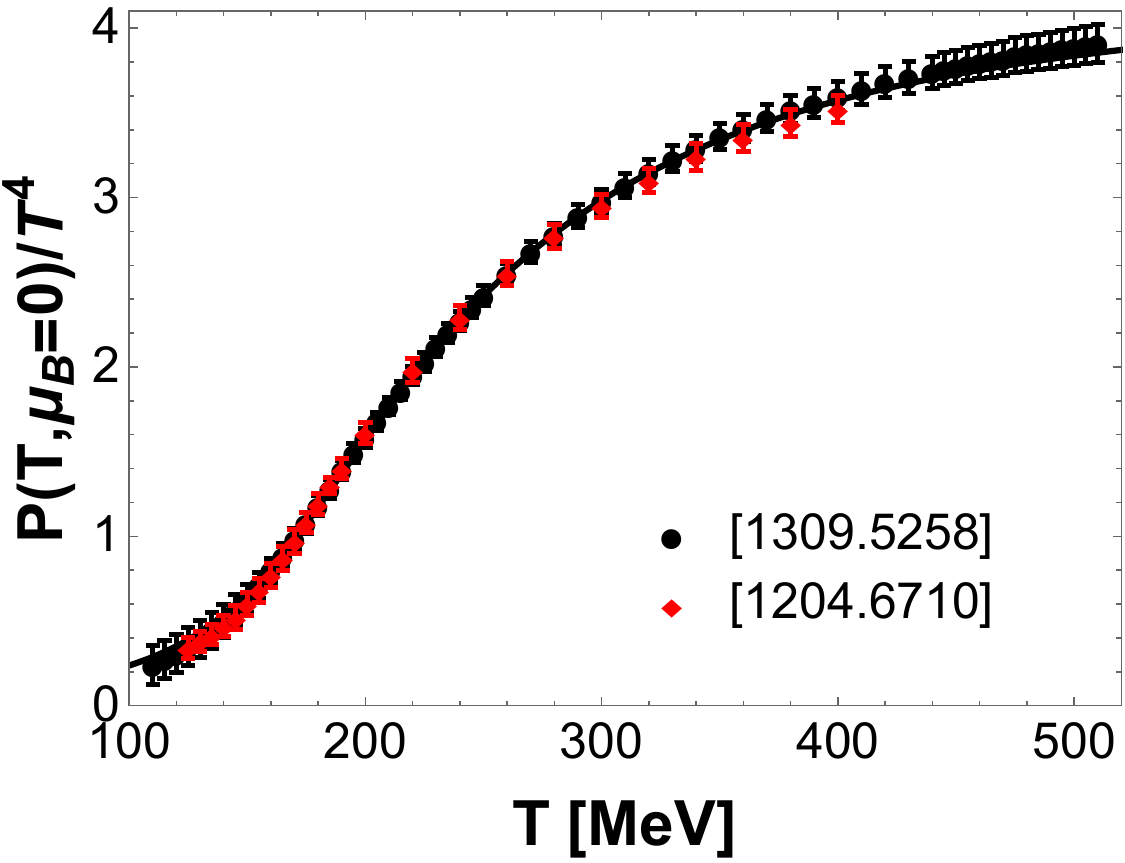}} 
\end{tabular}
\begin{tabular}{c}
\subfigure[]{\includegraphics[width=0.4\textwidth]{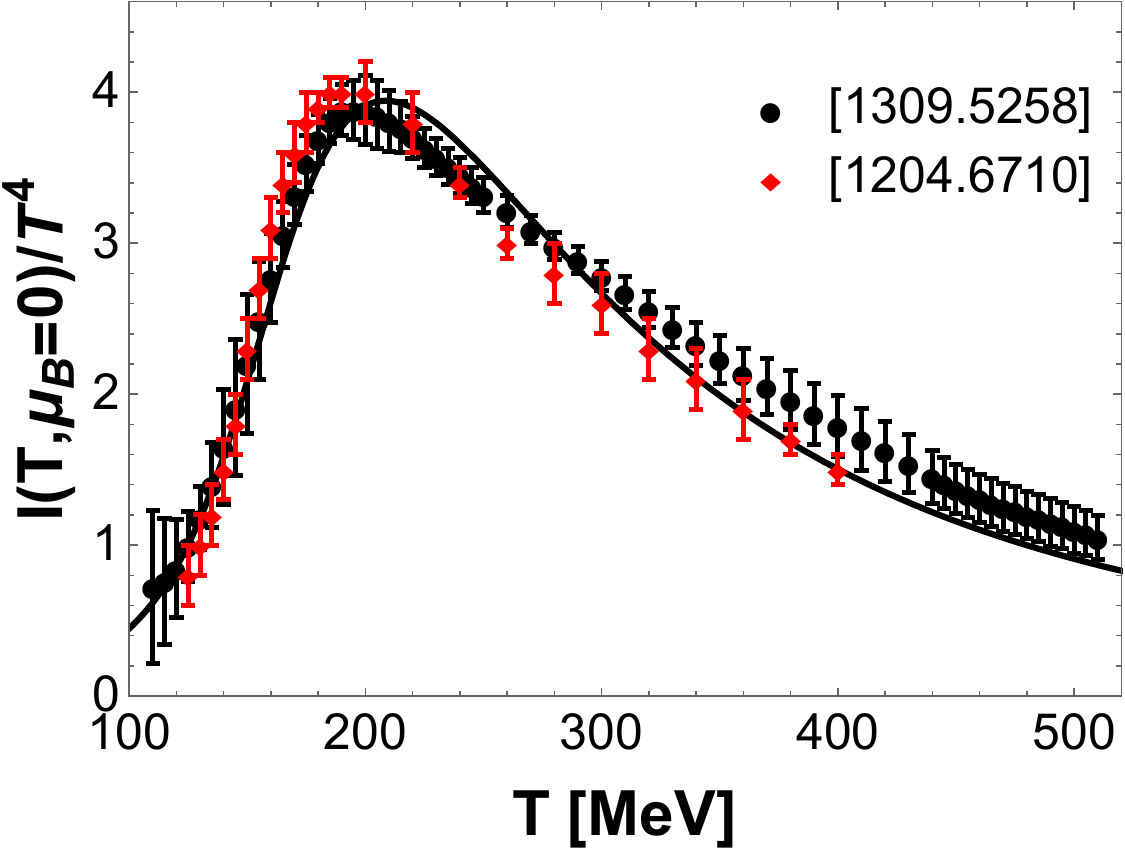}} 
\end{tabular}
\end{center}
\caption{\textbf{From Ref. \cite{Critelli:2017oub}.} Equation of state of the second generation improved EMD model of Refs. \cite{Critelli:2017oub,Grefa:2021qvt,Grefa:2022sav,Rougemont:2018ivt} at zero chemical potential, fitted to the state-of-the-art LQCD results from \cite{Borsanyi:2013bia} (black points). The older LQCD results from \cite{Borsanyi:2012cr} (red points) are also shown to explicitly display the improvements in the lattice results. The QCD equation of state is collectively represented here by the temperature dependence of (a) the entropy density, (b) the square of the speed of sound, (c) the energy density, (d) the pressure, and (e) the trace anomaly.}
\label{fig:EoS0}
\end{figure}

In Fig. \ref{fig:chi2B0}, we display the improvements in the holographic fits, from three different EMD models in the literature, taking as the target data to be described the LQCD results for the reduced second-order baryon susceptibility at vanishing chemical potential --- one can also notice the improvements in the lattice results (see the figure caption for the details). The profile for the Maxwell-dilaton coupling $f(\phi)$ in Eq. \eqref{eq:EMDf} was engineered to produce the result in the bottom panel of this figure, by using Eq. \eqref{eq:chi2B0} evaluated over the zero chemical potential, finite temperature EMD backgrounds. Those backgrounds, in turn, are generated with the choices of the EMD parameters in Eq. \eqref{eq:EMDV}, which were fixed in order to produce the results shown in Fig. \ref{fig:EoS0} for the holographic equation of state at $\mu_B=0$. In Fig. \ref{fig:EoS0}, the full set of LQCD results shown were used as inputs for the model. In particular, using the holographic model it seems very difficult to quantitatively reproduce the LQCD result for the trace anomaly over the entire temperature interval considered.

With the bottom-up EMD parameters for $V(\phi)$ fixed in Eq. \eqref{eq:EMDV} by the results displayed in Fig. \ref{fig:EoS0}, and the parameters for $f(\phi)$ fixed in Eq. \eqref{eq:EMDf} by the results displayed in the bottom panel of Fig. \ref{fig:chi2B0}, one can proceed to make holographic predictions for several observables relevant for the physics of the strongly coupled QGP. Aside from the specific set of LQCD results at $\mu_B = 0$ used to fix the free parameters of the EMD model, any other calculation follows as a legitimate \emph{prediction} of the holographic setup considered.

\begin{figure}[htp!]
\begin{center}
\begin{tabular}{c}
\subfigure[]{\includegraphics[width=0.4\textwidth]{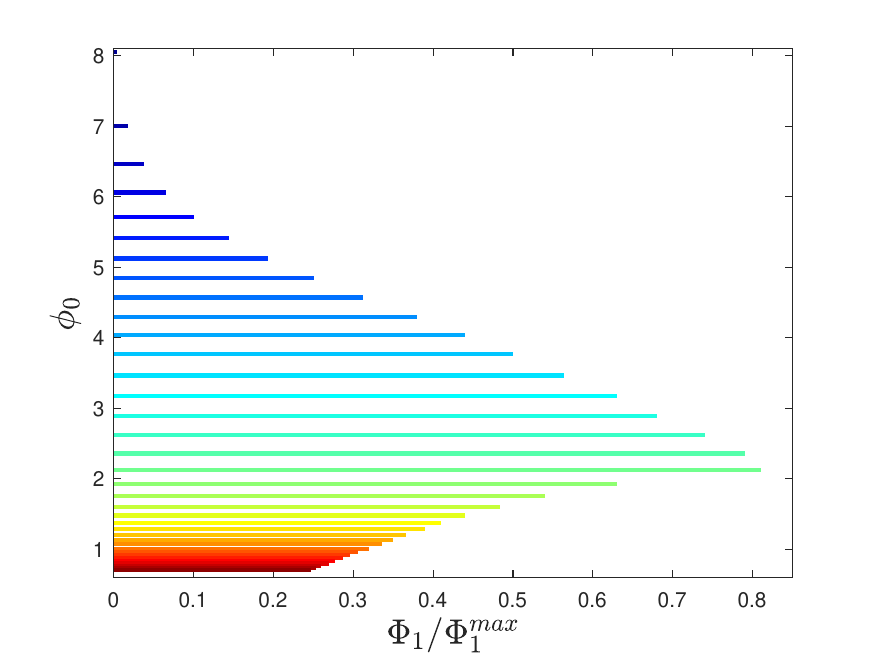}} 
\end{tabular}
\begin{tabular}{c}
\subfigure[]{\includegraphics[width=0.4\textwidth]{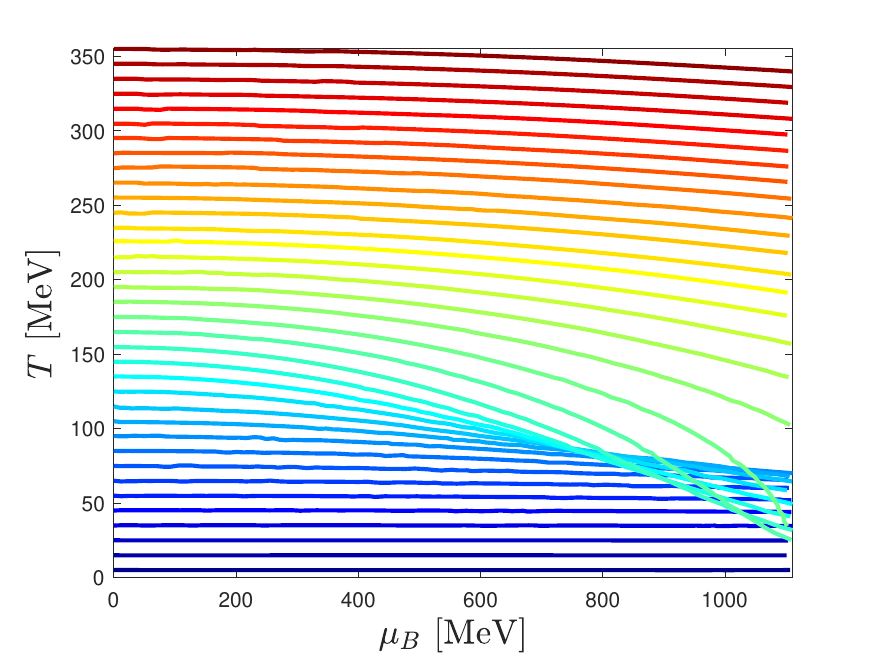}} 
\end{tabular}
\begin{tabular}{c}
\subfigure[]{\includegraphics[width=0.4\textwidth]{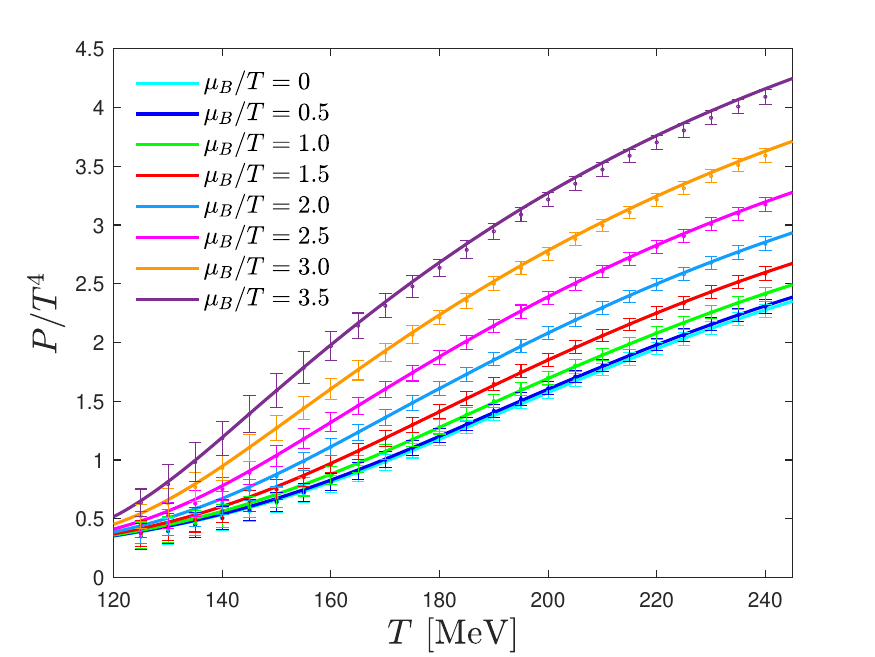}} 
\end{tabular}
\begin{tabular}{c}
\subfigure[]{\includegraphics[width=0.4\textwidth]{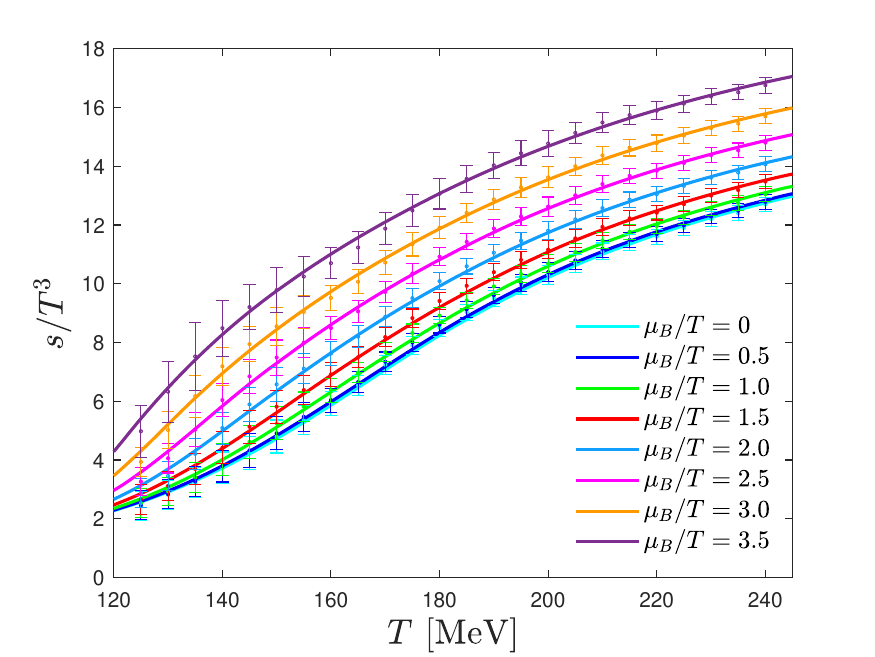}} 
\end{tabular}
\begin{tabular}{c}
\subfigure[]{\includegraphics[width=0.4\textwidth]{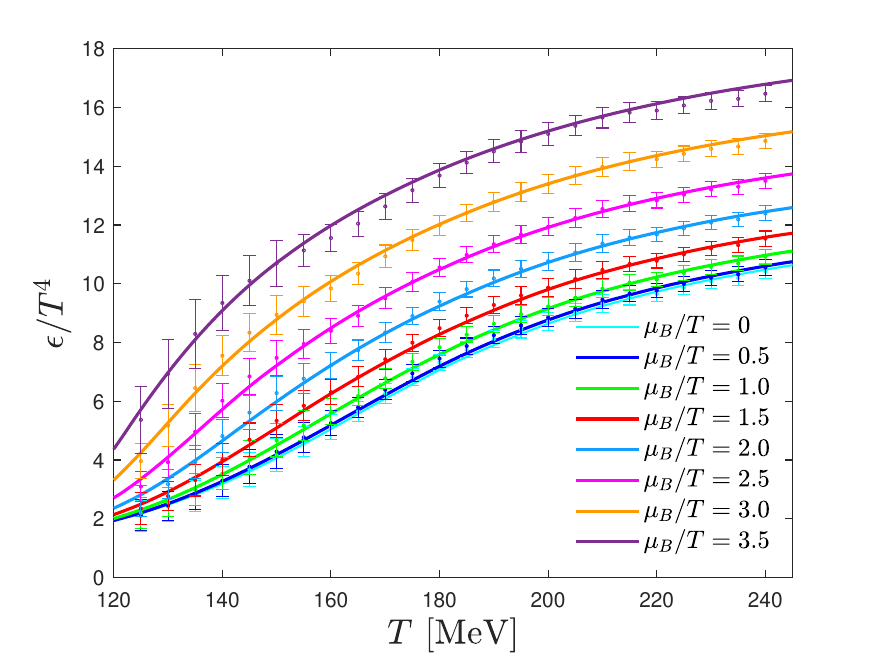}} 
\end{tabular}
\begin{tabular}{c}
\subfigure[]{\includegraphics[width=0.4\textwidth]{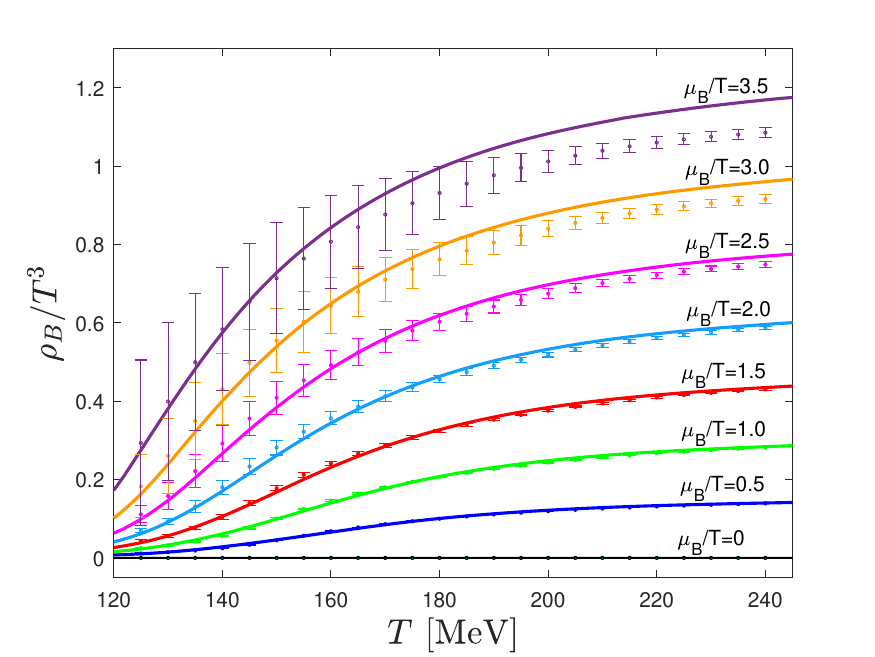}} 
\end{tabular}
\end{center}
\caption{\textbf{From Refs. \cite{Grefa:2021qvt,Grefa:2022sav}.} (a) Grid of initial conditions for the numerical EMD black hole solutions and (b) the corresponding grid of points in the $(T,\mu_B)$-plane of the dual QFT. Predictions for the holographic equation of state at finite temperature and baryon chemical potential compared to state-of-the-art LQCD results from \cite{Borsanyi:2021sxv}: (c) pressure, (d) entropy density, (e) energy density and (f) baryon charge density.}
\label{fig:EMD-ICs-EoS}
\end{figure}

\begin{figure}[h!]
    \centering
    \includegraphics[width=0.80\textwidth]{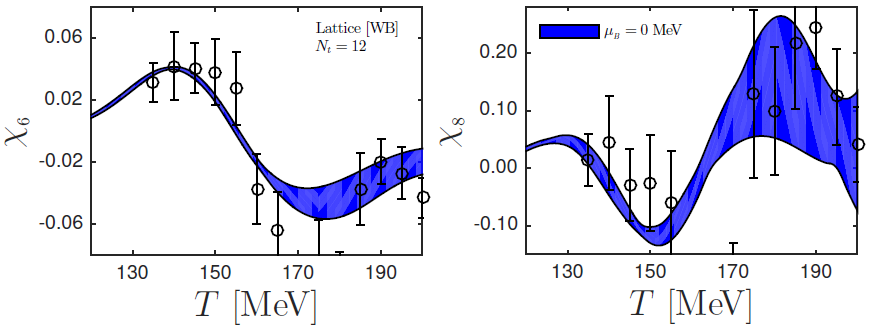}
    \includegraphics[width=0.50\textwidth]{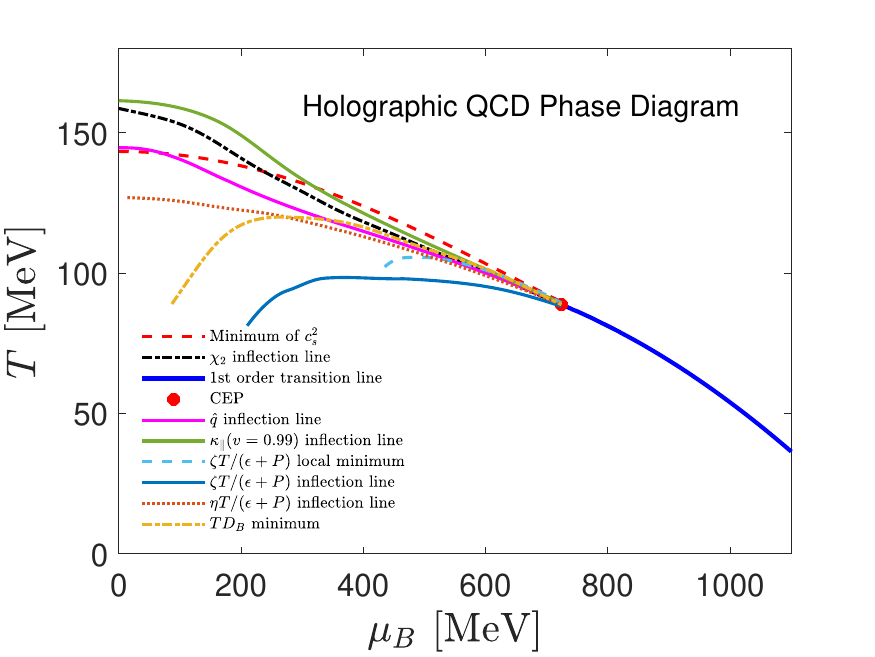}
    \caption{\textbf{Top panel (from Ref. \cite{Rougemont:2018ivt}):} EMD holographic predictions for the sixth and eighth order reduced baryon susceptibilities at zero chemical potential and the corresponding LQCD results from \cite{Borsanyi:2018grb} (still not extrapolated to the continuum, but calculated using a lattice with a temporal extent $N_\tau=12$). \textbf{Bottom panel (from Ref. \cite{Grefa:2022sav}):} EMD holographic prediction for the hot and baryon dense QCD phase diagram.}
    \label{fig:chi68-CEP}
\end{figure}

In order to populate the phase diagram of the model, several EMD black hole solutions are numerically generated with a set of initial conditions $(\phi_0,\Phi_1/\Phi_1^{\textrm{max}})$ chosen as indicated in the two top panels of Fig. \ref{fig:EMD-ICs-EoS} \cite{Grefa:2021qvt}, where $\Phi_1^{\textrm{max}} = \sqrt{-2V(\phi_0)/f(\phi_0)}$ is a bound on the maximum value of $\Phi_1$, given some $\phi_0>0$ (which produces only positive values for the dilaton field), such as to have asymptotically AdS$_5$ solutions \cite{DeWolfe:2010he}. The corresponding holographic EMD \emph{predictions} for the QCD equation of state at finite temperature and baryon chemical potential are also shown in Fig. \ref{fig:EMD-ICs-EoS} and compared to state-of-the-art LQCD results at finite baryon density (with $\mu_Q=\mu_S=0$, as in the holographic model) \cite{Borsanyi:2021sxv}. One notices a good quantitative agreement between the EMD holographic predictions and the lattice results for the QCD equation of state at finite $(T,\mu_B)$, except for the baryon charge density for $T\gtrsim 190$ MeV with $\mu_B/T\gtrsim 2$. It is important to emphasize that the holographic predictions shown in Fig. \ref{fig:EMD-ICs-EoS} were obtained from the holographic EMD model of Ref. \cite{Critelli:2017oub}, which was constructed in 2017, 4 years \textit{before} the publication of the lattice results of Ref. \cite{Borsanyi:2021sxv}. As far as we know, this was the first model in the literature, holographic or not, to correctly predict at the quantitative level the behavior of this state-of-the-art lattice QCD equation of state at finite temperature and baryon chemical potential.
In this regard, it is also important to point out that in the same 2017 paper \cite{Critelli:2017oub}, holographic predictions were put forward for higher-order baryon susceptibilities at zero chemical potential, which were quantitatively confirmed one year later by the LQCD simulations of Ref. \cite{Borsanyi:2018grb}, as depicted in the top panel of Fig. \ref{fig:chi68-CEP}. This is particularly relevant in order to show part of the predictive power of holographic EMD models since baryon susceptibilities higher than the second order one at zero chemical potential were not used to fix the free parameters and functions of the model. Therefore, results for higher order baryon susceptibilities follow as actual holographic predictions of the EMD model.

A broad scanning of the phase diagram of the EMD model of Ref. \cite{Critelli:2017oub}, comprising not only the crossover region and the CEP originally reported in this paper, but also the line of first-order phase transition ending at the CEP, was finally obtained in Ref. \cite{Grefa:2021qvt}, thanks to the significant algorithmic and numerical improvements achieved in that work, which also allowed the calculation of physical observables over the phase transitions regions in the phase diagram of the model. The EMD model prediction for the QCD phase diagram in the $(T,\mu_B)$-plane is displayed in the bottom panel of Fig. \ref{fig:chi68-CEP}, with the predicted CEP location lying around $(T,\mu_B)_{\textrm{CEP}}^{\textrm{[1706.00455]}}\approx(89,724)$ MeV. The different curves characterizing the crossover region refer to characteristic points (extrema or inflections) of different equilibrium and transport observables that evolve with  $\mu_B$ such that they merge at the CEP \cite{Grefa:2022sav}. The CEP location also coincides with the end of the coexistence region with multiple black hole solutions with the same values of $(T,\mu_B)$ in the phase diagram of the model, as displayed in Fig. \ref{fig:EMD-ICs-EoS} (b). Within this coexistence region, the thermodynamically stable branch of black hole solutions refers to the backgrounds with the largest pressure (or, equivalently, the smallest free energy). In Ref. \cite{Grefa:2021qvt}, also the discontinuity gaps for all the considered thermodynamic observables were calculated across the first-order phase transition line.

We remark that the functional forms of $V(\phi)$ and $f(\phi)$ are not uniquely fixed by current lattice QCD results. The very same set of LQCD results at $\mu_B=0$ \cite{Borsanyi:2013bia,Bellwied:2015lba}, which was used to fix the dilaton potential and the Maxwell-dilaton coupling function for the EMD model of Refs. \cite{Critelli:2017oub,Grefa:2021qvt,Grefa:2022sav,Rougemont:2018ivt}, was also employed to fix different functional forms for $V(\phi)$ and $f(\phi)$ in the EMD model proposed in Ref. \cite{Knaute:2017opk}. They also found a good quantitative fit to those set of LQCD results, and a very close result to that of \cite{Critelli:2017oub} ($\Delta\approx 2.73294$) for the scaling dimension of the QFT operator dual to the bulk dilaton field, namely $\Delta\approx 2.769$. Although the EMD model of Ref. \cite{Knaute:2017opk} had not been compared to LQCD results at finite $\mu_B$, it predicts a CEP in a different location in the phase diagram, $(T,\mu_B)_\textrm{CEP}^{\textrm{[1702.06731]}}=(111.5\pm 0.5,611.5\pm 0.5)$ MeV.

\begin{figure}[h]
    \centering
    \includegraphics[width=0.45\textwidth]{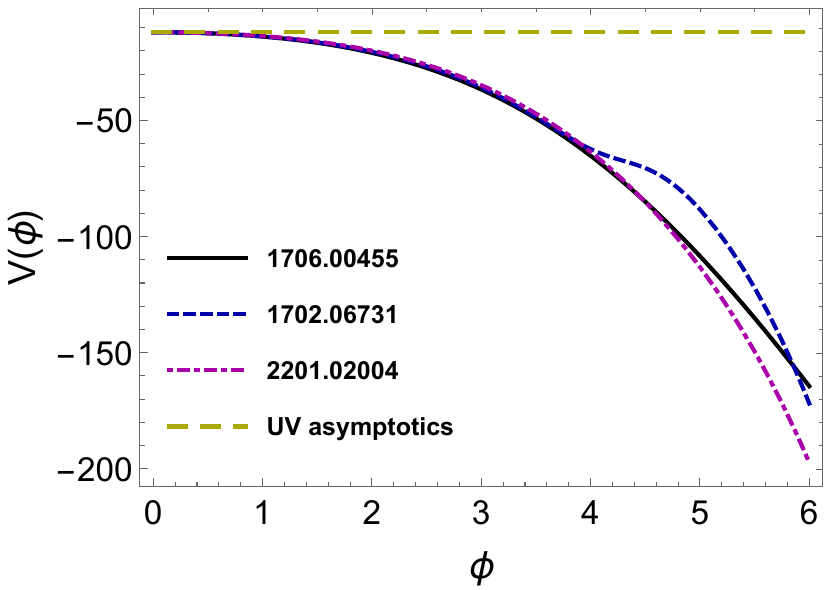}
    \includegraphics[width=0.45\textwidth]{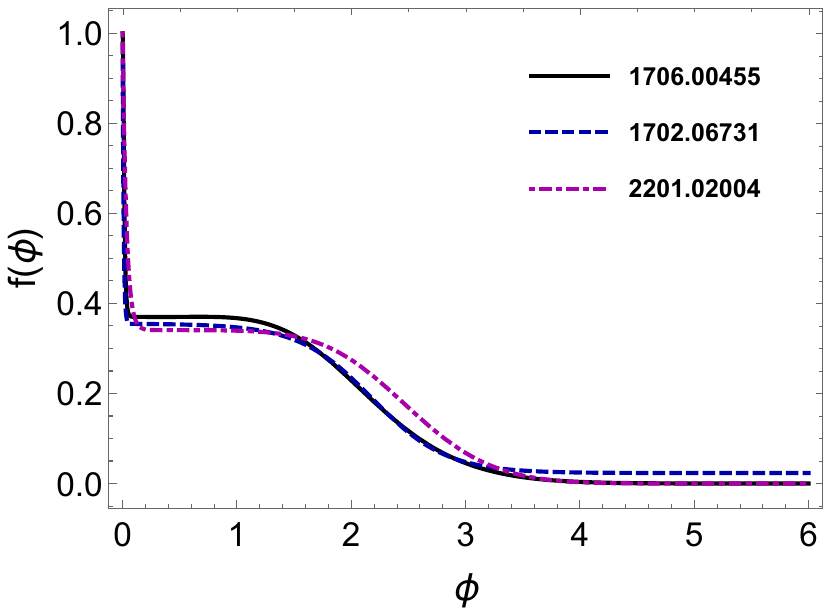}
    \includegraphics[width=0.45\textwidth]{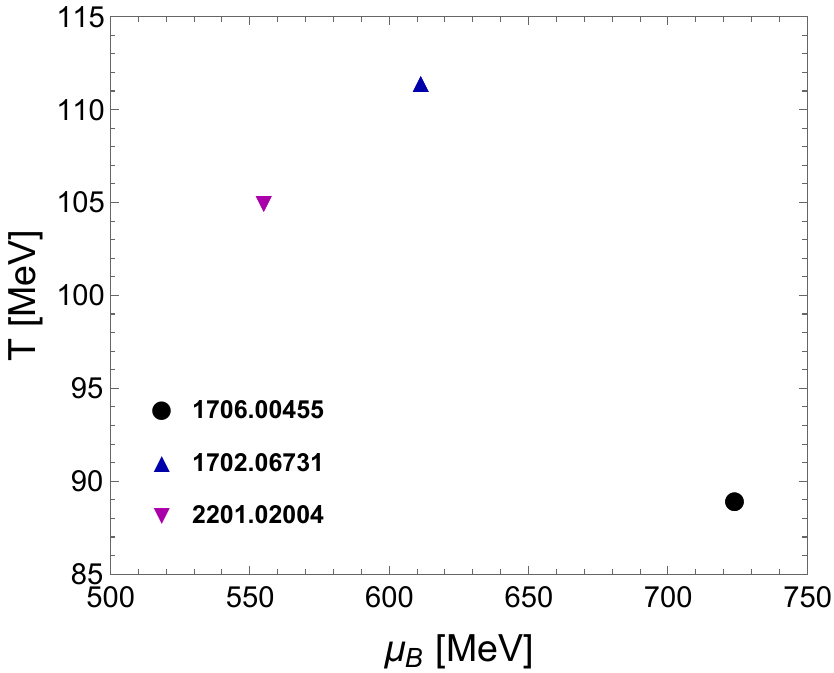}
    \caption{Holographic predictions, from three different and competing bottom-up EMD models \cite{Critelli:2017oub,Knaute:2017opk,Cai:2022omk}, for the location of the conjectured CEP in the QCD phase diagram at finite temperature and baryon chemical potential (bottom panel). We also display the corresponding model functions $V(\phi)$ and $f(\phi)$ (top panel), from where it is evident the very fast variation of the Maxwell-dilaton coupling near the boundary ($\phi\to 0$) for all three EMD models.}
    \label{fig:ceps}
\end{figure}

More recently, another competing EMD model was proposed in Ref. \cite{Cai:2022omk} that employed the LQCD results for the equation of state at finite temperature and $\mu_B=0$ from the HotQCD collaboration \cite{HotQCD:2014kol} to fix $V(\phi)$. For the baryon susceptibility they used the Wuppertal-Budapest results from \cite{Borsanyi:2021sxv} to fix $f(\phi)$, also imposing by construction $\Delta=3$ for the scaling dimension of the QFT operator dual to the dilaton. Up until this work from 2022 \cite{Cai:2022omk}, only the Wuppertal Budapest LQCD results were used.  While the  Wuppertal Budapest and HotQCD collaboration results predominately agree, there are still quantitative differences at large temperatures and the error bars from HotQCD are slightly larger.
Then, the results from \cite{Cai:2022omk} also produced a holographic equation of state in good quantitative agreement with the state-of-the-art LQCD results at finite temperature and baryon chemical potential from Ref. \cite{Borsanyi:2021sxv}, besides a good agreement with lattice results on higher order baryon susceptibilities \cite{Li:2023mpv}, but with yet another different location for the CEP, $(T,\mu_B)_\textrm{CEP}^{\textrm{[2201.02004]}}\approx(105,555)$ MeV.

In Fig. \ref{fig:ceps}, we display the holographic predictions for the QCD CEP from the three  competing bottom-up EMD models mentioned above, which were shown to be in quantitative agreement with available state-of-the-art LQCD results, while presenting different predictions for regions of the QCD phase diagram still out of the reach of first principles LQCD simulations. These three competing EMD models present a very fast variation in the behavior of the Maxwell-dilaton coupling function $f(\phi)$ near the boundary, which seems to be a rather robust feature connected to the holographic EMD description of LQCD results with $2+1$ flavors and physical values of the quark masses. These results motivated the need for a more systematic approach to investigate, in a quantitative way, the structure of the different EMD predictions for the location of the QCD critical endpoint. This can be accomplished through a Bayesian analysis of holographic EMD models, and initial results will be briefly mentioned in section \ref{sec:bayes}.

Before closing this section, we remark that strictly considering the EMD model, the aforementioned phase transitions have no clear order parameter associated to them. These are actual phase transitions in the usual thermodynamic sense, corresponding to regions of the $(T,\mu_B)$-plane where the pressure is non-analytic and does not match its Taylor expansion. In regions where first-order derivatives of the pressure like the entropy and charge densities change discontinuously there is a first-order phase transition, while in regions where these derivatives develop an infinite slope, such that second-order derivatives of the pressure (like the baryon susceptibility and the specific heat) diverge there is a second-order phase transition. In the bottom panel of Fig. \ref{fig:chi68-CEP} the blue line is the first-order phase transition line for the EMD model of Refs. \cite{Critelli:2017oub,Grefa:2021qvt,Grefa:2022sav,Rougemont:2018ivt}, which ends at the red point corresponding to the second-order critical point of that model. Since the EMD models reviewed here are associated to effective holographic descriptions of the hot and baryon dense QCD phase diagram, it is expected that such phase transitions should refer either to the deconfinement and/or to the chiral phase transition. Notice, however, that in QCD with dynamical fermions, as emulated by the EMD models reviewed here, the deconfinement transition has no clear order parameter, since the Polyakov loop is only an actual order parameter for pure Yang-Mills theory \cite{Greensite:2011zz}. Moreover, since the chiral symmetry is not exact in QCD, also the chiral condensate is not an exact order parameter for the chiral phase transition. Nonetheless, in order to compute it one needs to add an extra bulk scalar field to the model, which would play the role of being the source of the chiral condensate at the dual boundary QFT.\footnote{One possibility is to consider an action for an extra probe scalar field defined on top of the thermodynamic EMD backgrounds, as done e.g. in Ref. \cite{Cai:2022omk}. By following a similar reasoning, one can also add extra probe vector fields with the aim of describing some properties associated to hadron spectroscopy, as e.g. in Ref. \cite{Zollner:2021stb}. However, it is clear that such approaches amount to introduce extra fields and free functions/parameters to the holographic setup, which are not part of the EMD model itself. Also, a more general approach considering the fully backreacted interaction between the EMD and possible additional fields would constitute a much more difficult task to implement.}

	\subsubsection{Holographic transport coefficients}
         \label{sec:transport}

\hspace{0.42cm} One of the most attractive features of the holographic gauge-gravity duality, when applied to the strongly coupled QGP, is that, besides the evaluation of thermodynamic observables at finite temperature and baryon density, it also allows for the calculation of transport coefficients entering as microscopic inputs into hydrodynamic calculations and also the evaluation of other microscopic properties such as partonic energy loss. These transport observables, which are of fundamental relevance for the phenomenology of the QGP produced in relativistic heavy-ion collisions, are generally determined through the holographic duality by employing two kinds of approaches, namely,
\begin{enumerate}[i.]
\item Hydrodynamic coefficients (such as the first-order shear and bulk viscosity transport coefficients \cite{Policastro:2001yc,Kovtun:2004de,Policastro:2001yc,Buchel:2003tz,Gubser:2008yx,Gubser:2008sz,DeWolfe:2011ts,Cremonini:2011iq,Rougemont:2017tlu,Grefa:2022sav} and coefficients associated with higher-order derivative expansions of the energy-momentum tensor of the boundary QFT \cite{Baier:2007ix,Bhattacharyya:2007vjd,Finazzo:2014cna}, besides different conductivities and diffusion coefficients associated with the transport of conserved charges \cite{DeWolfe:2011ts,Rougemont:2015ona,Rougemont:2017tlu,Grefa:2022sav,Finazzo:2013efa}), and also the thermal production rates of photons and dileptons within the medium \cite{Finazzo:2015xwa,Caron-Huot:2006pee}, may be evaluated through the use of holographic Kubo formulas obtained via linear response theory. The Kubo formulas relate transport coefficients to the expectation values of retarded thermal correlators of gauge invariant operators at the dual QFT, which can be calculated by solving with some adequate boundary conditions linearized equations of motion for \textit{quadratic perturbations of the bulk fields defined at the level of the bulk action}, with these linearized equations of motion for the perturbations being evaluated over the equilibrium background geometries holographically associated with definite thermal states at the boundary QFT;\footnote{Alternatively, some of these hydrodynamic transport coefficients can also be calculated from the spectra of quasinormal modes in different channels of holographic gauge-gravity models, see e.g. \cite{Baier:2007ix}.}

\item Observables associated with momentum transport and the energy loss of partons within the strongly coupled quantum fluid are generally evaluated by employing the Nambu-Goto action for strings within different setups (which may be holographically associated with probe partons traversing the medium described by the background black hole solutions) \cite{Gubser:2006bz,Herzog:2006gh,Gubser:2006nz,Casalderrey-Solana:2007ahi,Liu:2006ug,Liu:2006he} (see also \cite{Rougemont:2015wca,Grefa:2022sav,Gursoy:2009kk,Gursoy:2010aa,Li:2014hja,Casalderrey-Solana:2014bpa,Brewer:2017fqy}).
\end{enumerate}

Let us first review some relevant EMD predictions for a few hydrodynamic transport coefficients, namely the shear viscosity, bulk viscosity, and baryon conductivity. Afterwards, we shall also briefly review some EMD results for transport observables associated with partonic energy loss.

Here we will consider the calculation of homogeneous hydrodynamic transport coefficients of the hot and baryon dense quantum fluid holographically dual to the EMD model close to thermal equilibrium.   The $SO(3)$ rotation symmetry of the isotropic medium classifies into different irreducible representations (also called ``channels'') the gauge and diffeomorphism invariant combinations of the \textit{linearized plane-wave EMD field perturbations at the level of the equations of motion}, evaluated at zero spatial momentum \cite{DeWolfe:2011ts}. The bulk viscosity of the boundary QFT is holographically dual to the diffeomorphism and gauge invariant bulk EMD perturbation transforming under the singlet (scalar) representation of $SO(3)$. The baryon conductivity is dual to the EMD perturbations transforming under the triplet (vector) representation, and the shear viscosity is dual to the EMD perturbations transforming under the quintuplet (tensor) representation of the $SO(3)$ rotation symmetry group of the isotropic medium. Indeed, due to the fact that these gauge and diffeomorphism invariant EMD perturbations transform under different irreducible representations of $SO(3)$, they do not mix at the linearized level and, consequently, one obtains a single decoupled equation of motion for each of these bulk perturbations \cite{DeWolfe:2011ts,Kovtun:2005ev}.

\begin{figure}
\begin{center}
\begin{tabular}{c}
\subfigure[]{\includegraphics[width=0.45\textwidth]{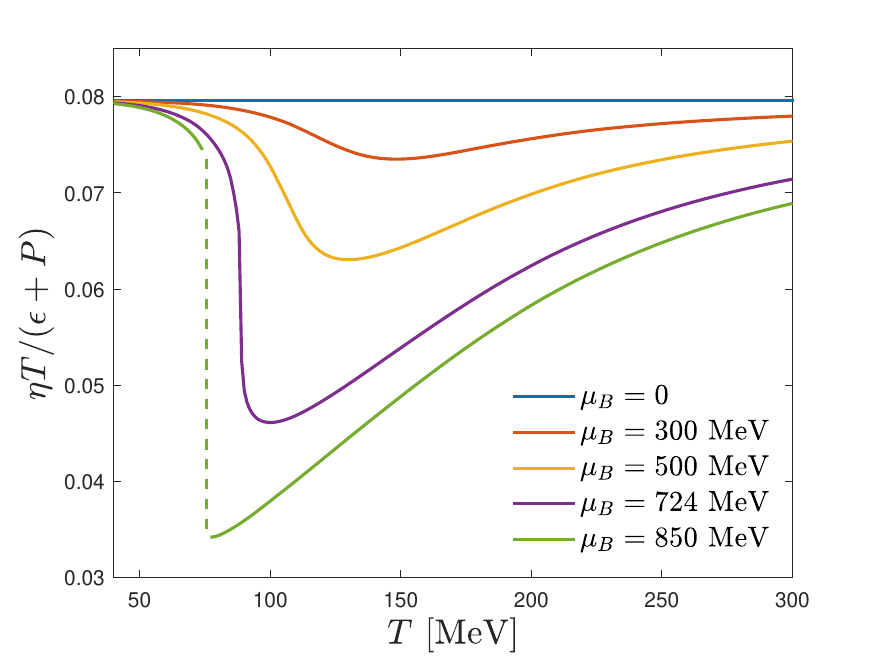}} 
\end{tabular}
\begin{tabular}{c}
\subfigure[]{\includegraphics[width=0.45\textwidth]{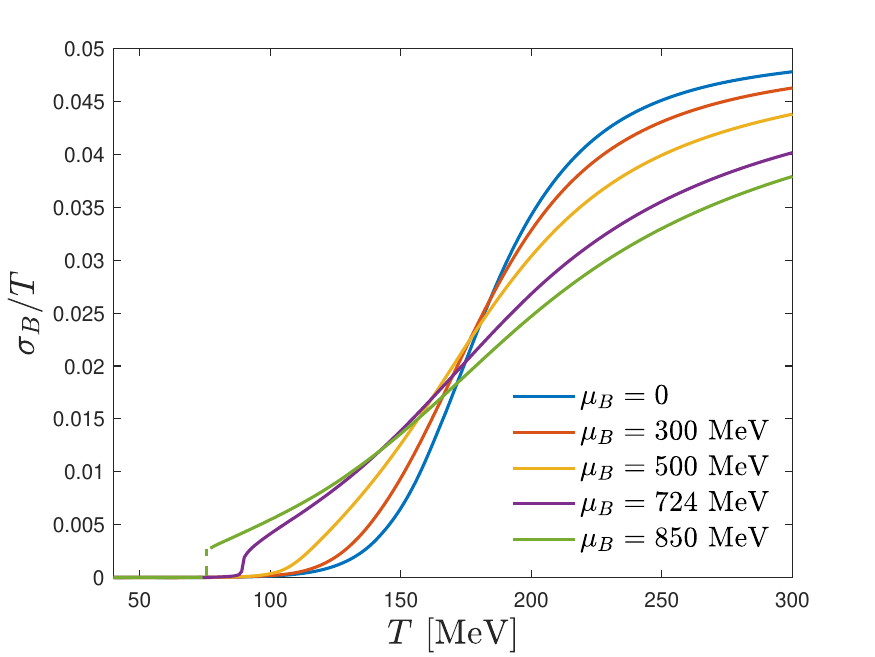}} 
\end{tabular}
\begin{tabular}{c}
\subfigure[]{\includegraphics[width=0.45\textwidth]{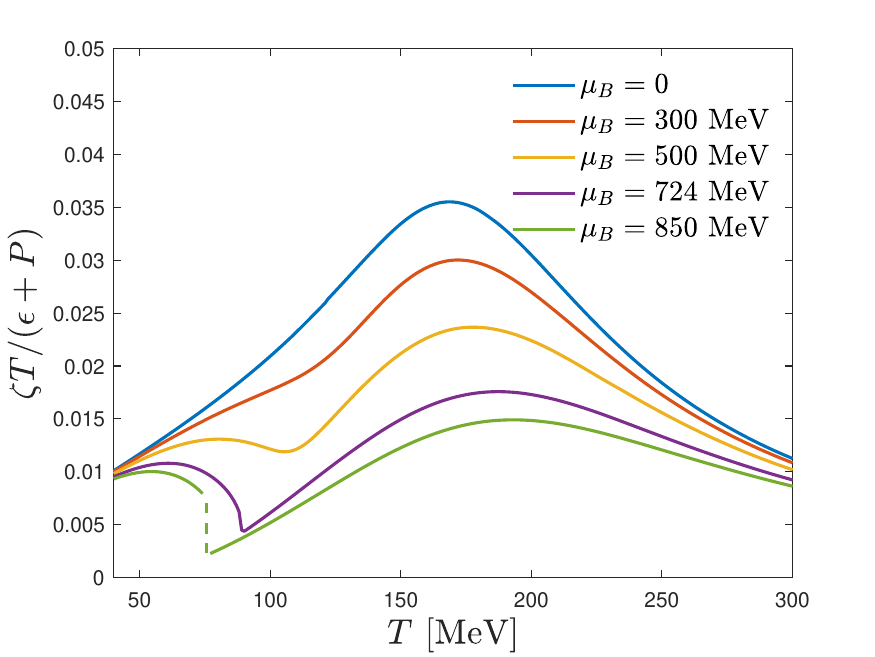}} 
\end{tabular}
\begin{tabular}{c}
\subfigure[]{\includegraphics[width=0.45\textwidth]{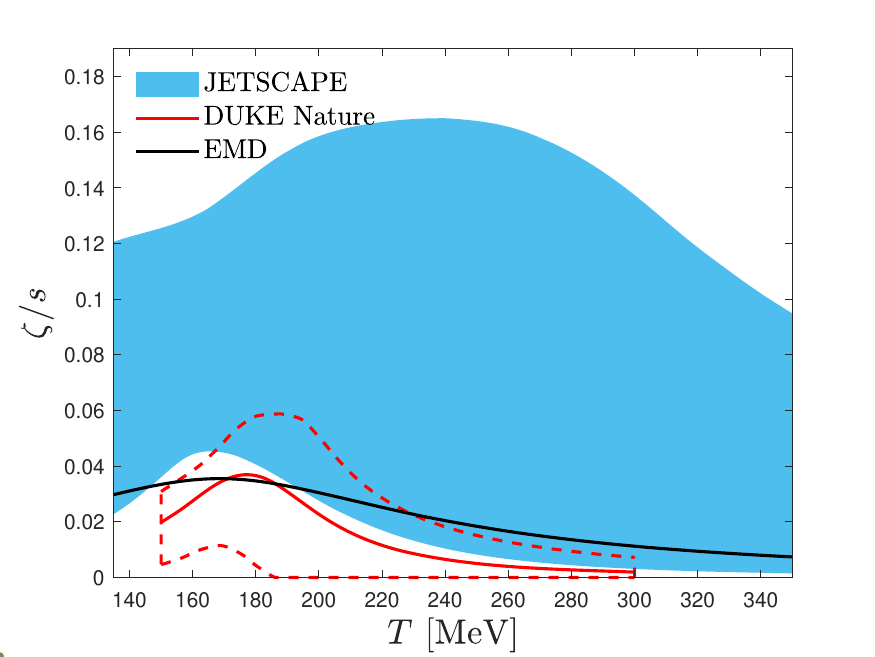}} 
\end{tabular}
\end{center}
\caption{\textbf{From Ref. \cite{Grefa:2022sav}.} Dimensionless combinations involving: (a) the shear viscosity, (b) the baryon conductivity, and (c) the bulk viscosity, calculated as functions of the temperature and the baryon chemical potential of the medium. The different observables develop an infinite slope at the CEP located at $(T,\mu_B)_{\textrm{CEP}}\approx(89,724)$ MeV while displaying discontinuity gaps across the line of first-order phase transition at larger values of $\mu_B$ and lower values of $T$. (d) We also show the comparison between the holographic EMD prediction for the specific bulk viscosity of the QGP at zero baryon chemical potential and the results favored by the phenomenological multistage models from Ref. \cite{JETSCAPE:2020shq} (blue band - the same one plotted on the left side of Fig. \ref{fig:bulkjetscape}) and from Ref. \cite{Bernhard:2019bmu} (red band).}
\label{fig:hydrotransport}
\end{figure}

The tensor components of the isotropic EMD $SO(3)$ quintuplet graviton perturbation are given by five independent combinations of components of the bulk metric field perturbation sourcing the piece of the boundary energy-momentum tensor which is traceless and transverse to the fluid flow. These components satisfy the same differential equation, corresponding to the equation of motion for a massless scalar perturbation over the background geometry considered. The equation of motion has the same form in the standard and in the numerical coordinates (as a consequence of the diffeomorphism invariance of these perturbations) \cite{DeWolfe:2011ts}. 
Then, it was shown \cite{DeWolfe:2011ts} that the shear viscosity satisfies $\eta/s = 1/4\pi\,\,\forall\,\,T>0,\,\mu_B\ge 0$, as expected since the isotropic EMD model fits into the very broad class of holographic gauge-gravity models which are translationally and rotationally invariant, besides having two derivatives of the metric field in the bulk action \cite{Kovtun:2004de,Policastro:2001yc,Buchel:2003tz}. However, the natural dimensionless ratio for the shear viscosity at finite baryon densities is no longer simply $\eta/s$, but rather $\eta T/(\epsilon+P)$ \cite{Liao:2009gb,Denicol:2013nua}. This dimensionless ratio reduces to $\eta/s$ when evaluated at $\mu_B=0$, developing a nontrivial behavior as a function of $(T,\mu_B)$ at nonzero baryon densities. $\eta T/(\epsilon+P)$  has been analyzed in detail across the phase diagram of the EMD model in Ref. \cite{Grefa:2022sav}, where it was shown that $\eta T/(\epsilon+P)$ decreases with increasing values of $\mu_B$.  In that work, $\eta T/(\epsilon+P)$ developed an inflection point and a minimum, with the former evolving toward the CEP of the model, where it acquires an infinite slope. 
For larger values of the baryon chemical potential and lower temperatures, $\eta T/(\epsilon+P)$ develops a discontinuity gap across the first-order phase transition line of the model, as depicted in Fig. \ref{fig:hydrotransport} (a). With the overall reduction in the value of $\eta T/(\epsilon+P)$ with increasing $\mu_B$, the EMD model predicts that the QGP becomes even closer to the perfect fluid limit in its baryon-dense regime.

The three vector components of the EMD $SO(3)$ triplet perturbation are associated with the spatial components of the perturbation of the bulk Maxwell field sourcing the baryon vector current at the dual boundary QFT. Again, due to the spatial isotropy of the medium, these vector components satisfy a single decoupled equation of motion.
One may consider the bulk spatial Maxwell perturbation, $a\equiv a_i$, $i\in\{x,y,z\}$,  to calculate in holography the baryon conductivity, which gives the same result in any direction. The equation of motion for the vector perturbation \cite{DeWolfe:2011ts} is
\begin{align}
a''(r,\omega)+\left[2A'(r)+\frac{h'(r)}{h(r)}+\frac{\partial_\phi f(\phi)}{f(\phi)}\phi'(r)\right]a'(r,\omega)+\frac{e^{-2A(r)}}{h(r)}\left[\frac{\omega^2}{h(r)}-f(\phi)\Phi'(r)^{2}\right]a(r,\omega)=0,
\label{eq:EOMvec}
\end{align}
where $\omega$ is the frequency of the plane-wave ansatz for the Maxwell perturbation and the prime denotes the radial derivative. One must solve Eq. \eqref{eq:EOMvec} imposing the infalling wave boundary condition for the Maxwell perturbation at the background black hole horizon. In holography,  this is equivalent to solving for the retarded thermal correlator of the boundary baryon vector current operator with the further requirement that the Maxwell perturbation is normalized to unity at the boundary \cite{DeWolfe:2011ts}. These two boundary conditions may be systematically implemented by writing the Maxwell perturbation as follows \cite{Grefa:2022sav},
\begin{align}
a(r,\omega)\equiv\frac{r^{-i\omega}P(r,\omega)}{r_{\textrm{max}}^{-i\omega}P(r_{\textrm{max}},\omega)},
\label{eq:aBdyCond}
\end{align}
where $r_\textrm{max}$ is a numerical parametrization of the boundary (see below Eq. \eqref{eq:chi2B0}), and $P(r,\omega)$ is a regular function at the black hole horizon, whose equation of motion is obtained by substituting \eqref{eq:aBdyCond} into \eqref{eq:EOMvec}. 
The holographic Kubo formula for the baryon conductivity in the EMD model in physical units of MeV \cite{DeWolfe:2011ts,Rougemont:2015ona,Rougemont:2017tlu,Grefa:2022sav} is given by 
\begin{align}
\sigma_{B}(T,\mu_{B})=-\frac{1}{2\kappa_{5}^{2}\phi_{A}^{1/\nu}}\lim_{\omega\rightarrow0}\frac{1}{\omega}\left(e^{2A(r)}h(r)f(\phi)\textrm{Im}[a^{*}(r,\omega)a'(r,\omega)]\right)\biggr|_\textrm{on-shell}\,\Lambda\,[\textrm{MeV}],
\label{eq:aKubo}
\end{align}
where the term between brackets in Eq. \eqref{eq:aKubo} is a radially conserved flux that can be calculated at any value of the radial coordinate. 
The details regarding the numerical procedure are discussed in \cite{Grefa:2022sav}. 

The dimensionless ratio $\sigma_B/T$ has been analyzed in detail in Ref. \cite{Grefa:2022sav} where it was shown that it generically increases with the temperature, as displayed in Fig. \ref{fig:hydrotransport} (b).  For $\sigma_B/T$ there is  a temperature window from $T\sim 150-180$ MeV where the different curves at fixed values of $\mu_B$ approximately cross.  For values of temperature above this crossing window, $T>180$ MeV, $\sigma_B/T$ decreases with increasing $\mu_B$, whereas the opposite behavior is observed for temperatures below the crossing window, $T<150$ MeV. One also notices that at the CEP of the model, the baryon conductivity is finite and develops an infinite slope, with a small discontinuity gap being observed across the first-order phase transition line at larger values of $\mu_B$ and lower values of $T$. In Ref. \cite{Grefa:2022sav}, it was also calculated the second-order baryon susceptibility, $\chi_2^B$, and the baryon diffusion coefficient, $D_B$ across the phase diagram of the EMD model. It was found that $\chi_2^B$ diverges at the critical point (a universal feature of all critical points) whereas $D_B\rightarrow 0$ at the CEP, since $D_B=\sigma_B/\chi_2^B$ and $\sigma_B$ remains finite at the critical point of the EMD model.

We remark that the baryon conductivity measures how the system, and more specifically its baryon current, responds to gradients of baryon chemical potential. Therefore, it is also directly proportional to the diffusion of baryon number --- that is, to how gradients of baryon number dissipate in the medium, with the proportionality factor corresponding to the baryon susceptibility. Observable effects of a diffusive baryon current in relativistic heavy-ion collisions were investigated, for instance, in Ref. \cite{Denicol:2018wdp}. Based on $3+1$ hydrodynamic simulations of heavy-ion collisions, effects of the baryon conductivity were found on the difference between proton and antiproton mean transverse momentum, as well as on the difference between proton and antiproton elliptic flow. Understanding the dissipative dynamics of baryon number is particularly important for the theoretical modeling of relativistic heavy-ion collisions at lower beam energies, and, therefore, for the experimental exploration of the QCD phase diagram.

The traceful and transverse piece of the boundary energy-momentum tensor $T^{\mu\nu}$ is associated with the bulk viscous pressure of the medium. Note that $\mathrm{Tr} \left[T^{\mu\nu}\right]\neq 0$ in nonconformal boundary QFTs, where the trace anomaly of $T^{\mu\nu}$ is related to the bulk dilaton field. The scalar EMD $SO(3)$ singlet perturbation is composed by the spatial trace of the graviton and the dilaton perturbation. The singlet perturbation sources the traceful part of $T^{\mu\nu}$, being holographically related to the bulk viscosity. Denoting the singlet perturbation by $\mathcal{H}$, its equation of motion \cite{DeWolfe:2011ts} is shown to be given by 
\begin{align}
\mathcal{H}''+\left[4A'+\frac{h'}{h}+\frac{2\phi''}{\phi}-\frac{2A''}{A'}\right]\mathcal{H}'+\left[\frac{e^{-2A}\omega^{2}}{h^{2}}+\frac{h'}{h}\left(\frac{A''}{A'}-\frac{\phi''}{\phi'}\right)+\frac{e^{-2A}}{h\phi'}\left(3A'\partial_\phi f(\phi)-f(\phi)\phi'\right)\Phi'^{2}\right]\mathcal{H}=0,
\label{eq:EOM-H}
\end{align}
which must be solved with infalling boundary condition at the background black hole horizon and normalized to unity at the boundary. In practice this is  implemented by setting,
\begin{align}
\mathcal{H}(r,\omega)\equiv\frac{r^{-i\omega}F(r,\omega)}{r_{\textrm{max}}^{-i\omega}F(r_{\textrm{max}},\omega)},
\label{eq:HBdyCond}
\end{align}
where $F(r,\omega)$ is a regular function at the black hole horizon, whose equation of motion is obtained by substituting \eqref{eq:HBdyCond} into \eqref{eq:EOM-H}.

The holographic Kubo formula for the bulk viscosity in the EMD model  \cite{DeWolfe:2011ts,Rougemont:2017tlu,Grefa:2022sav} is
\begin{align}
\frac{\zeta}{s}(T,\mu_{B})=-\frac{1}{36\pi}\lim_{\omega\rightarrow0}\frac{1}{\omega}\left(\frac{e^{4A(r)}h(r)\phi'(r)^{2}\textrm{Im}[\mathcal{H}^{*}(r,\omega)\mathcal{H}'(r,\omega)]}{A'(r)^{2}}\right)\biggr|_\textrm{on-shell},
\label{eq:HKubo}
\end{align}
where the term between brackets in Eq. \eqref{eq:HKubo} is a radially conserved flux that may be evaluated at any value of the radial coordinate. The details concerning the numerical calculations are discussed in \cite{Grefa:2022sav}. At $\mu_B=0$, the numerical results obtained using this holographic formula were checked to be the same as the holographic formula provided in \cite{Gubser:2008yx,Gubser:2008sz} by following a different approach based on the $r=\phi$ gauge. The latter approach, however, does not seem to be extensible to finite $\mu_B$ calculations.

Similarly to shear viscosity at $\mu_B>0$, one can no longer use $\zeta/s$ as the natural  hydrodynamic expression, but instead the dimensionless combination $\zeta T/(\epsilon+P)$ that reduces to $\zeta/s$ at $\mu_B=0$. $\zeta T/(\epsilon+P)$ was analyzed in detail in Ref. \cite{Grefa:2022sav}, where it was shown that $\zeta T/(\epsilon+P)$ develops a peak in the crossover region at $\mu_{B}=0$. In contrast to older versions of the EMD model from Refs. \cite{DeWolfe:2011ts,Rougemont:2017tlu}, this peak does not move toward the CEP of the model as one increases $\mu_B$.
Instead, in the EMD model of Ref. \cite{Grefa:2022sav}, the location of the peak in $\zeta T/(\epsilon+P)$ moves to slightly higher values of $T$ as the baryon density increases. 
While in the original EMD construction of Ref. \cite{DeWolfe:2011ts} the height of the peak of $\zeta T/(\epsilon+P)$ remains approximately constant as $\mu_{B}$ increases toward the CEP, both in the second generation improved EMD model of Ref. \cite{Grefa:2022sav} (see Fig. \ref{fig:hydrotransport} (c) ) and in the first generation improved model of Ref. \cite{Rougemont:2017tlu} the magnitude of the peak of $\zeta T/(\epsilon+P)$ reduces as one increases the value of $\mu_{B}$. Therefore, the behavior of the peak of $\zeta T/(\epsilon+P)$ is clearly model dependent within the class of holographic EMD constructions.

In Fig. \ref{fig:hydrotransport} (c) at different values of $\mu_B$, $\zeta T/(\epsilon+P)$ starts to develop both an inflection point and a minimum as a function of $T$, with both characteristic points evolving toward the CEP location as the baryon density of the medium is increased (see also the bottom panel in Fig. \ref{fig:chi68-CEP}). At the CEP, $\zeta T/(\epsilon+P)$ acquires an infinite slope, while further developing discontinuity gaps across the first-order phase transition line of the EMD model. Similarly to what happens with the shear viscosity, the magnitude of the bulk viscosity is also suppressed with increasing values of $\mu_B$. This overall suppression of viscous effects within the strongly coupled medium maybe constitutes a robust property of holographic EMD models seeded with lattice QCD inputs, since this same qualitative behavior has been also observed in the older versions of the EMD model of Refs. \cite{DeWolfe:2011ts,Rougemont:2017tlu}.

In Fig. \ref{fig:hydrotransport} (d), we show the comparison between the EMD prediction for $[\zeta/s](T)$ at $\mu_B=0$ to extracted values of $[\zeta/s](T)$ from recent Bayesian analyses \cite{Bernhard:2019bmu,JETSCAPE:2020shq} that simultaneously describe several experimental heavy-ion data. The holographic EMD prediction for $[\zeta/s](T)$ is in the ballpark of values favored by state-of-the-art phenomenological models. Considering that $\eta/s$ (for any holographic model) is in the correct magnitude for extracted $\eta/s$ from experimental data and that there is  quantitative agreement for the equation of state as well (see  Figs. \ref{fig:EMD-ICs-EoS} and \ref{fig:chi68-CEP}) between the EMD predictions and the QCD equation of state and susceptibilities at finite $(T,\mu_B)$, there is reasonable evidence for the practical and quantitative applicability of bottom-up EMD holography as an effective modeling of the strongly coupled QGP produced in heavy-ion collisions. This argument will be further strengthened in section \ref{sec:EMDmag}, when we will discuss the applicability of the magnetic version of the EMD model at finite temperature and magnetic fields to the physics of the hot and magnetized QGP.

The fact that at the CEP of the EMD model the baryon conductivity and also the shear and bulk viscosities remain finite indicates that the EMD model is compatible with the model B dynamical universality class \cite{Hohenberg:1977ym}. This seems to be a common feature of large $N_c$ gauge theories (as in any holographic gauge-gravity model) \cite{Natsuume:2010bs}, and it is different from general expectations for $N_c=3$ QCD, where these three observables are expected to diverge at the CEP \cite{Son:2004iv,Moore:2008ws,Monnai:2016kud}, in compatibility with the model H dynamical universality class \cite{Hohenberg:1977ym}.

It is also informative to briefly comment on some results obtained from the calculation of the spectra of homogeneous quasinormal modes (QNMs) in the $SO(3)$ quintuplet, triplet, and singlet channels of the EMD model \cite{Rougemont:2018ivt}. In fact, the QNMs of asymptotically AdS black holes \cite{Horowitz:1999jd,Kovtun:2005ev,Berti:2009kk,Heller:2014wfa,Janik:2015iry} encode a wide range of physical information concerning the holographic dual QFT linearly perturbed out of thermal equilibrium.

The near-boundary expansions of the perturbed bulk fields typically feature a leading order non-normalizable mode and a subleading normalizable mode for each field perturbation. The leading modes source the corresponding local and gauge invariant operators at the dual boundary QFT, while the subleading modes are associated with the expectation values of these operators. If one sets the subleading modes to zero at the boundary and imposes the infalling wave condition at the black hole horizon, the corresponding solutions to the linearized equations of motions for the bulk perturbations can be used to evaluate the on-shell action and obtain the retarded thermal correlators of the dual QFT, which are associated through Kubo formulas to transport coefficients of the strongly coupled quantum fluid. For transport coefficients extracted from the imaginary part of the Green's functions, this procedure is physically equivalent to the calculation of transport coefficients through the use of radially conserved fluxes, which has been discussed before.

On the other hand, since the retarded thermal correlators of the dual QFT are given by minus the ratio between the subleading and the leading modes of the bulk perturbations \cite{Son:2002sd}, by setting these leading modes to zero at the boundary and imposing the causal infalling wave condition at the black hole horizon, one gets the poles of these Green's functions. Since the frequency eigenvalue problem for QNMs defined on asymptotically AdS spacetimes is precisely defined by the Dirichlet boundary condition corresponding to the vanishing of these leading modes at the boundary \cite{Kovtun:2005ev},\footnote{Notice this is different from the calculation of transport coefficients discussed before, where these leading modes for the on-shell perturbations of the bulk fields were normalized to unity at the boundary.} one sees that the QNMs describing the exponential decay of linear perturbations of asymptotically AdS black holes holographically correspond to the poles of retarded thermal Green's functions at the dual QFT. These, in turn, describe hydrodynamic and non-hydrodynamic dispersion relations of collective excitations in the strongly coupled quantum fluid, in terms of which it is possible to calculate, respectively, some hydrodynamic transport coefficients \cite{Baier:2007ix,Janik:2016btb} (in an alternative way to the more direct method of holographic Kubo formulas previously discussed) and also some upper values for characteristic equilibration times of the dual QFT linearly perturbed out of equilibrium.

Indeed, as discussed in \cite{Horowitz:1999jd}, the non-hydrodynamic QNMs\footnote{Non-hydrodynamic QNMs are associated with collective excitations of the medium with nonvanishing frequencies even in the homogeneous regime of perturbations with zero wavenumber.} with the lowest absolute value of its imaginary part, corresponding to the longest-lived non-hydrodynamic excitations of the system, give upper bounds for different equilibration times of the medium close to thermal equilibrium. From the lowest homogeneous non-hydrodynamic QNMs in the $SO(3)$ quintuplet, triplet, and singlet channels of the EMD model of Ref. \cite{Rougemont:2018ivt}, it has been shown that the equilibration times in these different channels are very close to each other at high temperatures while developing a pronounced separation at the CEP. This result indicates that the energy-momentum tensor dual to the bulk metric field, the baryon current dual to the bulk Maxwell field, and the scalar condensate dual to the bulk dilaton field, equilibrate at considerably different rates in the critical regime of the EMD model, with the baryon current taking the longest time to approach thermal equilibrium, while the energy-momentum tensor generally equilibrates faster than the other observables, also within the regions of the phase diagram far from the criticality. Moreover, in most cases, the characteristic equilibration times of the medium decrease with increasing values of the baryon chemical potential, while strongly increasing with decreasing values of temperature.

There have been also various holographic calculations of transport coefficients associated with partonic energy loss within strongly coupled quantum fluids, such as the energy loss of heavy quarks due to the heavy quark drag force \cite{Gubser:2006bz,Herzog:2006gh}, the Langevin momentum diffusion coefficients for heavy quarks \cite{Gubser:2006nz,Casalderrey-Solana:2007ahi}, and the jet quenching parameter associated with the energy loss of light partons moving at the speed of light \cite{Liu:2006ug,Liu:2006he}.  
These energy loss transport coefficients are evaluated by considering different calculations done with a probe Nambu-Goto (NG) action for a classical string defined over the background solutions for the bulk fields. 
The NG action depends on $\sqrt{\lambda_t}$, where the `t Hooft coupling is typically considered in holographic calculations as an extra free parameter. In principle, this parameter may be fixed in different ways by considering holographic observables calculated with the NG action compared to different kinds of phenomenological data (see e.g. Refs. \cite{Rougemont:2015wca,Critelli:2016cvq}). For the class of isotropic EMD models at finite temperature and baryon density, the holographic formulas for these partonic energy loss observables were derived in Ref. \cite{Rougemont:2015wca}. The corresponding results for the improved EMD model \cite{Grefa:2022sav} were numerically calculated across its phase diagram, including the regions with the CEP and the line of first-order phase transition. It was found that the heavy quark drag force and energy loss, the Langevin momentum diffusion, and the jet quenching parameter are all enhanced by increasing the baryon density of the medium toward the critical region of the phase diagram.  In fact, faster partons are more sensitive to the temperature and baryon chemical potential of the medium. Those results indicate that there is more jet suppression and partonic energy loss in the baryon-dense regime of the fluid. All of these observables developed an infinite slope at the CEP, while displaying large discontinuity gaps across the line of first-order phase transition. In the bottom panel of Fig. \ref{fig:chi68-CEP} some crossover characteristic curves (made of sequences of inflection points or extrema) of these observables converging to a single location corresponding to the CEP are displayed with other characteristic curves for different observables of the model --- see Ref. \cite{Grefa:2022sav} for details.

\begin{table}[h]
    \centering
    \begin{tabular}{|c||c|c|}
    \hline
    \multicolumn{3}{|c|}{Prior}\\
    \hline\hline
         Parameter & min & max  \\
\hline\hline
        $\Lambda$ & 400 MeV & 1400 MeV\\
    \hline
        $\kappa_2$ & 9.0 & 15.0 \\
    \hline\hline
       $\gamma_1$ & 0.40 & 0.57 \\
       \hline
       $\gamma_2$ & 0.50 &0.68 \\
    \hline
      $\Delta \phi_V$ & 1.5 & 3.0 \\
      \hline\hline
       $A$ & 0.25 & 0.50\\
    \hline
       $\phi_1$ & -0.1 & 0.5\\
    \hline
       \phantom{(J)} $\delta\phi_1$ {(J)} & $10^{-5}$ & 0.3\\
    \hline
       $\phi_2$ & 0.8 & 4.5 \\
    \hline
       $\delta\phi_2$ & 0.2 & 4.0\\
    \hline    
    \end{tabular}\hspace{2cm}%
    \begin{tabular}{|c||c|c||c|}
    \cline{1-3}
     \multicolumn{3}{|c||}{Posterior 95\% CI}&\multicolumn{1}{c}{}\\
    \cline{1-3}\hline
         Parameter & min & max & best value \\
    \hline\hline
        $\Lambda$ & 862 MeV & 1043 MeV & 955 MeV\\
    \hline
        $\kappa_2$ & 11.3 & 11.5 & 11.4 \\
    \hline\hline
       $\gamma_1$ & 0.50 & 0.54 & 0.52 \\
       \hline
       $\gamma_2$ & 0.60 & 0.62 & 0.61 \\
    \hline
      $\Delta \phi_V$ & 1.6  & 2.1 & 1.8 \\
      \hline\hline
       $A$ & 0.369 & 0.374 & 0.371\\
    \hline
       $\phi_1$ & 0.000 & 0.025 & 0.002\\
    \hline
       \phantom{(J)}  $\delta\phi_1$  {(J)}  & 0.0001 & 0.0032 & 0.0003\\
    \hline
       $\phi_2$ & 2.1 & 2.3 & 2.2 \\
    \hline
       $\delta\phi_2$ & 0.65 & 0.73 & 0.69\\
    \hline
    \end{tabular}    
    \caption{Prior ranges (left) and 95\% confidence intervals for the posterior distribution (right) for parameters of the parametric ansatz for the EMD model, fit to reproduce  the baryon susceptibility and the entropy density from lattice QCD at $\mu_B=0$, within error bars \cite{Borsanyi:2013bia,Borsanyi:2021sxv}. The symbol `(J)' marks $\delta\phi_1$, for which Jeffreys prior, uniform over its logarithm, was taken instead of a uniform distribution.}
    \label{tab:bayesian}
\end{table}

\subsubsection{Holographic Bayesian analysis}
 \label{sec:bayes}

\begin{figure}[h]
\begin{center}
{\includegraphics[width=\textwidth]{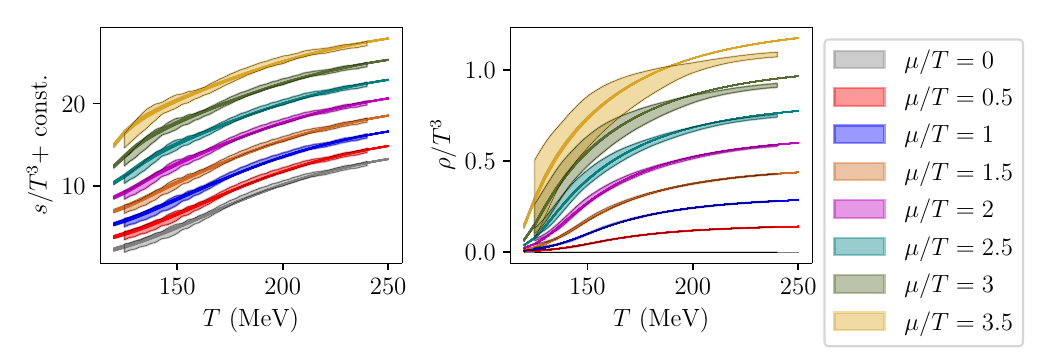}}
\end{center}
\caption{Comparison of the finite-density QCD equation of state from the lattice to Bayesian predictions derived within 
the parametric Ansatz for the EMD model. A representative sample of the posterior distribution is shown as colored solid lines, which accumulate in the finer bands. Lattice QCD results \cite{Borsanyi:2021sxv} are shown as the wider translucent bands. 
The  panel on the left shows a comparison to the entropy density, where results corresponding to different values of $\mu/T$ are shifted so as to avoid overlap between them. The panel on the right shows results for the baryon density. }
\label{fig:Bayesian}
\end{figure}

\hspace{0.42cm} The results for the EMD model discussed above rely on the choice of holographic potentials $V(\phi)$ and $f(\phi)$. That is, calculations require that suitable functional forms are provided, along with the corresponding parameters. As discussed above, several competing parametrizations for these functions can be found in the literature, but no systematic comparison between them has been performed thus far. 
A pressing question regarding any particular parametrization of the EMD model concerns how much of its predictions are informed by lattice QCD results used to fit the different parameters, and how robust they are against uncertainties in these results. 
 Such issues can only be addressed by quantifying uncertainties in $V(\phi)$ and $f(\phi)$ and systematically comparing different parametrizations. 

\hspace{0.42cm} The tools required for a systematic analysis of parameter sensitivity and uncertainty quantification in modeling the QCD equation of state can be found in the framework of Bayesian statistical inference \cite{sivia1996data,gregory2005bayesian}. 
In recent years, Bayesian statistics have become the state-of-the-art tool for systematically assessing models and hypotheses across high-energy physics, including neutron-star \cite{Lim:2019som,Huth:2021bsp,Zhou:2023hzu,Gorda:2023usm} and heavy-ion physics \cite{Bernhard:2019bmu,JETSCAPE:2020shq,Nijs:2020roc,Moreland:2018gsh,JETSCAPE:2020mzn,Nijs:2021clz}.  
The core tenet of Bayesian inference resides in Bayes' theorem:
\begin{equation}\label{eq:Bayes}
    P(M^{(\theta)}| D) = \frac{P(D| M^{( \theta)}) \times P(M^{(\theta)})}{P(D)},
\end{equation}
where $D$ represents the data and $M^{(\theta)}$ is a given model with parameters $\theta$. Equation~\eqref{eq:Bayes}  follows from expressing the joint probability $P(D\cap M^{(\theta)})$ in terms of the associated conditional probabilities $ P(M^{(\theta)}| D)$ and $P(D| M^{( \theta)})$. The conditional distribution $ P(M^{(\theta)}| D)$ is called the posterior and can be used to discriminate between different parameter sets $\theta$. It is the product of the likelihood $P(D| M^{( \theta)})$, quantifying agreement between model and data, and the prior $P(M^{(\theta)})$, which assigns a priori weights to the different parameter sets to reflect prior knowledge. 
The denominator $P(D)$ on the right-hand side of Eq.~\eqref{eq:Bayes} is known as the evidence and can be obtained as a normalization constant. 

Recently, an improved numerical implementation of the EMD model developed within the \href{https://muses.physics.illinois.edu/}{MUSES Collaboration} has enabled a Bayesian analysis over lattice QCD results for the zero-density equation of state \cite{Hippert:2023bel}. 
In Eqs.~\eqref{eq:EMDV} and \eqref{eq:EMDf}, the very nonlinear character of the potentials over $\phi$ make functional forms, such as seen in Fig.~\ref{fig:ceps}, highly sensitive to precise parameter values.  
A complete Bayesian analysis is presented in \cite{Hippert:2023bel}, while here, we briefly highlight and explain the results obtained from an initial analysis.

New parametric ansatze for the free functions $V(\phi)$ and $f(\phi)$ of the holographic EMD action \eqref{eq:EMDaction} are introduced to reproduce qualitative features of Eqs.~\eqref{eq:EMDV} and \eqref{eq:EMDf} in a way that depends more transparently on parameter values:
\begin{equation}\label{eq:parametric_V}
 V(\phi) = -12\cosh\left[\left(\frac{\gamma_1\,\Delta\phi_V^2 + \gamma_2 \,\phi^2}{\Delta \phi_V^2 + \phi^2}\right) \phi\right],
\end{equation}
\begin{equation}\label{eq:parametric_f}
 f(\phi) = 1 - (1-A_1) \left[\frac{1}{2} + \frac{1}{2}\tanh\left(\frac{\phi - \phi_1}{\delta \phi_1}\right)\right]  
 - A_1\left[\frac{1}{2} + \frac{1}{2}\tanh\left(\frac{\phi - \phi_2}{\delta \phi_2}\right)\right].
\end{equation}
Equation~\eqref{eq:parametric_V} interpolates between two different exponential slopes, $\gamma_1$ and $\gamma_2$, for $\phi\ll\Delta\phi_V$ and $\phi\gg\Delta\phi_V$, respectively. Equation~\eqref{eq:parametric_f}, on the other hand, goes from $f(\phi)\approx 1$, for $\phi_1-\phi\ll \delta\phi_1$ to a plateau of height $f(\phi)\approx A$,  for $\phi$ in the range $\phi_1-\phi_2$, before finally going to $f(\phi)\approx 0$, for $\phi-\phi_2\gg\delta\phi_2$. 

The prior distribution for parameter values was taken to be uniform within designated ranges, shown in Table~\ref{tab:bayesian}, on the left. In the case of a single parameter, marked in Table~\ref{tab:bayesian} with a `(J)', a uniform distribution over its logarithm was chosen instead (i.e., Jeffreys prior was employed).
Random samples from this prior distribution were then fed into a  Markov Chain Monte Carlo (MCMC) algorithm \cite{speagle2019conceptual,2006S&C....16..239T}. This MCMC implements random changes to parameters such that the equilibrium probability distribution, to be reached after a sufficiently large number of iterations, coincides with the posterior distribution given by Eq.~\eqref{eq:Bayes}. This algorithm can then be reiterated to generate a large sample of parameter sets from the posterior.  
Parameters are fit based on both the baryon susceptibility and the entropy density from lattice QCD at $\mu_B=0$ \cite{Borsanyi:2013bia,Borsanyi:2021sxv}. 
The agreement between model and lattice results is quantified by the likelihood $P(D| M^{( \theta)})$, chosen to be Gaussian.  
The corresponding covariance matrix is chosen according to  the lattice QCD error bars while implementing auto-correlation between neighboring  points. An extra parameter is introduced to gauge these correlations and is also estimated within the Bayesian inference \cite{sivia1996data}. 

The 95\% confidence interval obtained from lattice QCD results in this fashion is shown in Table~\ref{tab:bayesian}, on the right.  
Finally, parameter sets from the posterior can be used to compute predictions. The statistical distribution of predictions can then be used to quantify uncertainties stemming from the lattice QCD errorbars, as well as the sensitivity to different model parameters. As a check that these predictions are compatible with lattice QCD results, Fig.~\ref{fig:Bayesian} compares predictions for different values of $\mu_B/T$, shown as thin semitransparent lines, to the finite-density lattice QCD equation of state from \cite{Borsanyi:2021sxv}, shown as wide bands with the same color scheme. While it is not apparent at first sight, thousands of lines are shown over each band in Fig.~\ref{fig:Bayesian}. 
Remarkably, the zero-density equation of state constrains the model parameters so tightly that these lines accumulate in what appears to be a very thin band. 

Constraining the model with input from lattice QCD in this fashion, one is able to extract predictions at higher densities, and even around the QCD phase transition. 
Because it generates a large set of model realizations, this Bayesian analysis of the EMD model will also enable the investigation of the role of each different model parameter, both in predictions and in fitting lattice results. 
Perhaps even more importantly, this kind of analysis provides the possibility of assigning probabilities to predictions and hypotheses. 
In principle, Bayesian model selection can also be used to discriminate between different models. 
Overall, the combination of bottom-up holographic models with Bayesian tools thus provides a promising tool for extrapolating knowledge on the low-density and high-temperature QCD equation of state to higher densities in a partially systematic way. 
Because of its ability to capture the physics of the strongly coupled QGP in the crossover region, the EMD model is a particularly fitting candidate for this task.

	\subsection{Other holographic models}
         \label{sec:others}

\hspace{0.42cm} Although the focus of the present review is on the results from bottom-up holographic EMD models for the hot and baryon-dense QCD phase diagram, in this section, we briefly mention some results obtained from other kinds of holographic constructions.

Within the broad class of bottom-up Einstein-Dilaton constructions, but without considering the effects of flavor dynamics effectively enclosed in the form of the dilaton potential matched to the corresponding LQCD results, as originally proposed in Refs. \cite{Gubser:2008ny,Gubser:2008yx,DeWolfe:2010he,DeWolfe:2011ts}, there is the so-called class of ``Improved Holographic QCD'' (ihQCD) models originally devised in Refs. \cite{Gursoy:2007cb,Gursoy:2007er}, and further reviewed in \cite{Gursoy:2010fj}. Due to the fact that flavor dynamics are not taken into account in those ihQCD models, such a class of bottom-up holographic constructions actually refers to effective models for pure Yang-Mills systems, instead of QCD. In a pure YM system at $T=0$ at large color-charge separations, there is a linear confining potential for infinitely heavy probe quarks as well as a mass gap featured in the physical spectrum of glueball excitations, which are both well described by ihQCD models. 
In contrast to the deconfinement crossover observed in actual QCD with $2+1$ dynamical quark flavors, pure YM theory has a first-order phase transition  between a confining gas of glueballs and a deconfined phase corresponding to a pure gluon plasma. 
At finite temperature the  ihQCD models are able to achieve this first-order phase transition, just like what is seen in pure YM theory.  
 However, $\eta/s=1/4\pi$ in these ihQCD models that demonstrates the theory is  strongly coupled at all energy scales and, therefore, misses crucial properties related to asymptotic freedom in the ultraviolet. 
 As explicitly shown in Ref. \cite{Cremonini:2012ny}, higher curvature corrections to ihQCD models can provide a nontrivial temperature dependence for $[\eta/s](T)$, allowing this observable to acquire a similar profile to what is expected for pure YM and also QCD matter where $[\eta/s](T)$ is expected to largely increase with the temperature of the medium in the ultraviolet regime due to asymptotic freedom. 
 Simple Einstein's gauge-gravity models with two derivatives of the metric field in the bulk gravity action lack asymptotic freedom, while the consideration of higher curvature corrections for the bulk action is associated with corrections that reduce the value of the effective `t Hooft coupling of the dual QFT at the boundary.

 \begin{figure}[h]
\begin{center}
\begin{tabular}{c}
\subfigure[]{\includegraphics[width=0.45\textwidth]{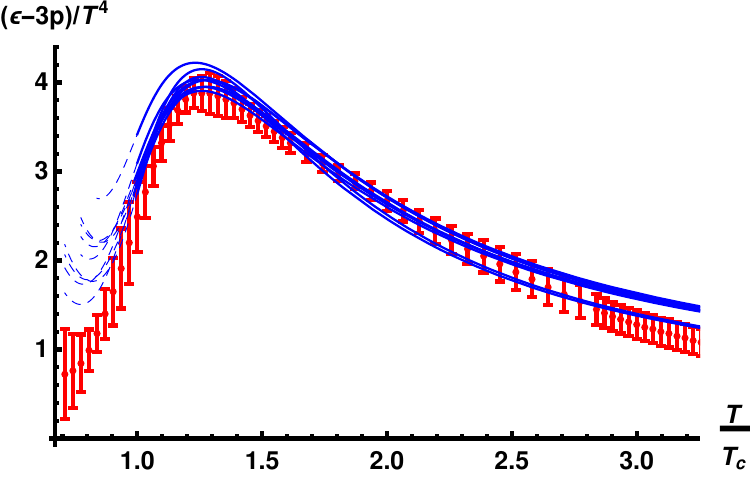}} 
\end{tabular}
\begin{tabular}{c}
\subfigure[]{\includegraphics[width=0.45\textwidth]{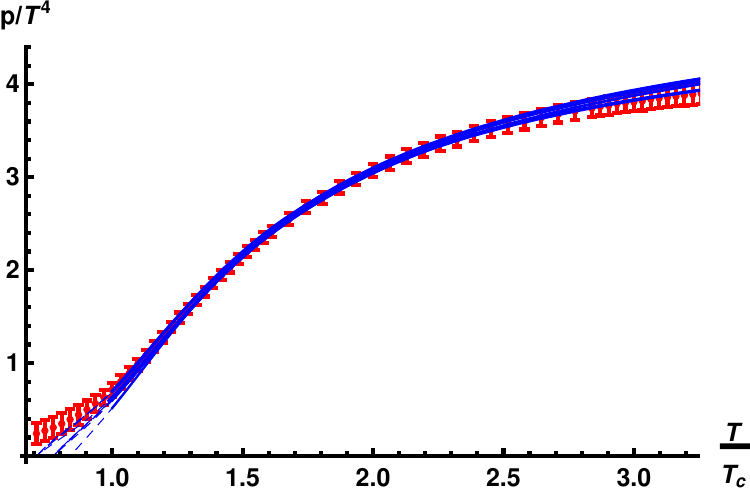}} 
\end{tabular}
\begin{tabular}{c}
\subfigure[]{\includegraphics[width=0.45\textwidth]{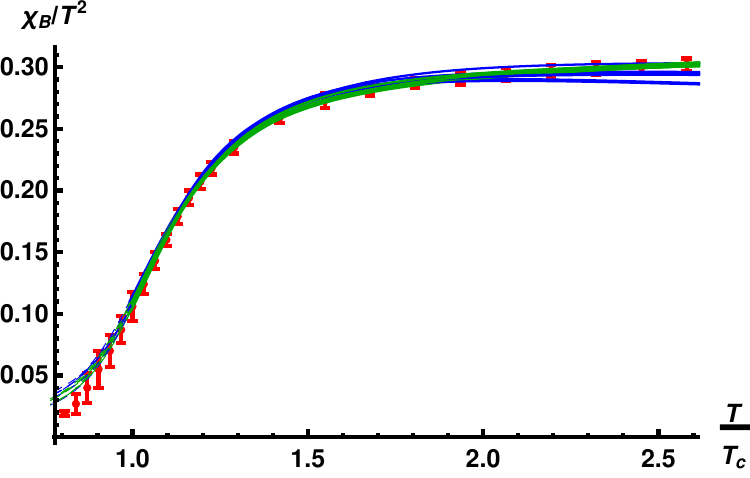}} 
\end{tabular}
\end{center}
\caption{\textbf{From Ref. \cite{Jokela:2018ers}.} Fits for several different holographic V-QCD models to: (a) the trace anomaly, (b) the pressure, and (c) the second order baryon susceptibility, taking as phenomenological inputs LQCD results with $2+1$ flavors at $\mu_B=0$ from Refs. \cite{Borsanyi:2013bia,Borsanyi:2011sw}.}
\label{fig:V-QCD}
\end{figure}

Generalizations of the original ihQCD constructions for pure YM systems that consider a very large number $N_f$ of quark flavors where the ratio $x\equiv N_f/N_c$ remains finite in the holographic setup\footnote{The number of colors $N_c$ is always very large.} are known in the literature as V-QCD models \cite{Jarvinen:2011qe,Alho:2013hsa,Jokela:2018ers,Jarvinen:2021jbd}.
The letter ``V'' stands for the so-called Veneziano limit of large $N_c,\,N_f$, with fixed $x=N_f/N_c$. In such bottom-up models, the flavor dynamics are taken into account by considering the full backreaction of tachyonic flavor D-branes on the gluonic backgrounds. The V-QCD models have been employed to calculate a large number of physical observables, ranging from spectroscopy \cite{Amorim:2021gat,Jarvinen:2022gcc} to thermodynamic quantities \cite{Jokela:2018ers} and transport coefficients \cite{Hoyos:2021njg}, and have been also used in some far-from-equilibrium calculations, see e.g. \cite{Ecker:2019xrw}. 
Most of these V-QCD models have been mainly applied in the literature to study the physics involving neutron stars and QCD matter at high densities, see also the recent review \cite{Hoyos:2021uff}.

The class of EMD models reviewed here may be viewed as Taylor-expanded versions of the more general class of V-QCD models with vanishing tachyon field (see e.g. the discussion in section 3.2 of Ref. \cite{Hoyos:2021uff}). 
However, it is important to stress that the details involved in the holographic constructions may lead to considerably different results. 
By comparing the fitting results for the EMD model of Refs. \cite{Critelli:2017oub,Grefa:2021qvt,Grefa:2022sav,Rougemont:2018ivt} with the LQCD results in Fig. \ref{fig:EoS0} and in the bottom panel of Fig. \ref{fig:chi2B0}, one can see that the EMD model provides a better description of first principles lattice results on the QCD thermodynamics than the several different V-QCD models considered in Fig. \ref{fig:V-QCD}. In particular, for the trace anomaly of the energy-momentum tensor, one notices in Fig. \ref{fig:V-QCD} (a) that the different V-QCD constructions miss even qualitatively the correct LQCD behavior for this observable below the pseudocritical temperature. Indeed, while in actual QCD with $2+1$ flavors there is no phase transition at $\mu_B=0$ between the hadron gas and QGP regimes, but just an analytical crossover \cite{Aoki:2006we,Borsanyi:2016ksw}, in the holographic V-QCD approach there is a first-order phase transition \cite{Jokela:2018ers}, which is reminiscent from the ihQCD backgrounds embedded in such constructions. Therefore, keeping in mind the limitations and shortcomings stated in section \ref{sec:EMD},  it is fair to say that the EMD class of holographic models discussed in this review remains the leading description to provide a quantitative description of lattice results on actual QCD thermodynamics with $2+1$ dynamical flavors with physical quark masses, both at zero and finite baryon density.

Another class of holographic models, but of top-down nature, which has been extensively studied in the literature, mainly connected to spectroscopic properties of QCD, is the so-called Witten-Sakai-Sugimoto model \cite{Witten:1998zw,Sakai:2004cn,Sakai:2005yt} --- see also \cite{Rebhan:2014rxa} for a review.\footnote{This top-down holographic construction stems from Type IIA instead of Type IIB superstring theory. Contrary to most gauge-gravity models, the background geometries in the Witten-Sakai-Sugimoto model are not asymptotically AdS and feature a dilaton field that diverges at the boundary, consequently, the Witten-Sakai-Sugimoto model has no ultraviolet fixed point \cite{Hoyos:2021uff}.} This kind of holographic model has not been shown to be able to provide an accurate quantitative description of first principles lattice results of hot QCD thermodynamics with dynamical quark flavors. On the other hand, in Ref. \cite{Kovensky:2021kzl} the Witten-Sakai-Sugimoto approach has been employed to provide a phenomenologically realistic description of cold and dense nuclear matter at zero temperature, which is in good agreement with some known theoretical and observational constraints regarding the physics of neutron stars. See also the recent review \cite{Hoyos:2021uff} for a broad discussion on the holographic modeling of compact stars.

	\newpage
	\section{Holographic models for the hot and magnetized quark-gluon plasma}
         \label{sec:magnetic}

\hspace{0.42cm} The QCD phase diagram is not just a function of ($T,\mu_B$) but is also dependent on the chemical potentials for strangeness ($\mu_S$) and electric charge ($\mu_Q$), electromagnetic fields, the number of flavors relevant for a given physical environment, etc. By varying the centrality class of heavy-ion collisions it is possible to investigate the phase diagram of QCD in the plane of temperature and magnetic field, $(T,eB)$. The most intense magnetic fields ever created by  humankind are reached in high-energy peripheral heavy-ion collisions at RHIC ($eB_\textrm{max}\sim 5\, m_\pi^2\sim 0.09$ GeV$^2$ for Au+Au collisions at center of mass energies of $\sqrt{s_\textrm{NN}}=200$ GeV with an impact parameter of $b\sim 12$ fm) and at the LHC ($eB_\textrm{max}\sim 70\, m_\pi^2\sim 1.3$ GeV$^2$ for Pb+Pb collisions at center of mass energies of $\sqrt{s_\textrm{NN}}=2.76$ TeV with an impact parameter of $b\sim 13$ fm) \footnote{We note that $eB = 1$ GeV$^2\Rightarrow B \approx 1.69\times 10^{20}$ G.} --- see e.g. Fig. 2 in \cite{Deng:2012pc}; see also Refs. \cite{Skokov:2009qp,Pang:2016yuh,Bali:2011qj,Bloczynski:2012en,Tuchin:2013ie}. The study of the QCD matter under the influence of strong magnetic fields is also relevant in the context of the physics of magnetars \cite{Duncan:1992hi} and of the early universe \cite{Vachaspati:1991nm,Grasso:2000wj}, making it a very active research field in recent years see, e.g.,  \cite{Kharzeev:2007jp,Fukushima:2008xe,Fraga:2012rr,Dexheimer:2011pz,Kharzeev:2012ph,Andersen:2014xxa,Miransky:2015ava,Rather:2021azv,Peterson:2023bmr}.

Even though very intense magnetic fields are produced in the early stages of noncentral heavy-ion collisions, being therefore important in those initial stages, due to the receding spectator hadrons fastly leaving the collision zone, one generally expects the magnitude of such strong magnetic fields to have significantly decayed by the time the QGP is formed \cite{Huang:2015oca}.
Early papers argued that by considering effects due to the electric conductivity induced in the medium \cite{Tuchin:2013apa,Gursoy:2014aka} and the quantum nature of the sources of such fields \cite{Holliday:2016lbx}, the  decay of the magnetic field may be considerably delayed within the medium. More recently in \cite{Wang:2021oqq} it was argued that an incomplete electromagnetic response of the medium to the decaying external magnetic field that is associated with an induced electric current that is lower than expected by Ohm's law, leads to a strong suppression in the magnitude of the induced magnetic field in the medium (two orders below previous estimates in the literature done by assuming the validity of Ohm's law). This argument may help to explain the consequences of the recent STAR isobar run \cite{STAR:2021mii} where it was originally thought that strong magnetic fields would lead to the chiral magnetic effect.

Nonetheless, it is interesting to investigate the structure of the QCD phase diagram in the $(T,eB)$-plane from a theoretical perspective. At low temperatures the magnitude of the chiral condensate is enhanced with increasing magnetic fields constituting the so-called \textit{magnetic catalysis} phenomenon \cite{Gusynin:1995nb}. However, for higher temperatures slightly above the QCD crossover region, the inverse effect is observed with a reduction in the magnitude of the chiral condensate and a decreasing pseudocritical crossover temperature for increasing values of the magnetic field, known as \textit{inverse magnetic catalysis} (or \textit{magnetic inhibition}) as found in the first principles lattice QCD simulations of Refs. \cite{Bali:2011qj,Bali:2012zg,Bali:2013esa,Bruckmann:2013oba,Bali:2014kia}, see also \cite{Fukushima:2012kc}. There is also  a prediction \cite{Cohen:2013zja} that a first-order phase transition line ending at a critical point exists in the $(T,eB)$-plane of the QCD phase diagram for very high values of the magnetic field, $eB\sim 4 - 10(2)$ GeV$^2$ \cite{Endrodi:2015oba,DElia:2021yvk}, although current lattice simulations \cite{Bali:2014kia} for the QCD equation of state with $2+1$ flavors and physical values of the quark masses only found an analytic deconfinement crossover for values of $110$ MeV $< T < 300$ MeV and $eB \lesssim 0.7$ GeV$^2$.

Various holographic models have been proposed in the literature to study different aspects of strongly coupled quantum systems under the influence of external magnetic fields, with either a more qualitative view towards different physical observables calculated from holographic methods --- see e.g. \cite{DHoker:2009mmn,DHoker:2009ixq,Basar:2012gh,Critelli:2014kra,Rougemont:2014efa,Li:2016bbh,Zhang:2018mqt,Fuini:2015hba,Demircik:2016nhr,Ammon:2017ded,Cartwright:2019opv,Cartwright:2021maz,Hoyos:2011us,Fukushima:2021got,Dudal:2018rki,Dudal:2015wfn,Gursoy:2016ofp,Arefeva:2020vae,Ballon-Bayona:2022uyy,Bohra:2019ebj,Dudal:2021jav,Jena:2022nzw}, or with a more quantitative perspective aimed towards direct comparisons with results from first principles LQCD calculations --- see, for instance, \cite{Critelli:2016cvq,Rougemont:2015oea,Finazzo:2016mhm,Rougemont:2020had}.

In the present section, we focus on quantitative holographic EMD predictions for some thermodynamic and transport observables of the hot and magnetized strongly coupled QGP.

         \subsection{Magnetic Einstein-Maxwell-Dilaton models}
         \label{sec:EMDmag}

\hspace{0.42cm} The first phenomenological magnetic holographic EMD model at finite temperature with a constant external magnetic field (and $\mu_B=0$) was \cite{Rougemont:2015oea}. This model generalized the isotropic approach considered in the previous section to anisotropic EMD backgrounds with the $SO(3)$ rotation symmetry broken down to $SO(2)$ in the transverse plane to the magnetic field. 
The general form of the bulk action in this case is the same as in Eq. \eqref{eq:EMDaction}, but the Maxwell-dilaton coupling function $f(\phi)$ must be different from the case at finite temperature and baryon chemical potential. 
In the EMD model at finite $(T,\mu_B)$, $f(\phi)$ effectively represents the coupling associated with the conserved baryon current, with this coupling being dynamically fixed in the holographic setup by matching the LQCD baryon susceptibility evaluated at finite temperature and zero chemical potential, as discussed in section \ref{sec:EoS}. 
In the case of the magnetic EMD model at finite $(T,eB)$ the coupling must be associated with the electric sector, instead of the baryon sector of the dual QFT at the boundary. Then, instead of the baryon susceptibility the phenomenological input seeded to the holographic model to ``teach'' the asymptotically AdS black hole backgrounds to behave as a hot and magnetized QGP, is the LQCD magnetic susceptibility evaluated at finite temperature and zero magnetic field.\footnote{In principle, one could also choose to use the electric susceptibility, instead of the magnetic susceptibility, in order to fix the Maxwell-dilaton coupling $f(\phi)$ for the electric sector of the dual QFT at the boundary. However, as discussed in Appendix A of Ref. \cite{Rougemont:2015oea}, a simple EMD model is not versatile enough to adequately cover the entire electromagnetic sector of the QGP, in the sense that by fixing $f(\phi)$ by matching the LQCD electric susceptibility, one obtains a holographic prediction for the magnetic susceptibility in disagreement with the corresponding LQCD result, and vice-versa. Therefore, it seems unfeasible to obtain a simultaneously good description of QCD magnetic and electric response functions using a single EMD model. Consequently, in order to describe magnetic field-related phenomena, one chooses the magnetic susceptibility as a phenomenological input to fix $f(\phi)$ within the holographic EMD approach.}

We shall review the main aspects of this endeavor in the next section, but before that, paralleling the discussion made in section \ref{sec:EoS} for the improvements done through the years regarding the EMD model at finite $(T,\mu_B)$, we briefly comment below on the improvements done also in the construction of the magnetic EMD model at finite $(T,eB)$.

The original construction at finite $(T,eB)$ presented in Ref. \cite{Rougemont:2015oea} has the same set of free parameters $\{G_5,\Lambda,V(\phi)\}$ of the first generation improved EMD model of Refs. \cite{Rougemont:2015wca,Rougemont:2015ona,Finazzo:2015xwa,Rougemont:2017tlu}, meaning that both models represent the same system at finite temperature when the baryon chemical potential and the magnetic field are turned off. On the other hand, as already mentioned, the Maxwell-dilaton coupling $f(\phi)$ for the magnetic EMD model is different from the baryon dense model. In Ref. \cite{Finazzo:2016mhm} it was constructed an improved version of the magnetic EMD model (with this improved version being also used in Refs. \cite{Critelli:2016cvq,Rougemont:2020had}), where the set of free parameters and functions $\{G_5,\Lambda,V(\phi),f(\phi)\}$ was updated by performing a better matching procedure to more recent lattice results on the QCD equation of state and magnetic susceptibility at finite temperature and zero magnetic fields. The set of improved free parameters $\{G_5,\Lambda,V(\phi)\}$, originally obtained in Ref. \cite{Finazzo:2016mhm} for the improved magnetic EMD model, was later employed also in the second generation improved EMD model of Refs. \cite{Critelli:2017oub,Grefa:2021qvt,Grefa:2022sav,Rougemont:2018ivt} describing a baryon dense medium. In what follows, we mainly review the results for physical observables calculated with the improved version of the magnetic EMD model at finite $(T,eB)$ from Refs. \cite{Critelli:2016cvq,Finazzo:2016mhm,Rougemont:2020had}.

         \subsubsection{Anisotropic holographic thermodynamics}
         \label{sec:EMDmag-EoS}

\begin{figure}
\begin{center}
\begin{tabular}{c}
\subfigure[]{\includegraphics[width=0.45\textwidth]{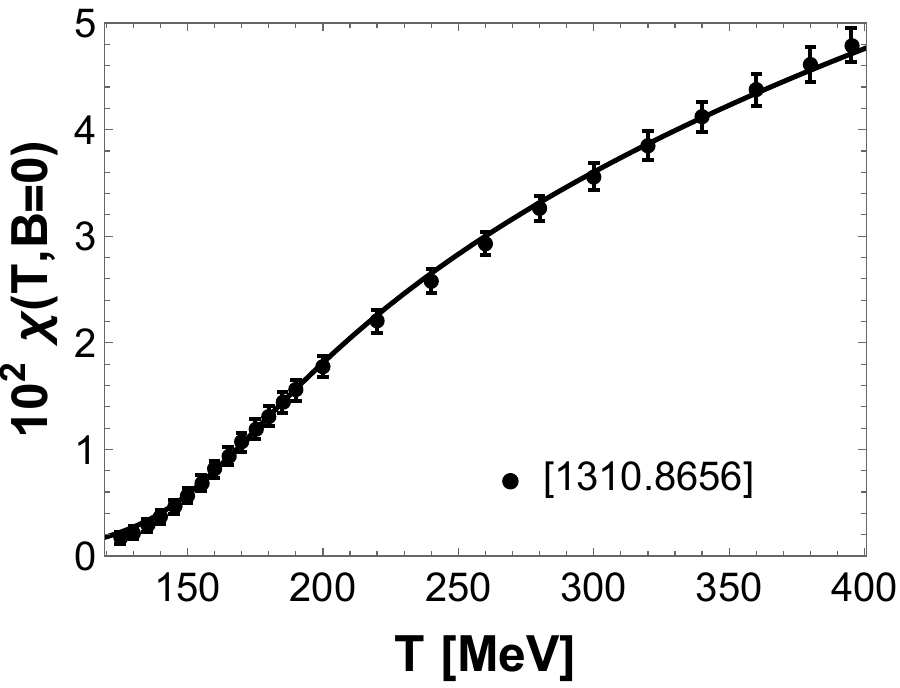}} 
\end{tabular}
\begin{tabular}{c}
\subfigure[]{\includegraphics[width=0.45\textwidth]{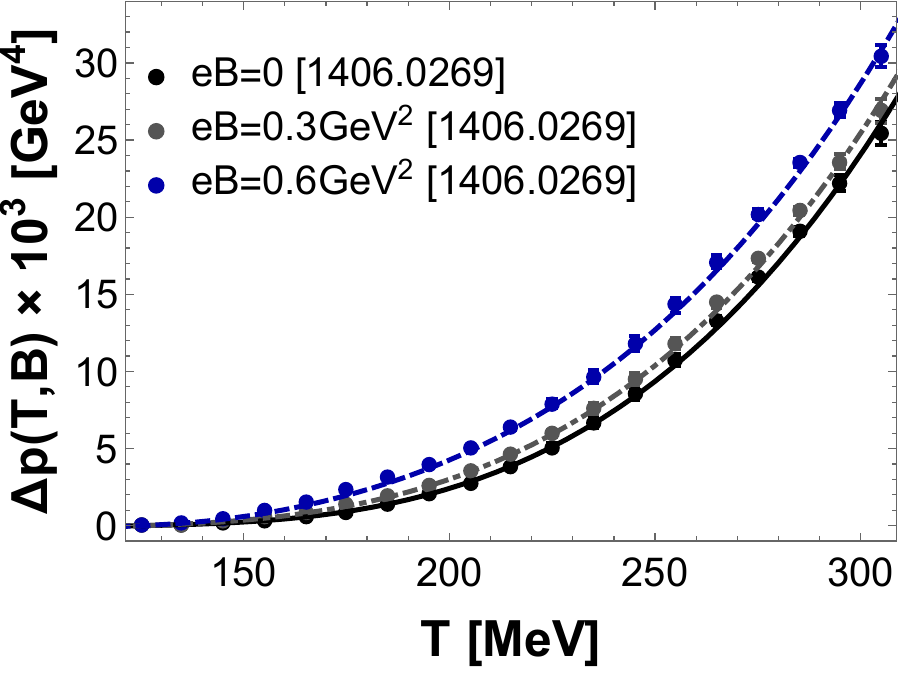}} 
\end{tabular}
\begin{tabular}{c}
\subfigure[]{\includegraphics[width=0.45\textwidth]{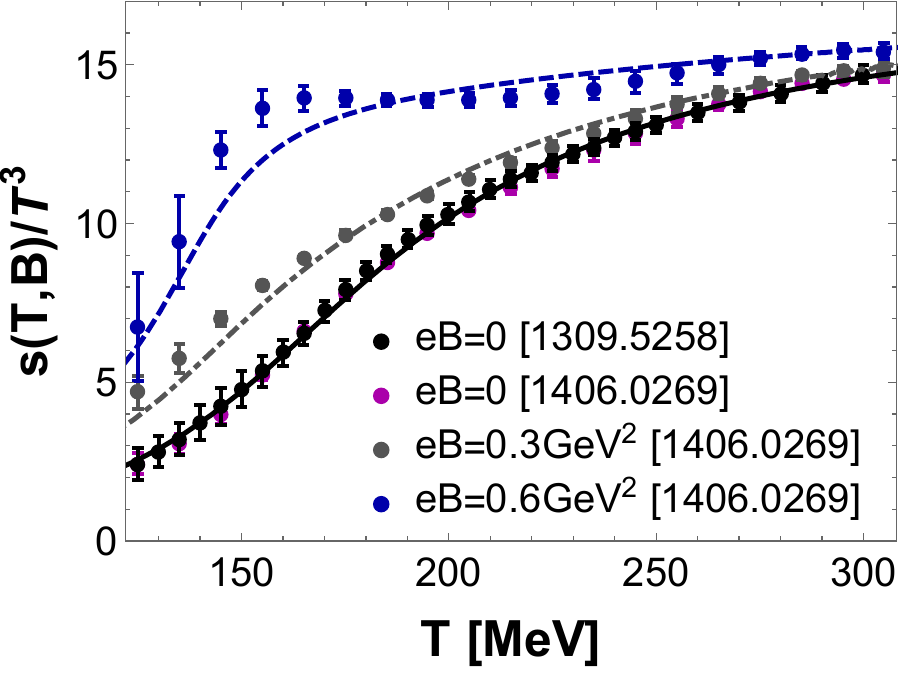}} 
\end{tabular}
\begin{tabular}{c}
\subfigure[]{\includegraphics[width=0.45\textwidth]{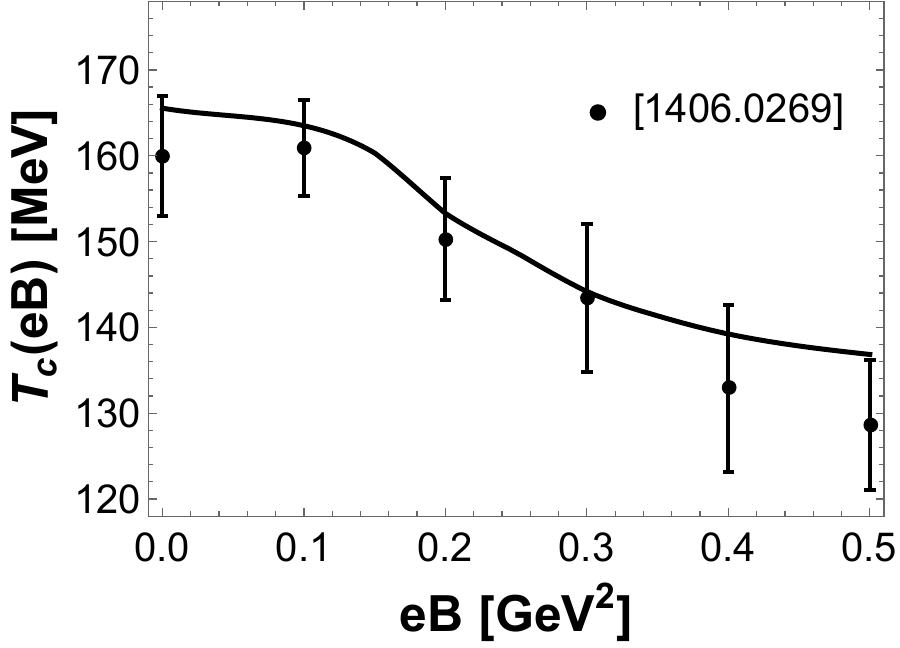}} 
\end{tabular}
\begin{tabular}{c}
\subfigure[]{\includegraphics[width=0.45\textwidth]{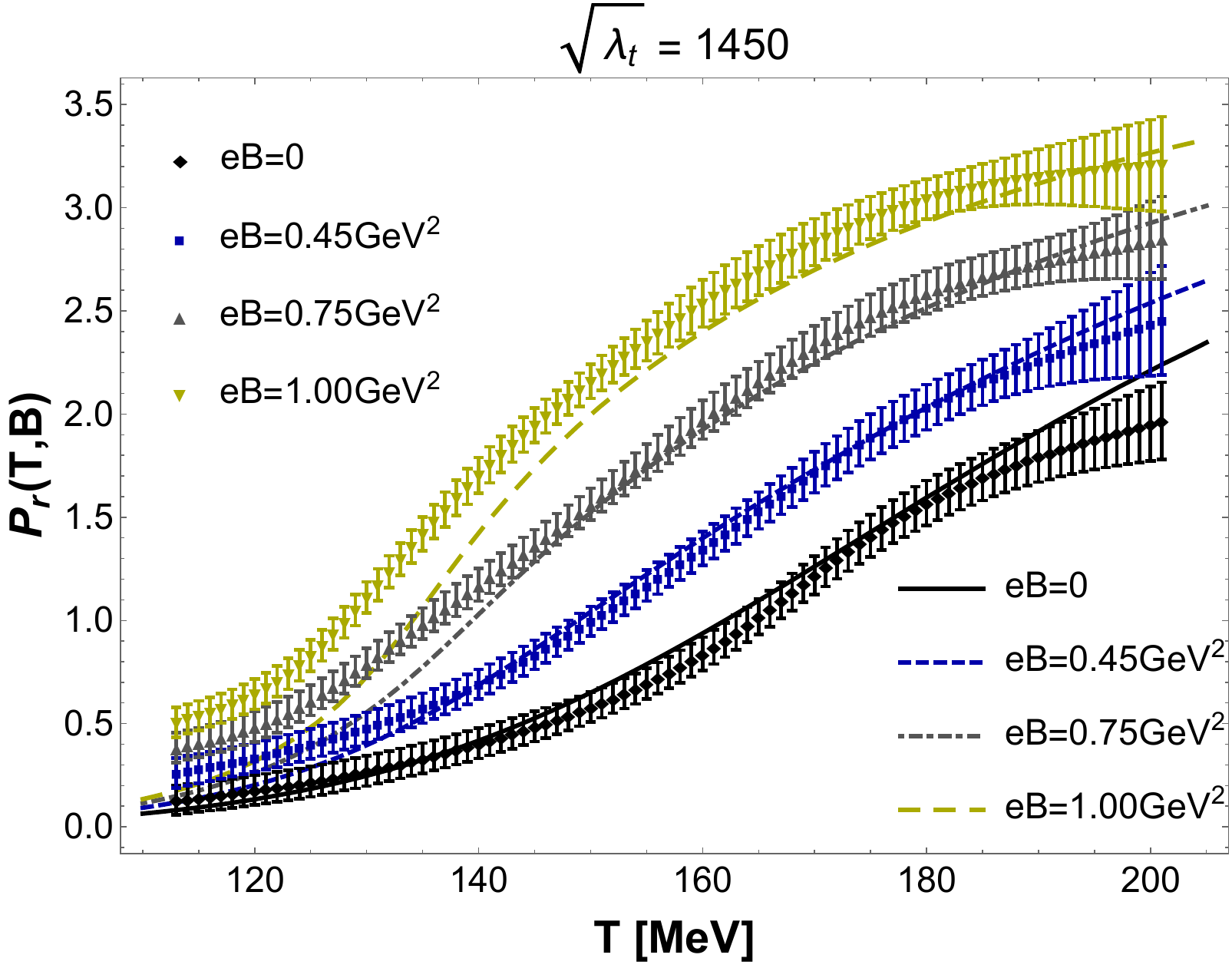}} 
\end{tabular}
\begin{tabular}{c}
\subfigure[]{\includegraphics[width=0.45\textwidth]{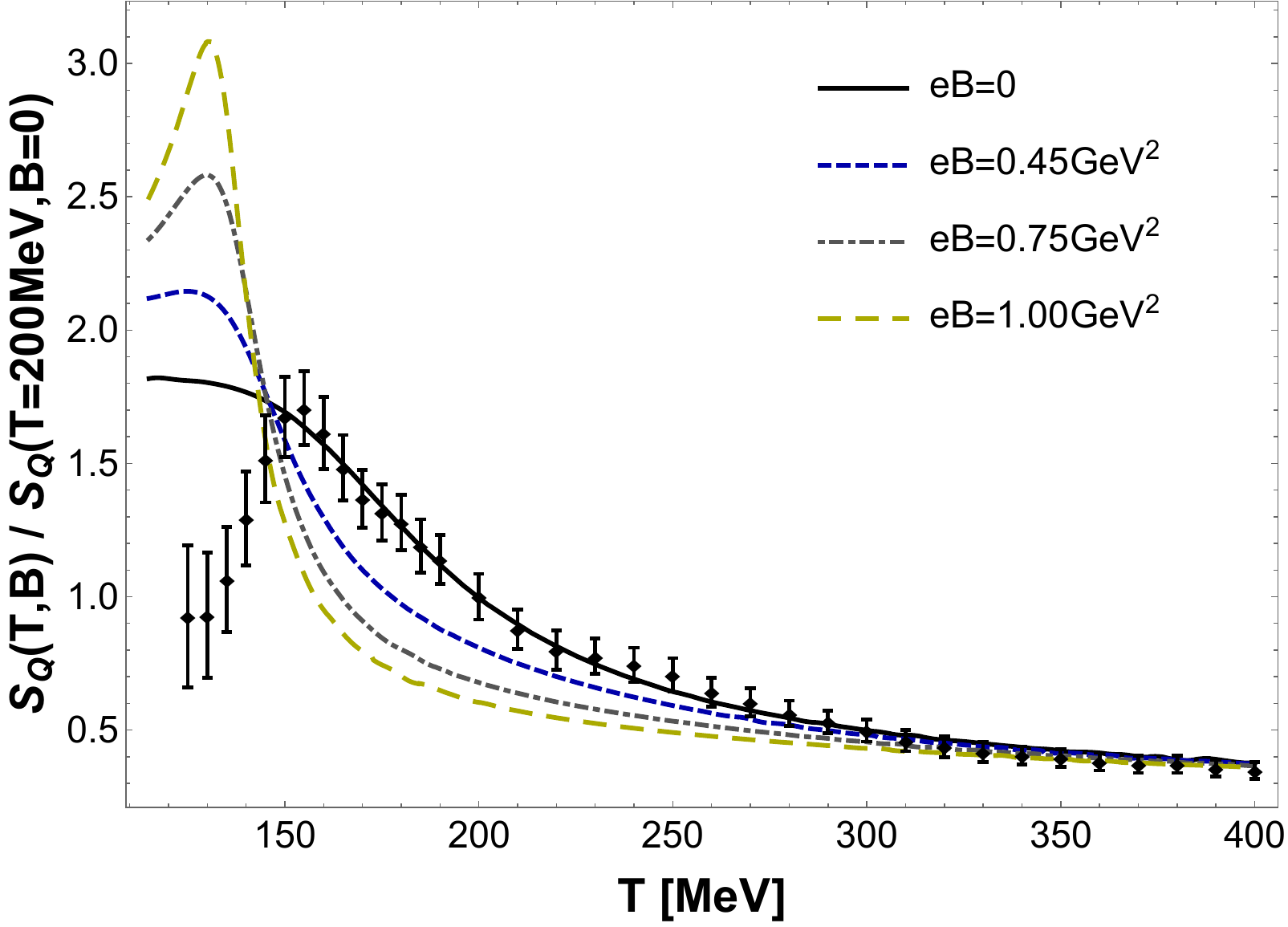}} 
\end{tabular}
\end{center}
\caption{\textbf{From Ref. \cite{Finazzo:2016mhm}:} (a) Holographic EMD magnetic susceptibility compared to LQCD results from \cite{Bonati:2013vba}. EMD predictions for (b) the pressure difference, $\Delta p(T,eB)\equiv p(T,eB)-p(T=125\textrm{MeV},eB)$, (c) the entropy density, and (d) the crossover temperature (extracted from the inflection of $s/T^3$), compared to LQCD results from \cite{Borsanyi:2013bia,Bali:2014kia}. \textbf{From Ref. \cite{Critelli:2016cvq}:} EMD predictions for: (e) the renormalized Polyakov loop (with a large `t Hooft coupling, $\sqrt{\lambda_t}=1450$) and (f) the heavy quark entropy, compared to LQCD results from \cite{Bruckmann:2013oba,Endrodi:2015oba,Bazavov:2016uvm}.}
\label{fig:EMD+Bthermo}
\end{figure}

\hspace{0.42cm} The general EMD equations of motion obtained from the bulk action \eqref{eq:EMDaction} are given by Eqs. \eqref{eq:EinsteinEqs} --- \eqref{eq:DilatonEq}. The presence of a constant external magnetic field, which we arbitrarily take to be directed along the $z$-axis, breaks the $SO(3)$ rotation symmetry of the dual QFT down to $SO(2)$ rotations around the direction of the magnetic field. This symmetry breaking implies that the ansatz for the bulk metric field must be anisotropic when the magnetic field is turned on. 
Thus, for the description of a hot and magnetized fluid in thermodynamic equilibrium, we take the following anisotropic and translationally invariant charged black hole ansatze for the bulk EMD fields \cite{Rougemont:2015oea,Finazzo:2016mhm},
\begin{align}
ds^2&= g_{\mu\nu}dx^\mu dx^\nu= e^{2a(r)}\left[-h(r)dt^2+dz^2\right]+e^{2c(r)}(dx^2+dy^2)+\frac{dr^2}{h(r)},\nonumber\\
\phi&=\phi(r), \qquad A=A_\mu dx^\mu=\mathcal{B}xdy\,\Rightarrow\, F=dA=\mathcal{B}dx\wedge dy,
\label{eq:anisoansatz}
\end{align}
where $\mathcal{B}$ is the constant magnetic field expressed in the numerical coordinates. By substituting the ansatze \eqref{eq:anisoansatz} into the general EMD field equations \eqref{eq:EinsteinEqs} --- \eqref{eq:DilatonEq}, one obtains the following set of coupled ordinary differential equations of motion \cite{Rougemont:2015oea,Finazzo:2016mhm},
\begin{align}
\phi''+\left(2a'+2c'+\frac{h'}{h}\right)\phi'-\frac{1}{h} \left(\frac{\partial V(\phi)}{\partial\phi}+\frac{\mathcal{B}^2e^{-4c}}{2}\frac{\partial f(\phi)}{\partial\phi}\right)&=0,\label{eq:anisophi}\\
a''+\left(\frac{14}{3}c'+\frac{4}{3}\frac{h'}{h}\right)a' +\frac{8}{3}a'^2+\frac{2}{3}c'^2+\frac{2}{3}\frac{h'}{h}c'
+\frac{2}{3h} V(\phi)-\frac{1}{6}\phi'^2&=0,\label{eq:anisoa}\\
c''-\left(\frac{10}{3}a'+\frac{1}{3}\frac{h'}{h}\right)c' +\frac{2}{3}c'^2-\frac{4}{3}a'^2-\frac{2}{3}\frac{h'}{h}a'
-\frac{1}{3h} V(\phi)+\frac{1}{3}\phi'^2&=0,\label{eq:anisoc}\\
h''+\left(2a'+2c'\right)h'&=0,\label{eq:anisoh}\\
a'^2+c'^2-\frac{1}{4}\phi'^2+\left(\frac{a'}{2}+c'\right)\frac{h'}{h}+4a'c'
+\frac{1}{2h}\left(V(\phi)+\frac{\mathcal{B}^2e^{-4c}}{2}f(\phi)\right)&=0,
\label{eq:anisoconstraint}
\end{align}
where Eq. \eqref{eq:anisoconstraint} is a constraint. The steps used to numerically solve the above equations of motion for a given pair of initial conditions $(\phi_0,\mathcal{B})$ are discussed in detail in Refs. \cite{Finazzo:2016mhm,Rougemont:2020had} (with algorithmic and numerical improvements regarding the original approach devised in \cite{Rougemont:2015oea}). 
Similarly to the EMD model at finite temperature and baryon density discussed in section \ref{sec:EoS}, one extracts the following set of ultraviolet expansion coefficients required for the holographic calculation of several thermodynamic observables: $\{h_0^\textrm{far},a_0^\textrm{far},c_0^\textrm{far},\phi_A\}$ from the numerical solutions for the background anisotropic EMD fields at finite temperature and magnetic field evaluated near the boundary. 
From these ultraviolet coefficients one can write down the following holographic formulas for the temperature $T$, the electric charge $e$ times the constant external magnetic field $B$ at the boundary (expressed in standard coordinates), and the entropy density $s$ (measured, respectively, in units of MeV, MeV$^2$, and MeV$^3$) \cite{Rougemont:2015oea,Finazzo:2016mhm,Rougemont:2020had},
\begin{align}
T=\frac{1}{4\pi\phi_A^{1/\nu}\sqrt{h_0^\textrm{far}}}\Lambda,\qquad
eB=\frac{e^{2\left(a_0^\textrm{far}-c_0^\textrm{far}\right)} \mathcal{B}}{\phi_A^{2/\nu}}\Lambda^2,\qquad
s=\frac{2\pi e^{2\left(a_0^\textrm{far}-c_0^\textrm{far}\right)}}{\kappa_5^2 \phi_A^{3/\nu}}\Lambda^3,\label{eq:anisothermo}
\end{align}
where the energy scale $\Lambda$, as well as the $5D$ Newton's constant and the dilaton potential are the same as given in Eq. \eqref{eq:EMDV}. 
In order to fix the Maxwell-dilaton coupling function $f(\phi)$ for the magnetic EMD model at finite temperature and magnetic field, one needs to dynamically match the holographic magnetic susceptibility at finite temperature and zero magnetic field with the corresponding LQCD result. 
As discussed in \cite{Rougemont:2015oea}, the holographic EMD formula for the regularized magnetic susceptibility evaluated at finite temperature and zero magnetic field may be written as follows in the numerical coordinates,\footnote{One should ideally take $T_{\textrm{low}}=0$, however, due to numerical difficulties in reaching exactly the vacuum geometry in the EMD model, we numerically subtract a zero magnetic field background geometry with a small but nonzero temperature, similarly to what was done in Eq. \eqref{eq:Papprox} for the calculation of the pressure.}
\begin{align}
\chi(T,B=0)=\chi_\textrm{bare}(T,B=0)-\chi_\textrm{bare}(T_\textrm{low},B=0)=-\frac{1}{2\kappa_5^2}\left[\left(\frac{1}{\sqrt{h_0^{\textrm{far}}}} \int_{r_{\textrm{start}}}^{r^{\textrm{var}}_{\textrm{max}}} dr f(\phi(r))\right)\biggr|_{T,B=0}-\left(\textrm{same}\right)\biggr|_{T_{\textrm{low}},B=0} \right]_{\textrm{on-shell}},
\label{eq:magsusc}
\end{align}
where $r^{\textrm{var}}_{\textrm{max}}\equiv\sqrt{h_0^{\textrm{far}}}\left[\tilde{r}^{\textrm{fixed}}_{\textrm{max}}- a_0^{\textrm{far}}+\ln\left(\phi_A^{1/\nu}\right)\right]$, with $\tilde{r}^{\textrm{fixed}}_{\textrm{max}}$ being a fixed ultraviolet cutoff in standard coordinates which must be chosen such that the upper limits of integration in Eq.\ \eqref{eq:magsusc} satisfy $r_{\textrm{conformal}}\le r^{\textrm{var}}_{\textrm{max}}\le r_{\textrm{max}}$ for all the background geometries under consideration. We remark that $r_{\textrm{conformal}}$ is a value of the radial coordinate\footnote{Typically, $r_{\textrm{conformal}}\sim 2$.} where the background geometry already reached the conformal AdS$_5$ ultraviolet fixed point (within some numerical tolerance), and $r_{\textrm{max}}\ge r_{\textrm{conformal}}$ is the maximum value of the radial coordinate up to which we perform the numerical integration of the bulk equations of motion. By taking as phenomenological input the LQCD magnetic susceptibility at finite temperature and zero magnetic field with $2+1$ flavors and physical values of the quark masses from Ref. \cite{Bonati:2013vba}, one may fix the form of the Maxwell-dilaton coupling function as follows \cite{Finazzo:2016mhm},
\begin{align}
f(\phi)=0.95\,\textrm{sech}(0.22\phi^2-0.15\phi-0.32),
\label{eq:fB}
\end{align}
with the result displayed in Fig. \ref{fig:EMD+Bthermo} (a).

Also in Fig. \ref{fig:EMD+Bthermo}, we show the predictions from the magnetic EMD model at finite $(T,eB)$ \cite{Finazzo:2016mhm} compared to the LQCD results  from \cite{Bali:2014kia} for (b) the pressure difference, $\Delta p(T,eB)\equiv p(T,eB)-p(T=125\textrm{MeV},eB)$, (c) the normalized entropy density $s/T^3$ (we also show the LQCD results from \cite{Borsanyi:2013bia} at $B=0$), and (d) the crossover temperature as a function of the magnetic field, as extracted from the inflection of $s/T^3$. For the values of the magnetic field considered there is no actual phase transition between the hadronic and partonic regimes of the hot and magnetized QCD matter, just an analytic crossover. Contrary to the EMD model at finite $(T,\mu_B)$ from Refs. \cite{Critelli:2017oub,Grefa:2021qvt,Grefa:2022sav}, whose phase diagram has been deeply investigated, the phase diagram of the magnetic EMD model at finite $(T,eB)$ from Refs. \cite{Critelli:2016cvq,Finazzo:2016mhm,Rougemont:2020had} still remains largely unexplored. One challenge, however, is that the magnetic EMD model typically requires a much larger set of background black hole solutions than the baryon dense model in order to allow for smooth interpolations of physical observables as functions of $T$ and $eB$.

Some comments akin to what was discussed before for the baryon dense EMD model are in order at this point. The holographic renormalization procedure for the magnetic EMD models is still not implemented in the literature, consequently, one faces limitations in what can be currently calculated with such models. As before, the pressure (and the energy density) cannot be extracted directly from the renormalized on-shell boundary action, since this quantity is still not available. However, similarly to what was done in Eq. \eqref{eq:Papprox} for the baryon dense EMD model, one may evaluate the pressure as the temperature integral of the entropy density in Eq. \eqref{eq:anisothermo} calculated with the magnetic field held fixed. As discussed in detail in Section 2 of \cite{Bali:2014kia}, such a procedure gives the isotropic pressure in the so-called ``$B$-scheme'', where the magnetic field is held fixed during compression, with the pressure being the response function of the system to such a compression. Correspondingly, this also gives the anisotropic longitudinal pressure in the direction of the magnetic field in the so-called ``$\Phi$-scheme'', where it is the magnetic flux that is held fixed during a compression. In the $\Phi$-scheme, the transverse pressures to the direction of the magnetic field depend on the magnetization of the medium, which requires holographic renormalization of the bulk action to be evaluated through the gauge-gravity duality, and that has not been calculated yet in Refs. \cite{Rougemont:2015oea,Finazzo:2016mhm}. Moreover, without the renormalized on-shell action, one is also currently missing the calculation of the energy density at finite magnetic field in the magnetic EMD models. Once the holographic renormalization procedure is implemented for the magnetic EMD models, it would be also interesting to check whether they satisfy a conjectured universal asymptotic scaling for the ratio between the transverse and longitudinal anisotropic pressures explicitly found in Ref. \cite{Endrodi:2018ikq} to hold for QCD and the magnetized SYM plasma.

In Ref. \cite{Critelli:2016cvq}, the holographic magnetic EMD model at finite $(T,eB)$ was further employed to calculate the magnitude of the expectation value of the renormalized Polyakov loop operator \cite{Polyakov:1978vu,tHooft:1977nqb,tHooft:1979rtg,Svetitsky:1982gs},\footnote{The holographic renormalization procedure for the calculation of the Polyakov loop involves only the on-shell Nambu-Goto (NG) action for a probe string extending from an isolated quark at the boundary up to the background black hole horizon deep into the bulk, and not the bulk action (which generates the black hole backgrounds, over which the probe string described by the NG action is defined) \cite{Bak:2007fk,Noronha:2009ud,Finazzo:2013rqy,Finazzo:2014zga}.} $P_r(T,eB)=|\langle\hat{L}_P\rangle_r|=e^{-F_Q^r(T,eB)/T}$, where $F_Q^r(T,eB)$ is the renormalized free energy of a single static heavy quark at the boundary.\footnote{The renormalization scheme at nonzero magnetic field employed in Ref. \cite{Critelli:2016cvq} was the same one used in the LQCD simulations of Refs. \cite{Bruckmann:2013oba,Endrodi:2015oba}.} In holography, this quantity depends on the `t Hooft coupling coming from the NG action, which in a bottom-up setup is taken as an extra free parameter. Since $\sqrt{\lambda_t}=L^2/\alpha'=\left(L/l_s\right)^2$,\footnote{See the discussion in section \ref{sec:preholo}.} where $l_s$ is the fundamental string length and $L$ is the asymptotic AdS radius (which is set here to unity), one expects that in the classical gauge-gravity regime of the holographic duality the `t Hooft coupling should be large, since in this limit, $l_s\ll L$. Indeed, by matching the overall magnitude of the holographic Polyakov loop, $P_r(T,eB)$, with the corresponding LQCD results from \cite{Bruckmann:2013oba,Endrodi:2015oba}, as illustrated in Fig. \ref{fig:EMD+Bthermo} (e), in Ref. \cite{Critelli:2016cvq} it was fixed the large value $\sqrt{\lambda_t}=1450$, which hints at a nontrivial consistency between top-down theoretical expectations and bottom-up phenomenological results within this holographic approach. Furthermore, one also notices that the magnetic EMD model provides a reasonable description of the LQCD results for the Polyakov loop in the deconfined regime of QCD matter corresponding to the strongly coupled hot and magnetized QGP, for magnetic fields up to $eB\lesssim 1$ GeV$^2$ with $T\gtrsim 150$ MeV.

Also in Ref. \cite{Critelli:2016cvq},  the holographic EMD prediction for the heavy quark entropy, $S_Q(T,eB)=-\partial F_Q^r(T,eB)/\partial T$ was computed. The ratio between any two different values of $S_Q$ is particularly interesting because it does not depend on the extra free parameter $\sqrt{\lambda_t}$ present in the holographic calculation of the Polyakov loop. Consequently, once the background black hole solutions are obtained, there are no extra free parameters to fix in such a calculation. In Fig. \ref{fig:EMD+Bthermo} (f), there are shown the EMD predictions for the ratio $S_Q(T,eB)/S_Q(T=200\textrm{MeV},eB=0)$, with the result at zero magnetic field being compared to the corresponding available LQCD result from \cite{Bazavov:2016uvm}. Interestingly enough, the EMD prediction at $B=0$ is in perfect quantitative agreement with the LQCD result in the deconfined regime for $T\gtrsim150$ MeV, while completely missing the correct behavior for the heavy quark entropy in the confined hadronic regime.

The disagreement found with the lattice results for the Polyakov loop and the heavy quark entropy below the crossover temperature in the hadronic regime, in clearly contrast to the quantitative agreement found above the crossover temperature in the partonic regime, is a consequence of the fact that the holographic EMD model is suited to describe the deconfined QGP phase of hot QCD matter but not the confined hadronic phase. It is also worthy to mention that the holographic results for these two physical observables compared to available LQCD data also attest the nature of the scalar field considered in the EMD action as being the dilaton field. This is so since in Ref. \cite{Critelli:2016cvq} it was also investigated the possibility that the scalar field $\phi$ in the EMD action is not the dilaton, in which case the bulk metric in string and Einstein frames would be the same for the EMD model. With such an approach, the holographic EMD results for the Polyakov loop and the heavy quark entropy become different from the corresponding LQCD data even at the qualitative level. As discussed in \cite{Critelli:2016cvq}, since the holographic EMD thermodynamics is matched to lattice data for the QCD equation of state and magnetic susceptibility at $B=0$, and since the only way to obtain results for the Polyakov loop and the heavy quark entropy in compatibility with LQCD data is by interpreting the scalar field in the EMD action as being the dilaton, such an analysis clearly sets the physical meaning of the scalar field $\phi$ in the bulk of the holographic correspondence for phenomenological EMD models.

The results from the magnetic EMD model at finite $(T,eB)$ displayed in Fig. \ref{fig:EMD+Bthermo}, and the results from the EMD model at finite $(T,\mu_B)$ shown in Figs. \ref{fig:EMD-ICs-EoS}, \ref{fig:chi68-CEP}, \ref{fig:hydrotransport} (d), compared to first principles LQCD results on several thermodynamic observables and transport coefficients posteriors from Bayesian analyses using heavy-ion data, comprise the main argument for the actual phenomenological applicability of EMD holography in the description of many aspects of the  hot QGP produced in heavy-ion collisions. 
These results are interesting mainly due to the following reasons:
\begin{itemize}
    \item The class of relatively simple bottom-up holographic EMD constructions reviewed here may be used to make physically reasonable predictions for the QGP, providing not only qualitative insight but also some quantitatively reliable results, which may extend beyond the current reach of first principles approaches in QCD.
    \item As a class of bottom-up holographic constructions, the phenomenological EMD models reviewed here provide further evidence that the holographic dictionary may be useful in practice even when the precise form of the holographic dual QFT at the boundary of the higher dimensional bulk spacetime is unknown.
    \item  Even though the precise holographic dual is unknown, the results reviewed here show that this holographic dual must be some effective $4D$ strongly coupled QFT which very closely mimics several aspects of QCD. While EMD holography differs from QCD in several aspects (e.g., concerning the lack of asymptotic freedom and the thermodynamic behavior in the confining hadronic regime), it is still able to capture several other key features of QCD.
\end{itemize}

         \subsubsection{Anisotropic holographic transport coefficients}
         \label{sec:EMDmag-transport}

\hspace{0.42cm} The presence of an external magnetic field (or, more generally, of any source of anisotropy) in the medium splits the transport coefficients into several anisotropic components, when compared to the more simple case of an isotropic medium. 
Holographic analyses regarding the anisotropic heavy quark drag forces and the Langevin momentum diffusion coefficients, and also the anisotropic jet quenching parameters involving light partons, were done e.g. in Refs. \cite{Li:2016bbh,Finazzo:2016mhm,Rougemont:2020had,Giataganas:2012zy,Ammon:2012qs,Jahnke:2015obr,Giataganas:2013zaa,Giataganas:2013hwa,Chakrabortty:2013kra}. 
Additionally, anisotropic shear and bulk viscosities were analyzed in \cite{Critelli:2016ley,Critelli:2014kra,Finazzo:2016mhm,Rebhan:2011vd,Jain:2014vka,Jain:2015txa}. 
At this time systematic checks of these transport coefficients have not yet been performed, since the field of relativistic magnetohydrodynamics is currently under intense development \cite{Huang:2009ue,Huang:2011dc,Roy:2015kma,Chandra:2015iza,Pu:2016ayh,Inghirami:2016iru,Grozdanov:2016tdf,Roy:2017yvg,Hernandez:2017mch,Denicol:2018rbw,Armas:2018atq,Armas:2018zbe,Denicol:2019iyh,Biswas:2020rps,Most:2021rhr,Most:2021uck,Panda:2020zhr,Panda:2021pvq,Vardhan:2022wxz}.
The purpose of the present section is to briefly review some of the main results obtained in Refs. \cite{Finazzo:2016mhm,Rougemont:2020had} regarding the magnetic EMD predictions at finite $(T,eB)$ for some transport coefficients of the strongly coupled hot and magnetized QGP.

The holographic formulas of the anisotropic heavy quark drag forces and Langevin momentum diffusion coefficients can be found in Appendix A of \cite{Finazzo:2016mhm}, and then applied to the magnetic EMD model at finite $(T,eB)$ as done in sections III.B and III.C of the same reference. The general conclusion\footnote{Which was shown, in section II of \cite{Finazzo:2016mhm}, to also hold for the top-down magnetic brane model of Ref. \cite{DHoker:2009mmn}.} is that energy loss and momentum diffusion for heavy quarks traversing a strongly coupled anisotropic plasma are enhanced by the presence of an external magnetic field, being larger in transverse directions than in the direction of the magnetic field. In Ref. \cite{Rougemont:2020had} it was found that also the anisotropic jet quenching parameters for light partons display an overall enhancement with increasing values of the external magnetic field, with the phenomenon of transverse momentum broadening being larger in transverse directions than in the direction of the magnetic field.\footnote{These conclusions were shown in \cite{Rougemont:2020had} to also hold for the top-down magnetic brane model of Ref. \cite{DHoker:2009mmn}.} Consequently, one generally predicts more energy loss for heavy and light partons traversing a strongly coupled quantum medium in the presence of an external magnetic field.

\begin{figure}
\begin{center}
\includegraphics[width=0.45\textwidth]{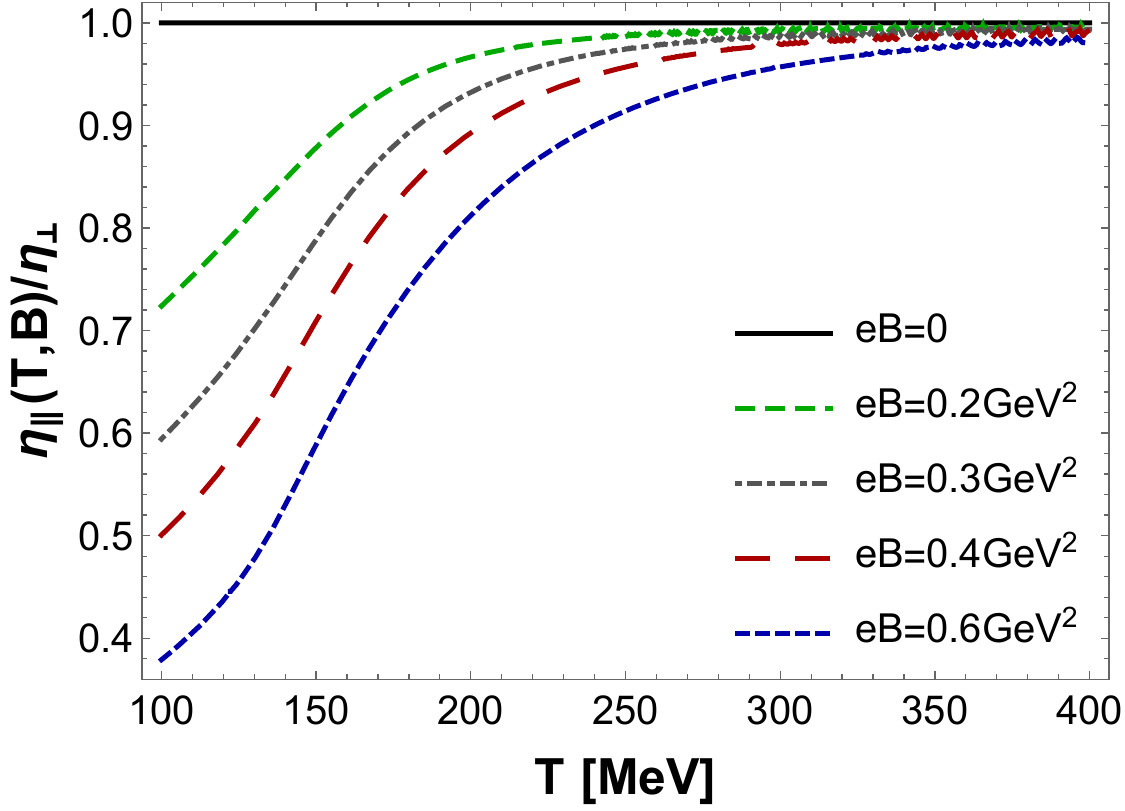} 
\end{center}
\caption{\textbf{From Ref. \cite{Finazzo:2016mhm}:} Ratio between the shear viscosities parallel and perpendicular to the direction of the external magnetic field.}
\label{fig:anisoshear}
\end{figure}

The holographic formulas for the anisotropic shear viscosities in the plane transverse to the magnetic field, $\eta_\perp$, and along the direction of the magnetic field, $\eta_\parallel$, were derived in \cite{Critelli:2014kra} and reviewed in Appendix A of \cite{Finazzo:2016mhm}. The anisotropic $\eta/s$ ratios are then
\begin{align}
\frac{\eta_\perp}{s} =  \frac{1}{4\pi},\qquad \frac{\eta_\parallel}{s} =  \frac{1}{4\pi}\, \frac{g_{zz}(r_H)}{g_{xx}(r_H)},
\label{eq:anisoshear}
\end{align}
from which one recovers the isotropic result $\eta_\perp/s=\eta_\parallel/s\equiv\eta/s=1/4\pi$ when $B=0$, since in this case the background metric is isotropic $g_{zz}=g_{xx}$. 
At nonzero magnetic fields, only $\eta_\parallel/s$ varies with the value of the external magnetic $B$ while $\eta_\perp/s=1/4\pi$ is constant.  
In Fig.\ \ref{fig:anisoshear} the results for the ratio $\eta_\parallel/\eta_\perp$ in the magnetic EMD model at finite $(T,eB)$ are shown.
 The anisotropic shear viscosity is lower in the direction parallel the magnetic field than in the transverse plane, with its magnitude being reduced as one increases the value of $B$. 
 Along the external magnetic field direction, a strongly coupled magnetized medium becomes progressively closer to the idealized perfect fluid limit field by enhancing the value of the magnetic field.\footnote{See e.g. \cite{Giataganas:2017koz} for a holographic confining dilatonic model where the anisotropy is driven by an axion field, also leading to a reduction of the component of the specific shear viscosity parallel to the anisotropic direction. See also \cite{Baggioli:2023yvc} for a discussion about the breaking of rotational invariance and its effects on the calculation of the shear viscosity of a p-wave superfluid model, where the rotational symmetry breaking does not lead to a value of $\eta/s$ below $1/4\pi$.}

	\newpage
	\section{Summary and outlook}
         \label{sec:outlook}

\hspace{0.42cm} In this work, we provided an up-to-date review of quantitative holographic EMD models for the hot and strongly coupled QGP produced in relativistic heavy-ion collisions. 
We reviewed both isotropic EMD constructions at finite temperature and baryon chemical potential with vanishing electromagnetic fields and anisotropic EMD backgrounds at finite temperature and magnetic field with zero chemical potential. 
Evidence that the holographic duality can quantitatively provide reliable predictions for the hot and deconfined QGP phase of QCD, depending on the class(es) of gauge-gravity models considered and how their free parameters are fixed by phenomenological inputs, was discussed. These key results highlight precisely this evidence for the reliability of the EMD predictions:

\begin{enumerate}[i)]
\item \textbf{EMD model for the $(T,\mu_B)$-plane of QCD:} in Figs. \ref{fig:EMD-ICs-EoS} and \ref{fig:chi68-CEP} we displayed, respectively, the holographic predictions for the equation of state at finite temperature and baryon chemical potential, and for the 6th and 8th order baryon susceptibilities at $\mu_B=0$, compared to state-of-the-art first principles LQCD results; and in Fig. \ref{fig:hydrotransport} (d), we have shown the EMD prediction for the bulk viscosity to entropy density ratio at vanishing baryon density compared to the profiles favored by the latest phenomenological multistage models that simultaneously describes several different experimental data from relativistic heavy-ion collisions. As an isotropic and translationally invariant holographic model with two derivatives of the metric field in the bulk gravity action, the model naturally encompasses a small  shear viscosity, $\eta/s=1/4\pi$, compatible with the overall magnitude estimated for the strongly coupled QGP produced in heavy-ion collisions. A number of other holographic EMD models are currently available in the literature which have been also shown to successfully describe LQCD results at the quantitative level, such as the works presented in Refs. \cite{Knaute:2017opk,Cai:2022omk,Li:2023mpv}.

\item \textbf{Magnetic EMD model for the $(T,eB)$-plane of QCD:} in Fig. \ref{fig:EMD+Bthermo} we displayed the holographic predictions for the anisotropic equation of state, the crossover transition temperature, the renormalized Polyakov loop, and the heavy quark entropy at finite temperature and magnetic field compared to the available first principles LQCD results.
\end{enumerate}

The holographic EMD model allows one to go beyond the current capabilities of LQCD simulations. For instance, one prediction of this model is the existence of a critical end point. While different competing EMD models do provide differences in the predicted location of this critical point after fitting to LQCD results for $\mu_B=0$, they all lead to the existence of a critical point in an approximately similar region of the QCD phase diagram. Such a spread of critical points clearly motivates a more systematic investigation of different parametrizations of the free functions and parameters of the bottom-up class of holographic EMD models through Bayesian statistical inference. 

A detailed Bayesian analysis of such models is presented in \cite{Hippert:2023bel}, but preliminary results were discussed in section \ref{sec:bayes}. 
This Bayesian analysis considered uniform prior distributions of the free parameters. 
Using the LQCD results for the entropy density and the baryon susceptibility at $\mu_B=0$ as constraints, the posterior distributions for the free parameters of the holographic EMD setup become strongly constrained, as shown in Table \ref{tab:bayesian}. 
Thousands of different EMD models were generated  within the constrained posterior distributions that provided holographic predictions for the behavior of the QCD equation of state at finite temperature and baryon density. 
 The resulting equation of state  has remarkably  thin bands, as shown in Fig. \ref{fig:Bayesian}, which are  in quantitative agreement with state-of-the-art lattice results for the QCD equation of state also at finite baryon density.\footnote{Although some deviations exist for the baryon charge density at high temperatures and high baryon chemical potentials, as depicted in Fig. \ref{fig:Bayesian}. However, that also corresponds to the regime where the expansion scheme may begin to break down in lattice QCD and/or weaker coupling may be relevant.} For a complete analysis considering regions of the phase diagram beyond the reach of current lattice simulations and the distribution of critical points predicted by a broader class of holographic EMD models see \cite{Hippert:2023bel}.

A critical assessment of the most relevant limitations and the drawbacks of holographic approaches to the description of hot QCD phenomenology were also discussed in detail. First, classical holographic gauge-gravity models with two derivatives of the metric field in the bulk gravity action lack asymptotic freedom, with the dual effective QFT at the boundary of the higher dimensional bulk spacetime being strongly coupled at all energy scales. This is explicitly manifest in the temperature-independent value of $\eta/s=1/4\pi$ found in these models, which is in contrast to the gas-like pQCD results at asymptotically high temperatures. 
Instead of a trivial ultraviolet fixed point, classical holographic gauge-gravity models which are asymptotically AdS feature a strongly coupled ultraviolet fixed point, being asymptotically safe but not asymptotically free. 
The lack of asymptotic freedom and the constant value $\eta/s=1/4\pi$ are presumably tied to the neglected contributions from massive string states and quantum string loops in the classical gravity bulk theory. 
This can be possibly improved by considering higher derivative corrections associated with massive string states in the bulk action, which in the presence of a nontrivial dilaton background has already been shown in the literature \cite{Cremonini:2012ny} to produce temperature-dependent profiles for $\eta/s$ in holographic models. However, the systematic construction of phenomenologically realistic and fully-backreacted dilatonic models with higher-order derivative corrections is a challenging task still not accomplished in the literature. 

Another very general limitation of classical holographic gauge-gravity models regards their inability to describe the thermodynamic and transport properties of the confining hadron resonance gas phase of QCD. 
This limitation is related to the large $N_c$ character of classical gauge-gravity models, in which the pressure in the confining phase is largely suppressed by a multiplicative factor of $\sim N_c^{-2}$ relatively to the deconfined QGP phase.\footnote{One very clear manifestation of such a limitation has been shown in Fig. \ref{fig:EMD+Bthermo} (f), where the holographic prediction for the heavy quark entropy was found to be in perfect agreement with the corresponding LQCD results above the pseudocritical crossover temperature, while for temperatures below the crossover region the holographic heavy quark entropy suddenly completely misses the correct LQCD behavior.} In principle, this situation can be improved by considering quantum string loops contributions to the dilatonic bulk theory. However, this task is considerably more complicated than the one discussed in the previous paragraph.

Specific limitations and drawbacks of the holographic EMD models reviewed here have been also identified in the literature. For instance, the strangeness neutrality condition realized in heavy-ion collisions is not implemented in the EMD model, as it only features a single chemical potential (in the case considered here, the baryon chemical potential). Moreover, in the investigation of the phase diagram of the EMD model of Refs. \cite{Critelli:2017oub,Grefa:2021qvt,Grefa:2022sav,Rougemont:2018ivt}  no regions were found where the square of the speed of sound exceeds its conformal limit ($c_s^2|_\textrm{CFT}=1/3$), strongly indicating that such models are inadequate to describe the dense QCD equation of state of the most massive neutron stars \cite{Bedaque:2014sqa,Tews:2018kmu,Miller:2019cac,Tan:2020ics,Landry:2020vaw,Tan:2021ahl,Altiparmak:2022bke,Mroczek:2023eff}. Furthermore, as mentioned in section \ref{sec:EMDmag}, the magnetic EMD model is not versatile enough to simultaneously describe the magnetic and the electric sectors of the QGP with a single Maxwell-dilaton coupling function $f(\phi)$.

For future work, it is important to extend dilatonic holographic approaches  to simultaneously include fully backreacted effects from conserved baryon, electric, and strangeness charges. Such an endeavor would enable the implementation of strangeness neutrality, which is relevant for applications in heavy-ion collisions. In order to pursue this task within a consistent implementation of QCD flavor symmetry in the holographic setup, the EMD class of holographic models should be substituted by a more general class of (fully backreacted) Einstein-Yang-Mills-Dilaton (EYMD) models.

Still within the class of holographic EMD models, the more complicated magnetic EMD setups at finite temperature and magnetic field remain largely unexplored. Most of its phase diagram has yet to be investigated and holographic renormalization still needs to be implemented, which will allow for the calculation of several physical observables not addressed in the present review. Additionally, a Bayesian analysis would be another important next step to understand properties at large $B$ fields (as in the case of the Bayesian analysis implemented for the isotropic setup at finite baryon density in Ref. \cite{Hippert:2023bel}).

Other important developments to be pursued in the future include the consideration of rotation effects for the strongly coupled dual plasma by taking into account more general ansatze for the bulk fields allowing for rotating and charged asymptotically AdS black holes. Also numerical simulations of far-from-equilibrium holographic dynamics \cite{Chesler:2013lia} should be further pursed, such as the consideration of holographic Bjorken flow and holographic collisions of shockwaves in the context of the phenomenologically realistic EMD models reviewed in this manuscript.
	

	\newpage
	\section*{Acknowledgements}  
	This material is based upon work supported in part by the
National Science Foundation under grants No. PHY-2208724 and No. PHY-2116686 and in part by the U.S. Department of Energy, Office of Science, Office of Nuclear Physics, under Award Number DE-SC0022023, DE-SC0021301, DE-SC0020633, DE-SC0023861. This work was supported in part by the National Science Foundation (NSF) within
the framework of the MUSES collaboration, under grant
number No. OAC-2103680. This research was supported in part by the National Science Foundation under Grant No. PHY-1748958. 


	\bibliography{mybibfile,noninspire}

	
	
		
\end{document}